\documentclass[12pt]{article}

\usepackage{graphicx}
\usepackage{amsmath}
\usepackage{amssymb}
\usepackage{amsthm}
\usepackage{pstricks}
\usepackage{color}
\usepackage{wrapfig}
\usepackage{hyperref}
\begin{document}

\begin{flushright}
CYCU-HEP-13-05 \\
KUNS-2441
\end{flushright}

\vspace{4ex}

\begin{center}

{\LARGE\bf Gauge Symmetries in Heterotic Asymmetric Orbifolds}

\vskip 1.4cm

{\large  
Florian Beye$^{1}$,
Tatsuo Kobayashi$^{2}$
and
Shogo Kuwakino$^{3}$
}
\\
\vskip 1.0cm
{\it $^1$Department of Physics, Nagoya University, Nagoya 464-8602, Japan} \\
{\it $^2$Department of Physics, Kyoto University, Kyoto 606-8502, Japan} \\
{\it $^3$Department of Physics, Chung-Yuan Christian University, 
Chung-Li 32023, Taiwan} \\

\vskip 3pt
\vskip 1.5cm

\begin{abstract}
We study heterotic asymmetric orbifold models. By utilizing the lattice engineering technique, we classify (22,6)-dimensional Narain lattices with right-moving non-Abelian group factors which can be starting points for ${\bf Z}_3$ asymmetric orbifold construction. We also calculate gauge symmetry breaking patterns.
\end{abstract}

\end{center}

\newpage


\section{Introduction}

String theory is expected to be the fundamental theory which unifies elementary particle physics and quantum gravity. Its compactification provides four-dimensional string theories and their low energy effective theories, in which required properties such as gravity and gauge interactions, as well as chiral matter can be realized. However, further properties of these effective theories depend on the geometry of the internal extra dimensions.

One of the promising compactification methods is the orbifold construction \cite{orbifold} of hetetoric string theory \cite{hetero}. In this framework, we quantize closed strings on a six-dimensional toroidal orbifold in the internal dimensions. Model building towards supersymmetric standard models and some GUT-like extended models are vastly investigated in ${\bf Z}_N$ and ${\bf Z}_N \times {\bf Z}_M$ orbifold compactification \cite{heteroph,Katsuki:1989bf,Kobayashi:2004ud,Buchmuller,Kim,Lebedev:2006kn,Blaszczyk:2009in} (for a review \cite{Nilles:2008gq}). Since heterotic string theory has a microscopic description by a Lagrangian, it is possible to investigate phenomenological properties of the four-dimensional effective theory, such as the Yukawa couplings and higher order couplings, by considering selection rules of string interactions and explicitly computing string amplitudes \cite{Hamidi:1986vh,Burwick:1990tu,Choi:2007nb,Kobayashi:2011cw}. 

We can consider an extension of the orbifolding procedure without losing string consistency: in the asymmetric orbifold construction \cite{Narain:1986qm}, orbifold actions for the left and right-movers on the string world-sheet are generalized to be independent. The generalization will expand the scope of realistic model building in heterotic string vacua. In the literature \cite{Ibanez:1987pj}, ${\bf Z}_3$ asymmetric orbifold models with one Wilson line were studied. In addition, GUT models with higher Kac-Moody levels were constructed \cite{Kakushadze:1996hi,Ito:2010df}. However, systematic searches for realistic models, e.g. supersymmetric standard models, have not been done for asymmetric orbifold vacua. Such systematic studies are quite important in  ${\bf Z}_3$ asymmetric orbifold model building as well as for e.g. ${\bf Z}_6$ or ${\bf Z}_{12}$ orbifolds. 

In this paper, we study the ${\bf Z}_3$ asymmetric orbifold compactification systematically. A starting point for model building is a Narain compacification, i.e. a Lorentzian even self-dual lattice, with (22,6) dimensions \cite{Narain:1985jj}. Unfortunately, there are infinitely many possible Narain lattices. But if one restricts to a certain class of ${\bf Z}_3$ actions, the number of lattices compatible with the action is finite. Yet, it is difficult to cover all possibilities by a blind search. However, all Euclidean even self-dual lattices in 8, 16 and 24 dimensions are classified. From these one can construct Lorentzian even self-dual lattices by using the lattice engineering technique \cite{Lerche:1988np}. For example in \cite{Ito:2010df}, the lattice engineering technique is applied in a GUT model construction. In the right-mover part of the Narain lattice we require either the $E_6$ or the $A_2^3$ lattice in order to realize the ${\bf Z}_3$ automorphism \cite{Katsuki:1989bf,Kobayashi:1991rp}. The $D_4 \times A_2$ lattice can also realize the ${\bf Z}_3$ twist by using the outer automorphism of $D_4$. Thus, we classify the (22,6)-dimensional Lorentzian even self-dual lattices with an  $E_6$ or $A_2^3$ right-mover lattice, which can be constructed using the 8, 16 and 24 dimensional Euclidean lattices through the lattice engineering technique. We also give a comment on (22,6)-dimensional Lorentzian even self-dual lattices containing a $D_4$ right-mover lattice. This allows us to fix possible Narain lattices including gauge symmetries. Then, we can break gauge symmetries by embedding the ${\bf Z}_3$ orbifold twist into the gauge space. In symmetric orbifold models, gauge symmetries after such breaking are classified by studying all breaking patterns of the $E_8$ root lattice through ${\bf Z}_N$ shifts \cite{Katsuki:1989kd}. In our case, the Narain lattices used as starting point include root lattices of several simple Lie groups possibly shifted by fundamental weights. Thus, the analysis of breaking patterns becomes more involved, but we can classify all possible gauge groups after such ${\bf Z}_3$ shift breaking. Our procedure for classification and model building can be extended to other ${\bf Z}_N$ and ${\bf Z}_N \times {\bf Z}_M$ asymmetric orbifold models.

This paper is organized as follows. In section 2, we construct (22,6)-dimensional Narain lattices which include an $E_6$ or $A_2^3$ right-moving lattice from 24-dimensional Euclidean lattices through the lattice engineering technique. These can be used for the ${\bf Z}_3$ asymmetric orbifold construction. In section 3, we consider gauge symmetry breaking patterns caused by orbifold actions. We also show the construction of a (22,6)-dimensional Narain lattice in a more detailed manner in section 4, together with a simple ${\bf Z}_3$ asymmetric orbifold model. In section 5 we present (22,6)-dimensional Narain lattices that have a right-moving $D_4 \times A_2$ group factor and comment on their gauge breaking patterns. Section 6 is devoted to conclusions and discussion.

\section{Asymmetric orbifold construction of the heterotic string}

\subsection{Heterotic asymmetric orbifolds} \label{SecHetorb}

In this work we consider the heterotic string in its bosonic construction \cite{hetero} compactified to four dimensions on an asymmetric toroidal orbifold. In the light-cone formalism, the original world-sheet CFT contains 24 bosonic left-movers $X_{\rm L}$, eight bosonic right-movers $X_{\rm R}$ and eight right-moving fermions $\Psi_{\rm R}$. The asymmetric orbifolding procedure \cite{Narain:1986qm} consists of the following steps:
\begin{description}
\item[Toroidal compactification.] The most general toroidal compactification is obtained by specifying periodic boundary conditions for $X_{\rm L}$ and $X_{\rm R}$, except for the extended dimensions we keep. These boundary conditions are encoded in a $(22, 6)$-dimensional Lorentzian lattice $\Gamma$ which represents the quantized momenta $(p_{\rm L}, p_{\rm R})$. In this context $\Gamma$ is called a Narain lattice  \cite{Narain:1985jj} and modular invariance requires it to be even and self-dual. This toroidal theory possesses $\mathcal{N}=4$ SUSY in four dimensions and has a gauge symmetry $G_{\rm L}$ of rank $22$ which is determined by the left-mover part of the Narain lattice. Henceforth, $X_{\rm L}$, $X_{\rm R}$ and $\Psi_{\rm R}$ shall denote only the compactified directions.

\item[Asymmetric twisting/shifting.] Next, a twist $\theta = (\theta_{\rm L}, \theta_{\rm R})$ in combination with a shift $V = (V_{\rm L}, V_{\rm R})$ is applied to the toroidally compactified directions, acting as follows:
\begin{align}\label{eq:twisting}
	(\theta, V) : \;\; \begin{array}{lll} X_{{\rm L}} & \to & 
	\theta_{{\rm L}} X_{{\rm L}} + V_{\rm L} \\
	X_{{\rm R}} & \to & \theta_{{\rm R}} X_{{\rm R}} + V_{\rm R}\\
	\Psi_{\rm R} & \to & \theta_{{\rm R}} \Psi_{\rm R} \end{array} \;\;.
\end{align}
The shift vector $V$ is only defined modulo a lattice vector. Here, $X_{{\rm R}}$ and $\Psi_{\rm R}$ are twisted in the same manner in order to retain world-sheet supersymmetry. Furthermore, ($\theta$,$V$) is chosen such that the above action generates the cyclic group $\mathbf{Z}_N$. For consistency, $\theta$ must be a lattice automorphism, i.e. it has to conserve the scalar product defined on $\Gamma$. The orbifold action given by eqn.~\eqref{eq:twisting} is called symmetric if $\theta_{\rm L}$ can be written as $\theta_{\rm R} \oplus \theta_{\rm 16}$ for some twist $\theta_{16}$. In our case we explicitly consider twists which are asymmetric, i.e. do not obey this requirement.
\end{description}

Here, we restrict our considerations to a class of asymmetric orbifold models that fulfills the following requirements:

\begin{enumerate}
\item In order to obtain phenomenologically viable models we only consider twists which retain $\mathcal{N}=1$ space-time SUSY. This is achieved when all eigenvalues $\lambda$ of $\theta_{\rm R}$ fulfill $\lambda \neq 1$ \cite{orbifold}. Also, as an implication $V_{\rm R}$ can be set zero. 
\item The left-movers are not twisted but only shifted, i.e. $\theta_{\rm L} = 1$. This shifting breaks the gauge group $G_{\rm L}$ of the toroidal theory to a subgroup $G_{\rm orb} \subset G_{\rm L}$ of the same rank. 
\item For simplicity, we restrict to $\mathbf{Z}_3$ asymmetric orbifolds.
\end{enumerate}

A model which obeys the above constraints is fully specified by
\begin{enumerate}
\item A suitable (22,6)-dimensional Narain lattice $\Gamma$.
\item An automorphism $\theta$ of $\Gamma$ of order 3 that acts as the identity on the 22-dimensional left-mover part, i.e. only twists the right-movers.
\item A shift vector $V$ that is zero in the right-mover direction and fulfills $3 V \in \Gamma$. 
\end{enumerate}

In the following, a Narain lattice $\Gamma$ is described by left-right combined momenta $p = (p_{{\rm L}}, p_{{\rm R}} )$ with the Lorentzian metric, namely, $p^2 =  p_{{\rm L}}^2 - p_{{\rm R}}^2 $. Modular invariance of the closed string theory restricts the momentum lattice to be even and self-dual. If we denote the Narain lattice as $\Gamma = \sum_i n_i \gamma_i$ by using a basis $\gamma_i $ and integers $n_i$, the even and self-dual conditions are given by $( n_i \gamma_i )^2 \in 2 {\bf Z}$ for all $n_i$ and $ \Gamma = \tilde{\Gamma} $ with $\tilde{\Gamma} \equiv  \sum_i m_i \tilde{\gamma}_i$. Here $\tilde{\gamma}_i$ satisfies $\gamma_i \cdot \tilde{\gamma}_j = \delta_{ij}$ and the $m_i$ are integers. 

Also, $\Gamma$ can be described by a suitable combination of root lattices and their fundamental weights, the conjugacy classes. For a simple example, the $E_8$ even self-dual lattice which has only left-moving degrees of freedom is described by the momenta $p \in \Gamma_{E_8} \equiv \sum_{i=1}^{8} n_i\alpha^{E_8}_i$ with simple roots $\alpha^{E_8}_i$ and integers $n_i$. In terms of conjugacy classes, $\Gamma_{E_8}$ is written as $0_{E_8}$. Also, the ${\rm Spin(32)}/{\bf Z}_2$ lattice is given by the 16-dimensional momenta $p \in \Gamma_{D_{16}} \cup (\Gamma_{D_{16}} + \omega_{15}^{D_{16}})$, where $\Gamma_{D_{16}} \equiv \sum_{i=1}^{16} n_i\alpha^{D_{16}}_i$ with simple roots $\alpha^{D_{16}}_i$, and with the fifteenth fundamental weight $\omega_{15}^{D_{16}}$ which corresponds to the spin representation (see figure \ref{fig:Dynkin} for the definition of the Dynkin diagram). In terms of conjugacy classes, $\Gamma_{{\rm Spin(32)}/{\bf Z}_2}$ is written as $0_{D_{16}} \cup s_{D_{16}}$.

\subsection{Lattice engineering technique}

As seen in the previous subsection, the starting point for an asymmetric orbifold construction is a $(22, 6)$-dimensional Narain lattice $\Gamma$ possessing a suitable discrete symmetry $\theta$. For the purpose of classifying such lattices, the lattice engineering technique \cite{Lerche:1988np} is useful. By this method we can construct new lattices from known ones. The lattice engineering procedure allows us to replace one of the left-moving group factors in the Narain lattice, say  $G$, by some dual group factor $\overline{G}_{{\rm dual}}$ in the right-movers, and vice versa. Here, we denote a right-moving factor by using the bar sign. The dual group factor $G_{{\rm dual}}$ is maximal so that $G \times G_{{\rm dual}} \subset G_{{\rm esd}}$, where $G_{{\rm esd}}$ is the group factor of an even self-dual lattice $\Gamma_{G_{{\rm esd}}}$. For example, the 8-dimensional even self-dual lattice $\Gamma_{E_8}$ has $G_{{\rm esd}} = E_8$. The lattice $\Gamma_{G_{{\rm esd}}}$ is described by conjugacy classes of $G_{{\rm esd}}$, and, by a suitable decomposition, also in terms of conjugacy classes of $G \times G_{{\rm dual}}$. The left-right replacement gives rise to a new lattice with different dimensions. If we start from a modular invariant lattice, the resulting lattice is also modular invariant since the two group factors $G$ and $\overline{G}_{{\rm dual}}$ have the same modular transformation properties due to their conjugacy classes. 

We show a simple example of the lattice engineering technique. Let us denote the $E_8$ even self-dual lattice in terms of $E_6 \times A_2 \subset E_8$. This leads to the decomposition
\begin{align}
	0_{E_8} = (0_{E_6},0_{A_2}) \cup (1_{E_6},1_{A_2}) \cup (2_{E_6},2_{A_2}) \;,
\end{align}
where $1$ and $2$ denote the conjugacy classes corresponding to the fundamental and the anti-fundamental weights of $E_6$ and $A_2$, respectively. From this, we find that $E_6$ and $A_2$ are dual to each other. Here, let us consider the lattice engineering $A_2\to \overline{E}_6$. We replace each conjugacy class of $A_2$ as $0_{A_2} \to 0_{\overline{E}_6}$, $1_{A_2} \to 1_{\overline{E}_6}$ and $2_{A_2} \to 2_{\overline{E}_6}$. The resulting lattice is a Narain lattice in (6,6) dimensions that we call $\Gamma_{E_6 \times \overline{E}_6}$. It contains the conjugacy classes
\begin{align}
	(0_{E_6}, 0_{\overline{E}_6}) \cup (1_{E_6}, 1_{\overline{E}_6}) 
	\cup (2_{E_6}, 2_{\overline{E}_6}). 
\end{align}
These conjugacy classes are generated by $ (1_{E_6}, 1_{\overline{E}_6})$. This lattice gives rise to an $E_6$ gauge symmetry from the left-moving $E_6$ root lattice. In this way $(d, d)$-dimensional $G_{{\rm L}} \times \overline{G}_{{\rm R}}$ lattices with $G_{{\rm L}} = G_{{\rm R}}$ can be constructed. We will use this technique iteratively to construct (22,6)-dimensional lattices. Further examples are shown in section \ref{Examples}.

We list group factors and their conjugacy classes which are relevant to  ${\bf Z}_3$ orbifold model construction in table \ref{Tab.LET}. The conjugacy classes for the left-movers and the right-movers are denoted as $c_{{\rm L}}$ and $c_{{\rm R}}$. We also list normalizations for $U(1)$s. In the same way, tables of group factors and conjugacy classes relevant for other orbifold models, e.g. ${\bf Z}_6$, can be evaluated.

\begin{table}[h]
\begin{center}
\begin{tabular}{|c|c|c|c|c|}
\hline
$G_{{\rm L}}$ & $c_{{\rm L}} $ & $\overline{G}_{{\rm R}}$ 
& $c_{{\rm R}}$ & $U(1)$ unit \\
\hline
\hline
$E_{6}$ & $(1)$ & $\overline{A}_2$ & $(1)$ & --- \\
\hline
$D_{4}$ & \begin{tabular}{c} $(v)$ \\$(s)$ \\\end{tabular} 
& $\overline{D}_4$ & \begin{tabular}{c} $(v)$ \\$(s)$ \\\end{tabular} & --- \\
\hline
$A_{2}$ & $(1)$ & $\overline{E}_6$ & $(1)$ & --- \\
\hline
$A_{2}^2$ 
& \begin{tabular}{c} $(1, 0)$ \\$(1, 2)$ \\\end{tabular}
& $\overline{A}_2^2$ 
& \begin{tabular}{c} $(1, 2)$ \\$(2, 0)$ \\\end{tabular} 
& --- \\
\hline
$U(1)^2$ 
& \begin{tabular}{c} $(1/3, 1/2)$ \\$(1/4, 1/4)$ \\\end{tabular}
& $\overline{D}_4 \times \overline{A}_2$ 
& \begin{tabular}{c} $(s, 1)$ \\$(c, 0)$ \\\end{tabular}
& $2\sqrt{3}, 2$ \\
\hline
\end{tabular}
\caption[smallcaption]{Group factors and conjugacy classes relevant
for the lattice engineering in the ${\bf Z}_3$ orbifold construction.}
\label{Tab.LET}
\end{center}
\end{table}

\subsection{24-dimensional Euclidean lattices}

By the lattice engineering technique, we can construct (22,6)-dimensional Narain lattices from some known even self-dual lattices. As starting point we can choose a $8n$-dimensional Euclidean lattice that has only left-moving degrees of freedom. The eight-dimensional lattice ($\Gamma_{E_8}$) and the 16-dimensional lattices ($\Gamma_{E_8} \oplus \Gamma_{E_8}$ and $\Gamma_{{\rm Spin}(32)/{\bf Z}_2}$) are well known. In 24 dimensions all Euclidean even self-dual lattices have been classified, and there are 24 types thereof, the Niemeier lattices \cite{LeechNiemeier}. Except for the Leech lattice, we summarize group factors and generators for conjugacy classes for these lattices in table \ref{Tab.24DimLattice}. The underline indicates cyclic permutations, and all conjugacy classes for each lattice can be obtained by calculating the closure of the generators under addition. It is in principle also possible to apply the lattice engineering technique to 32-dimensional Euclidean lattices or even higher dimensional ones. However, these lattices are not fully classified and it is known that their number increases rapidly with $n$. Therefore these possibilities are beyond our scope.

\begin{table}[h]
\begin{center}
\begin{tabular}{|c|c|}
\hline
Group & Conjugacy class generators \\
\hline
\hline
$D_{24}$ & $(s)$ \\
\hline
$D_{16} \times E_8$ & $(s, 0)$ \\
\hline
$E_8^3$ & $(0, 0, 0)$ \\
\hline
$A_{24}$ & $(5)$ \\
\hline
$D_{12}^2$ & $(s, v), (v, s)$ \\
\hline
$A_{17} \times E_7$ & $(3, 1)$ \\
\hline
$D_{10} \times E_7^2$ & $(s, 1, 0), (c, 0, 1)$ \\
\hline
$A_{15} \times D_9$ & $(2, s)$ \\
\hline
$D_{8}^3$ & $( \underline{s, v, v})$ \\
\hline
$A_{12}^2$ & $(1, 5)$ \\
\hline
$A_{11} \times D_7 \times E_6$ & $(1, s, 1)$ \\
\hline
$E_{6}^4$ & $(1, \underline{0, 1, 2} )$ \\
\hline
$A_{9}^2 \times D_6$ & $(2, 4, 0), (5, 0, s), (0, 5, c)$ \\
\hline
$D_{6}^4$ & Even permutations of $(0, s, v, c)$ \\
\hline
$A_{8}^3$ & $( \underline{1, 1, 4} )$ \\
\hline
$A_{7}^2 \times D_5^2$ & $(1, 1, s, v), (1, 7, v, s)$ \\
\hline
$A_{6}^4$ & $(1, \underline{2, 1, 6})$ \\
\hline
$A_{5}^4 \times D_4$ & $(0, \underline{0, 2, 4}, 0), 
(3, 3, 0, 0, s), (3, 0, 3, 0, v), (3, 0, 0, 3, c)$ \\
\hline
$D_{4}^6$ & $(s, s, s, s, s, s), (v, \underline{v, 0, c, c, 0})$ \\
\hline

$A_{4}^6$ & $(1, \underline{0, 1, 4, 4, 1})$ \\
\hline
$A_{3}^8$ & $(3, \underline{2, 0, 0, 1, 0, 1, 1})$ \\
\hline
$A_{2}^{12}$ & $(2, \underline{1, 1, 2, 1, 1, 1, 2, 2, 2, 1, 2})$ \\
\hline
$A_{1}^{24}$ & 
$(1, \underline{0, 0, 0, 0, 0, 1, 0, 1, 0, 0, 1, 1, 
0, 0, 1, 1, 0, 1, 0, 1, 1, 1, 1})$ \\
\hline
\end{tabular}
\caption[smallcaption]{The 23 Niemeier lattices with non-empty root system.}
\label{Tab.24DimLattice}
\end{center}
\end{table}

\subsection{(22,6)-dimensional Narain lattices}

In this section, we construct (22,6)-dimensional Narain lattices which can be starting points for the ${\bf Z}_3$ asymmetric orbifold construction. Such kind of Narain lattice can be made from 8, 16 and 24-dimensional Euclidean lattices by the lattice engineering technique. 

In order to realize $\mathcal{N}=1$ SUSY in ${\bf Z}_3$ orbifold models, the right-moving twist vector should be $t_{{\rm R}} = ( 0, 1, 1, -2)/3$, and the right-moving part of the Narain lattice has to possess a ${\bf Z}_3$ discrete rotation symmetry. In general, in the asymmetric orbifold construction, we can also twist the left-movers. However, in this work we consider the case of not twisting the left-movers since this decreases the degeneracy factor that appears in the twisted sector. From these restrictions it turns out that we have to concentrate on the classification of (22,6)-dimensional Narain lattices with right-moving group factors $\overline{E}_6$ or $\overline{A}_2^3$. If we consider the case $\theta_{{\rm L}} \neq 1$ and demand a non-Abelian group factor in the right-movers, Narain lattices with $\overline{D}_4 \times \overline{A}_2$ can additionally be used for model building. We will comment on this possibility in section 5. 

In general, (22,6)-dimensional Narain lattices are composed of certain building blocks that are $(d_{{\rm L}}, d_{{\rm R}})$-dimensional Narain lattices with $d_{{\rm L}} \leq 22$ and $d_{{\rm R}} \leq 6$, e.g. $\Gamma = \Gamma_{E_6 \times \overline{E}_6} \oplus \Gamma_{E_8} \oplus \Gamma_{E_8} $. In this paper, we construct these building blocks from 8, 16 and 24-dimensional Euclidean lattices by the lattice engineering technique as in table \ref{Tab.LET}. Narain lattices for ${\bf Z}_3$ asymmetric orbifold models are constructed as follows. First, we construct (22,6)-dimensional lattices with $\overline{E}_6$ from 24-dimensional Euclidean lattices by the lattice engineering $A_2 \to \overline{E}_6$. This way, 31 lattices are obtained. In fact, these are all relevant lattices with $\overline{E}_6$ because reverse lattice engineering $\overline{E}_6 \to A_2$ has to lead back to one of the lattices in table \ref{Tab.24DimLattice}. The obtained lattices are listed in tables \ref{Tab.226DimLatticeWithE6bar1}-\ref{Tab.226DimLatticeWithE6bar5} and numbered sequentially as \#1 $\ldots$ \#31. Next, for lattices with $\overline{A}_2^3$, we have the following ways. From 24-dimensional Euclidean lattices, we construct 
\begin{description}
\item[(i)] (18,2)-dimensional lattices with $\overline{A}_2$ 
by the lattice engineering $E_6 \to \overline{A}_2$ 
together with two $A_2 \times \overline{A}_2$ lattices 
(lattice number \#32 $\ldots$ \#37),
\item[(ii)] (20,4)-dimensional lattices with $\overline{A}_2^2$ 
by the lattice engineering $A_2 \times A_2 \to \overline{A}_2 
\times \overline{A}_2$ together with $A_2 \times \overline{A}_2$ 
lattice (lattice number \#38 $\ldots$ \#83),
\item[(iii)] (12,4)-dimensional lattices with $\overline{A}_2^2$ 
by the lattice engineering $E_6 \to \overline{A}_2$ twice together 
with $A_2 \times \overline{A}_2$ lattice and $E_8$ lattice,
\item[(iv)] (6,6)-dimensional lattices with $\overline{A}_2^3$ 
by the lattice engineering $E_6 \to \overline{A}_2 $ three times
together with two $E_8$ lattices or ${\rm Spin}(32)/{\bf Z}_2$ lattice,
\item[(v)] (14,6)-dimensional lattices with $\overline{A}_2^3$ 
by the lattice engineering $A_2 \times A_2 \to \overline{A}_2 
\times \overline{A}_2$ and $E_6 \to \overline{A}_2 $ 
together with $E_8$ lattice (lattice number \#84 $\ldots$ \#90).
\end{description}
We find that there are six types of (18,2)-dimensional lattices 
with $\overline{A}_2$ in (i), 46 types of (20,4)-dimensional 
lattices with $\overline{A}_2^2$ in (ii) and seven (14,6)-dimensional 
lattices with $\overline{A}_2^3$ in (v) as listed in tables 
\ref{Tab.182DimLatticeWithA2bar}, 
\ref{Tab.204DimLatticeWithA2A2bar1}-\ref{Tab.204DimLatticeWithA2A2bar7}
and 
\ref{Tab.146DimLatticeWithA2A2A2bar1}, 
respectively. Among the lattices constructed above, some lattices are identical to each other, in that case we take only independent lattices. For example, it turns out that the (6,6)-dimensional lattice with $\overline{A}_2^3$ in (iv) is the same as the $E_6 \times \overline{E}_6$ lattice which is already known, so it can be omitted. In total we find 90 types of (22,6)-dimensional lattices with right-moving $\overline{E}_6$ or $\overline{A}_2^3$. We will show the construction of a Narain lattice by the lattice engineering technique in more detail in section \ref{Examples}.


\section{Group breaking patterns}

As we saw in the previous section, the conjugacy classes of the Narain lattice can be described by suitable combinations of conjugacy classes of each group factor. In orbifold models, a shift vector $V$ is defined by $V = p/N$, where $p$ is a momentum mode on the Narain lattice and $N$ is the order of the ${\bf Z}_N$ twist. In the case of symmetric $E_8 \times E_8$ orbifold models, $p \in ( \Gamma_{E_8}, \Gamma_{E_8}) $, so only one conjugacy class, $0_{E_8}$, is relevant. In general, however, (22,6)-dimensional Narain lattices are described by some combinations of root lattices possibly shifted by fundamental weights. This means that, in asymmetric models, we have to take into account group breaking patterns that are caused by conjugacy class combinations which are no longer a combination of root lattices only. 

First, we study group breaking patterns for the simple group factors $A_n, D_n, E_6, E_7$ and $E_8$. Namely we consider shift vectors that are given by root lattices shifted by fundamental weights of the groups divided by the order $N$. By extending the argument of the classification of $E_8$ group breaking patterns \cite{Katsuki:1989kd} to the cases of the other simple groups, we can classify inequivalent breaking patterns. After obtaining these building blocks, we can specify a shift vector by taking a suitable combination that is compatible with the conjugacy classes of the Narain lattice.

\subsection{Group breaking patterns by a shift vector}

\begin{figure}[htbp]
\begin{center}
\includegraphics[scale=0.4]{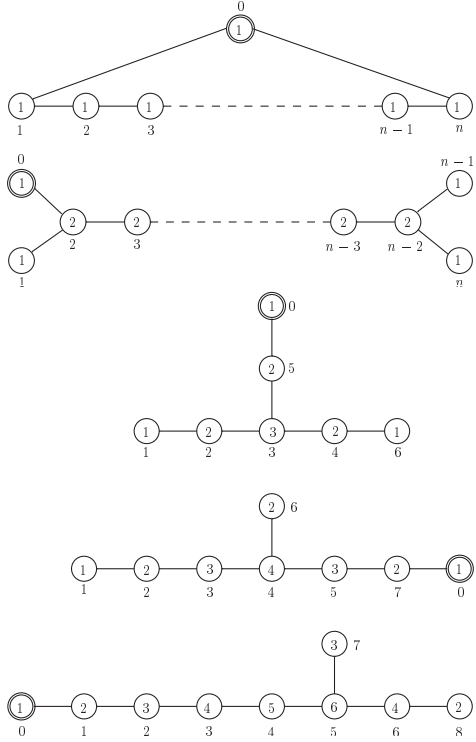}
 \caption{Extended Dynkin diagrams of the $A_n, D_n, E_6, E_7$ and $E_8$ groups.}
 \label{fig:Dynkin}
\end{center}
\end{figure}

Here we consider the calculation of group breaking patterns caused by a ${\bf Z}_N$ shift action. We define simple roots for $G \in \lbrace A_n, D_n, E_6, E_7, E_8\rbrace$ as $\alpha^{G}_i$ where $i = 1 \ldots {\rm Rank}(G) $. The lowest root for each group $\alpha_0^{G}$ can be described by
\begin{align}
	-\alpha_0^{G} &= \sum_i k_i^{G} \alpha_i^{G},
\end{align}
where $k_i^{G}$ is the Dynkin label of $G$ (see figure \ref{fig:Dynkin}),
\begin{align}
	k^{A_n} &= ( 1, 1, \cdots, 1 ), \\
	k^{D_n} &= ( 1, 2, 2, \cdots, 2, 2, 1, 1 ), \\
	k^{E_6} &= ( 1, 2, 3, 2, 2, 1 ), \\
	k^{E_7} &= ( 1, 2, 3, 4, 3, 2, 2 ), \\
	k^{E_8} &= ( 2, 3, 4, 5, 6, 4, 3, 2 ). 
\end{align}
A root lattice shifted by the $k$-th fundamental weight of group $G$ is given by
\begin{align} \label{eq:Gammagk}
	\sum_{i} n_i \omega_i^{G} + \omega_k^{G},
\end{align}
where $\omega_k^{G}$ is a fundamental weight which satisfies $\alpha_i^{G} \cdot \omega_j^{G} = \delta_{ij}$, and $n_i$ satisfies
\begin{align} \label{eq:Constraintforni}
	\sum_i n_i (C^{G})_{ij}^{-1} \in {\bf Z}.
\end{align}  
Here $(C^{G})_{ij}^{-1}$ is the inverse Cartan matrix. Note that, in the case of $k=0$, $\omega_0^{G}$ means the highest root, $\omega_0^{G} = - \alpha_0^{G}$. In this paper, we describe the root lattice of $G$ as
\begin{align}
	\sum_{i} n_i \omega_i^{G} + \omega_0^{G} ,
\end{align}
with \eqref{eq:Constraintforni}. Furthermore we can rewrite \eqref{eq:Gammagk} as
\begin{align}
	\sum_{i} n'_i \omega_i^{G},
\end{align}
with 
\begin{align}
	n'_{i} &=
	\left\{
	\begin{array}{l}
	n_i + \delta_{ik} \ \ ( k \neq 0 ) \\
	n_i + \sum_{j} k_j^{G} C_{ij} \ \ ( k = 0 ) 
	\end{array}
	\right.,
\end{align}
and \eqref{eq:Constraintforni}. Using Weyl reflections, it is sufficient to consider the case $n_i' \geq 0$.

Without loss of generality, a shift vector for a ${\bf Z}_N$ orbifold model can thus be expressed as 
\begin{align}
	V_k &= \frac{ \sum_{i} n_i \omega_i^{G} + \omega_k^{G} }{N},
\end{align}
with
\begin{align}
	n'_i \geq 0, \ \ \sum_i n_i (C^{G})_{ij}^{-1} \in {\bf Z}.
\end{align}
We can also show
\begin{align}
	0 \leq V_k \cdot ( - \alpha_0^{G} ) \leq 1,
\end{align}
which leads to 
\begin{align}
	0 &\leq \sum_i n_i k_i^{G} + k_k^{G} \leq N \ \ ( k \neq 0 ), \\
	0 &\leq \sum_i n_i k_i^{G} + 2 \leq N \ \ ( k = 0 ).
\end{align}
The shift actions for the simple roots and the highest root read
\begin{align}   \label{eq:Vaj}
	V_k \cdot \alpha_j^{G} 
	&= 
	\left\{
	\begin{array}{l}
	\frac{n_j}{N} + \frac{\delta_{kj}}{N} 
	\equiv \frac{ n'^{(k)}_j }{N} \ \ ( k \neq 0 ) \\
	\frac{n_j}{N} + \frac{ k_i^{G} C_{ij}^{G} }{N} 
	\equiv \frac{ n'^{(0)}_j }{N} \ \ ( k = 0 ) 
	\end{array}
	\right.,
\end{align}
and
\begin{align}   \label{eq:Va0}
	V_k \cdot ( - \alpha_0^{G} ) 
	&= 
	\left\{
	\begin{array}{l}
	( 1- \frac{n_0}{N} ) + \frac{k_k^{G}}{N} 
	\equiv 1 - \frac{ n'^{(k)}_0 }{N} \ \ ( k \neq 0 ) \\
	( 1- \frac{n_0}{N} ) + \frac{ 2 }{N} 
	\equiv 1 - \frac{ n'^{(0)}_0 }{N} \ \ ( k = 0 ) 
	\end{array}
	\right.,
\end{align}
where we define $n_0$ so that
\begin{align}
	N = \sum_{i=0}^{{\rm Rank}(G)} n_i k_i^{G},
\end{align}
and $k_0^{G} = 1$. If the orbifold phases of \eqref{eq:Vaj} and \eqref{eq:Va0} are non-integer, the corresponding adjoint states of $G$ are projected out, so $G$ is broken to a subgroup. Then, by calculating $n'^{(k)}_i$ $(i=0\ldots{\rm Rank}(G))$, we can read off the corresponding group breaking patterns by analyzing the extended Dynkin diagram.

In the following, we summarize group breaking patterns of each group $G$ for ${\bf Z}_3$ asymmetric orbifold models. We list $E_8$ group breaking patterns in table \ref{Tab:E8breaking}. These breaking patterns are the same as the result for symmetric $E_8 \times E_8$ orbifold models. $E_7$ group breaking patterns are listed in table \ref{Tab:E7breaking}. We find that certain two shift vectors which correspond to different conjugacy classes ($0$ and $1$) give rise to the same group breaking patterns. For example, shift vectors No.2 and No.8 lead to the $E_6 \times U(1)$ group. Note that, even though the breaking patterns are the same, different shift vectors may give us different orbifold models. In table \ref{Tab:E7breaking}, we list $E_6$ breaking patterns. There, we omit breaking patterns caused by shift vectors which correspond to the conjugacy class $2$ since they can be reproduced by considering suitable rotations of shift vectors for the conjugacy class $1$. We also see that some shift vectors lead to the same breaking pattern, e.g., shift vectors No.9, No.10 and No.11 all give rise to the $D_5 \times U(1) $ group. We list all shift vectors since their contributions to the modular invariance condition are different and may lead to different models. 

For $A_n (n = 1\ldots 22)$ and $D_n (n = 4\ldots 22)$, we list group breaking patterns in tables \ref{Tab.AnGroupBreaking1}, \ref{Tab.AnGroupBreaking2}, \ref{Tab.AnGroupBreaking3} and \ref{Tab.DnGroupBreaking}. By calculating $n_i'^{(k)}$, it turns out that there are many possible shift vectors and corresponding breaking patterns. Also we find the following patterns: for $A_n$ with $n = 2 \bmod 3$, we can see that shift vectors which correspond to the conjugacy class $c_{A_n} = 0 \bmod 3$ lead to the same series of group breaking patterns. Shift vectors for $c_{A_n} = 1 \bmod 3$ or $c_{A_n} = 2 \bmod 3$ also lead to the same series of breaking patterns. Furthermore, all conjugacy classes for $A_n$ with $n = 0 \bmod 3$ or $n = 1 \bmod 3$ lead to the same situations. For $D_n (n =4\ldots 22)$, there are conjugacy classes $0, s, v$ and $c$. It turns out that each of them leads to the same series of group breaking patterns. In the tables \ref{Tab.AnGroupBreaking1}-\ref{Tab.AnGroupBreaking3} and \ref{Tab.DnGroupBreaking}, we list only group breaking patterns for the representative conjugacy classes without showing the corresponding $n_i'^{(k)}$ and shift vectors since there are too many possibilities. However, it is not difficult to reproduce them.

\begin{table}[h]
\begin{center}
\begin{tabular}{|c|c|c|c|c|c|}
\hline
No. & C.C. & Group breaking & $n_i'^{(k)}$ & Shift vector ($3V$) & $(3V)^2$\\
\hline
\hline
$1$ & $0$ & $E_8$ & $(0,0,0,0,0,0,0,0;3)$ & $0$ & $0$ \\
\hline
$2$ & $0$ & $D_7 \times U(1)$ & $(0,0,0,0,0,0,0,1;1)$ 
& $\omega_8^{E_8}$ & $4$ \\
\hline
$3$ & $0$ & $A_8$ & $(0,0,0,0,0,0,1,0;0)$ & $\omega_7^{E_8}$ & $8$ \\
\hline
$4$ & $0$ & $E_6 \times A_2$ & $(0,1,0,0,0,0,0,0;0)$ & $\omega_2^{E_8}$ & $6$ \\
\hline
$5$ & $0$ & $E_7 \times U(1)$ & $(1,0,0,0,0,0,0,0;1)$ 
& $\omega_1^{E_8}$ & $2$ \\
\hline
\end{tabular}
\caption[smallcaption]{$E_8$ group breaking patterns and shift vectors. The last component of $n_i'^{(k)}$ corresponds to $n_0'^{(k)}$.}
\label{Tab:E8breaking}
\end{center}
\end{table}

\begin{table}[h]
\begin{center}
\begin{tabular}{|c|c|c|c|c|c|}
\hline
No. & C.C. & Group breaking & $n_i'^{(k)}$ & Shift vector ($3V$)& $(3V)^2$ \\
\hline
\hline
$1$ & $0$ & $E_7$ & $(0,0,0,0,0,0,0;3)$ & $0$ & $0$ \\
\hline
$2$ & $0$ & $E_6 \times U(1)$ & $(2,0,0,0,0,0,0;1)$ & $2\omega_1^{E_7}$ & $6$ \\
\hline
$3$ & $0$ & $A_6 \times U(1)$ & $(1,0,0,0,0,1,0;0)$ 
& $\omega_1^{E_7} + \omega_6^{E_7} $ & $8$ \\
\hline
$4$ & $0$ & $D_6 \times U(1)$ & $(0,0,0,0,0,0,1;1)$ & $\omega_7^{E_7}$ & $2$ \\
\hline
$5$ & $0$ & $A_5 \times A_2$ & $(0,0,0,0,1,0,0;0)$ & $\omega_5^{E_7} $ & $6$ \\
\hline
$6$ & $0$ & $D_5 \times A_1 \times U(1)$ & $( 0,1,0,0,0,0,0;1)$ 
& $\omega_2^{E_7} $ & $4$\\
\hline
\hline
$7$ & $1$ & $E_7$ & $(3,0,0,0,0,0,0;0)$ & $3 \omega_1^{E_7} $ 
& $\frac{27}{2}$ \\
\hline
$8$ & $1$ & $E_6 \times U(1)$ & $(1,0,0,0,0,0,0;2)$ 
& $\omega_1^{E_7}$ & $\frac{3}{2}$ \\
\hline
$9$ & $1$ & $A_6 \times U(1)$ & $(0,0,0,0,0,1,0;1)$ 
& $\omega_6^{E_7}$ & $\frac{7}{2}$ \\
\hline
$10$ & $1$ & $D_6 \times U(1)$ & $(1,1,0,0,0,0,0;0)$ 
& $\omega_1^{E_7} + \omega_2^{E_7}$ & $\frac{19}{2}$ \\
\hline
$11$ & $1$ & $A_5 \times A_2$ & $(0,0,1,0,0,0,0;0)$ 
& $\omega_3^{E_7} $ & $\frac{15}{2}$ \\
\hline
$12$ & $1$ & $D_5 \times A_1 \times U(1)$ 
& $(1,0,0,0,0,0,1;0)$ & $\omega_1^{E_7} + \omega_7^{E_7}$ & $\frac{11}{2}$ \\
\hline
\end{tabular}
\caption[smallcaption]{$E_7$ group breaking patterns and shift
  vectors. 
The last component of $n_i'^{(k)}$ corresponds to $n_0'^{(k)}$.}
\label{Tab:E7breaking}
\end{center}
\end{table}

\begin{table}[h]
\begin{center}
\begin{tabular}{|c|c|c|c|c|c|}
\hline
No. & C.C. & Group breaking & $n_i'^{(k)}$ & Shift vector ($3V$) & $(3V)^2 $ \\
\hline
\hline
$1$ & $0$ & $E_6$ & $(0,0,0,0,0,0;3)$ & $0$ & $0$ \\
\hline
$2$ & $0$ & $E_6$ & $(0,0,0,0,0,3;0)$ & $3 \omega_6^{E_6}$ & $12$ \\
\hline
$3$ & $0$ & $E_6$ & $(3,0,0,0,0,0;0)$ & $3 \omega_1^{E_6}$ & $12$ \\
\hline
$4$ & $0$ & $A_5 \times U(1)$ & $(0,0,0,0,1,0;1)$ & $ \omega_5^{E_6} $ & $2$ \\
\hline
$5$ & $0$ & $A_5 \times U(1)$ & $(0,0,0,1,0,1;0)$ 
& $ \omega_4^{E_6} +  \omega_6^{E_6} $ & $8$ \\
\hline
$6$ & $0$ & $A_5 \times U(1)$ & $(1,1,0,0,0,0;0)$ 
& $  \omega_1^{E_6} +  \omega_2^{E_6} $ & $8$ \\
\hline
$7$ & $0$ & $D_4 \times U(1) \times U(1)$ & $(1,0,0,0,0,1;1)$ 
& $  \omega_1^{E_6} +  \omega_6^{E_6} $ & $4$ \\
\hline
$8$ & $0$ & $A_2 \times A_2 \times A_2$ & $(0,0,1,0,0,0;0)$ 
& $  \omega_3^{E_6} $ & $6$ \\
\hline
\hline
$9$ & $1$ & $D_5 \times U(1)$ & $(0,0,0,0,0,2;1)$ 
& $ 2 \omega_6^{E_6} $ & $\frac{16}{3}$ \\
\hline
$10$ & $1$ & $D_5 \times U(1)$ & $(1,0,0,0,0,0;2)$ 
& $ \omega_1^{E_6} $ & $\frac{4}{3}$ \\
\hline
$11$ & $1$ & $D_5 \times U(1)$ & $(2,0,0,0,0,1;0)$ 
& $ 2  \omega_1^{E_6} +  \omega_6^{E_6} $ & $\frac{28}{3}$ \\
\hline
$12$ & $1$ & $A_4 \times A_1 \times U(1)$ & $(0,0,0,1,0,0;1)$ 
& $  \omega_4^{E_6}  $ & $\frac{10}{3}$ \\
\hline
$13$ & $1$ & $A_4 \times A_1 \times U(1)$ & $(0,1,0,0,0,1;0)$ 
& $  \omega_2^{E_6} +  \omega_6^{E_6} $ & $\frac{22}{3}$ \\
\hline
$14$ & $1$ & $A_4 \times A_1 \times U(1)$ & $(1,0,0,0,1,0;0)$ 
& $  \omega_1^{E_6} +  \omega_5^{E_6} $ & $\frac{16}{3}$ \\
\hline
\end{tabular}
\caption[smallcaption]{$E_6$ group breaking patterns 
and shift vectors. 
The last component of $n_i'^{(k)}$ corresponds to $n_0'^{(k)}$.}
\label{Tab:E6breaking}
\end{center}
\end{table}

\subsection{Group breaking patterns of ${\bf Z}_3$ asymmetric orbifold models}

Next, we consider gauge symmetries of four-dimensional ${\bf Z}_3$ orbifold models. We can calculate the breaking patterns by combining the results for each group factor from the previous subsection. For the (22,6)-dimensional lattices constructed in section 2, we list possible group patterns for the SM group or typical grand unified groups in tables \ref{Tab:SMGUT1} and \ref{Tab:SMGUT2}. In the tables, ``SM'', ``Pati-Salam'' and ``left-right symmetric'' denote the $SU(3) \times SU(2) \times U(1)$, $SU(4) \times SU(2) \times SU(2)$ and $SU(3) \times SU(2) \times SU(2) \times U(1)$ gauge groups, respectively. Note that, at this stage, we do not care about the modular invariance condition.


\section{${\bf Z}_3$ asymmetric orbifold models}\label{Examples}

\begin{figure}[t] 
\begin{center}
\setlength{\unitlength}{10.5mm} 
\begin{picture}(15,12) 
\put(4.3,9){\framebox(6,1){24-dimensional Euclidean lattice}} 
\put(7.2,9){\vector(0,-1){1.5}} 
\put(7.7,8.1){Lattice engineering technique} 

\put(3.1,6.5){\framebox(8,1){(22,6)-dimensional Narain lattice $G_{{\rm L}} \times \overline{G}_{{\rm R}}$}} 
\put(7.2,6.5){\vector(0,-1){1.0}} 

\put(0,3.5){\framebox(14.7,2)} 
\put(0.2,4.8){Calculate $n_{i}'^{(k)}$ and shift vectors for each $G_{{\rm L}}^i$, where $G_{{\rm L}} = \prod_i G_{{\rm L}}^i$} 
\put(0.2,4.3){Calculate group breaking patterns of $G_{{\rm L}}$} 
\put(0.2,3.8){Choose a suitable combination of shift vectors for $G_{{\rm L}}^i$ to satisfy modular invariance} 
\put(7.2,3.5){\vector(0,-1){1.0}} 

\put(0.2,0.5){\framebox(14,2)} 
\put(0.4,1.8){Read off massless spectrum} 
\put(0.4,1.3){Untwisted sector: cancelling orbifold phases} 
\put(0.4,0.8){${\bf Z}_3$ twisted sector: finding momentum modes that satisfy the massless condition} 

\end{picture} 
\end{center} 
\caption{The procedure for ${\bf Z}_3$ asymmetric orbifold model building.} 
\label{fig:Procedure}
\end{figure}
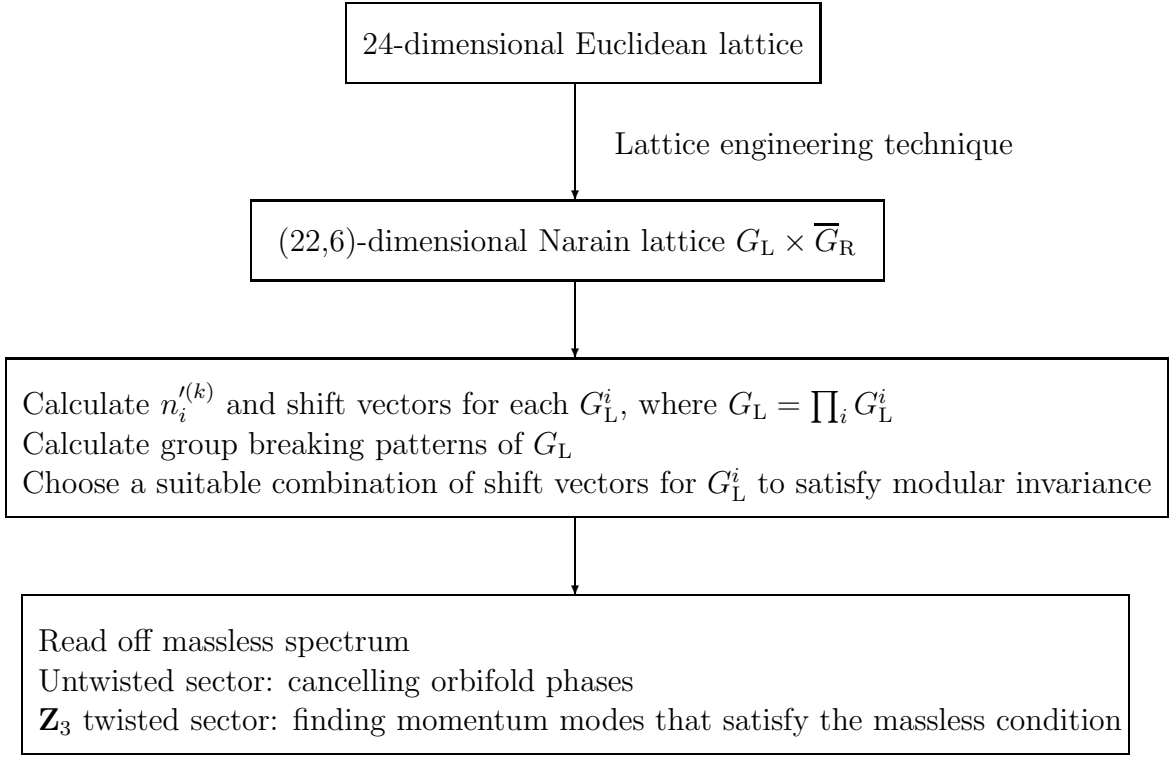

In this section, we demonstrate the construction of a (22,6)-dimensional Narain lattice with right-moving $E_6$ by the lattice engineering technique and show an example ${\bf Z}_3$ asymmetric orbifold model built from that lattice. This model fulfills the requirement given in subsection \ref{SecHetorb}, i.e. there is no twist in the left-movers. The following model building procedure (see figure \ref{fig:Procedure}) can also be applied to the other (22,6)-dimensional lattices with $\overline{E}_6$ or $\overline{A}_2^3$.   

Let us start with the 24-dimensional $A_{11} \times D_7 \times E_6$ lattice which contains the conjugacy classes generated by $(1, s, 1)$. All conjugacy classes of that lattice are given by
\begin{align}
	&(0, 0, 0), (1, s, 1), (2, v, 2), (3, c, 0), (4, 0, 1), (5, s, 2),
	\nonumber \\
	&(6, v, 0), (7, c, 1), (8, 0, 2), (9, s, 0), (10, v, 1), (11, c, 2).
\end{align}
We apply the lattice engineering $A_2 \to \overline{E}_6$ to the $A_2$ factor in $A_8 \times U(1) \times A_2 \subset A_{11}$, i.e.,
\begin{align}
	&A_{11} \times D_7 \times E_6 
	\xrightarrow[{\rm decompose}]{} A_{8} \times U(1) \times A_2 
	\times D_7 \times E_6 
	\xrightarrow[{\rm replace}]{} A_{8} \times U(1) 
	\times \overline{E}_6 \times D_7 \times E_6.
	\nonumber
\end{align}
Under the decomposition $A_8 \times U(1) \times A_2 \subset A_{11}$, the conjugacy classes $i_{A_{11}}$ are described as 
\begin{align}
	i_{A_{11}} = (i, i/36, 0) + \bigcup_{j=0}^8 (j, j/9, j \bmod 3).
\end{align} 
Here we take the $U(1)$ normalization $6$. Using this, we can decompose $i_{A_{11}}$ for all conjugacy classes in (22) and replace $0_{A_2} \to 0_{\overline{E}_6}, 1_{A_2} \to 1_{\overline{E}_6}$ and $2_{A_2} \to 2_{\overline{E}_6}$. The resulting lattice corresponds to the (22,6)-dimensional lattice \#15 in table \ref{Tab.226DimLatticeWithE6bar3}, whose conjugacy classes are generated by $(0 ,0, 1, 1/9, 1)$ and $(s ,1, 1, 1/36, 0)$ of $D_7 \times E_6 \times A_8 \times U(1) \times \overline{E}_6$. At this stage the four-dimensional heterotic string has $\mathcal{N}=4$ SUSY and an $SO(14) \times E_6 \times SU(9) \times U(1)$ gauge symmetry.

Let us consider gauge group breaking patterns for this lattice in the framework of ${\bf Z}_3$ asymmetric orbifolds. Since we twist the right-movers in order to obtain $\mathcal{N}=1$ SUSY, conjugacy classes of the lattice that are relevant for shift vectors have to include $0_{\overline{E}_6}$ in the right-mover. This is because the right-moving zero momentum mode which survives the twist action is contained in $0_{\overline{E}_6}$. The relevant conjugacy classes are 
\begin{align}
	\bigcup_{k=0}^{35} k ( s ,1, 1, 1/36, 0) \;,
\end{align}
i.e. the closure of $(s ,1, 1, 1/36, 0)$. Among the conjugacy classes, it turns out that, if we concentrate on only group breaking patterns, independent conjugacy classes are given by $(0, 0, 0, 0, 0)$ and $(s, 1, 1, 1/36, 0)$. This means that shift vectors for each conjugacy class are given by $V =p/3$ with
\begin{align} \label{ShiftVectorInRoot}
	p \in (\Gamma_{D_7}, \Gamma_{E_6}, \Gamma_{A_8}, 6n, 0)
\end{align}
or
\begin{align} \label{ShiftVectorInWeight}
	p \in (\Gamma_{D_7} + \omega_6^{D_7} ,\Gamma_{E_6}+ \omega_1^{E_6}, \Gamma_{A_8} + \omega_1^{A_8}, 6n + 1/6, 0),
\end{align}
respectively. Here $n$ is an integer. The shift vector breaks the left-moving $D_7, E_6$ and $A_8$ groups. We summarize corresponding group breaking patterns in table \ref{GroupBreakingForD7E6A8U1Lat}. Note that we do not take account of a modular invariance condition in this stage. Suitable combinations of shift vectors lead to modular invariance. Although here we do not list all breaking patterns for the other (22,6)-dimensional lattices since there are many possibilities, following the same procedure, it is easy to make corresponding tables.

\begin{table}[h]
\begin{center}
\begin{tabular}{|c||c|c|}
\hline
Group & Group breaking patterns &  Group breaking patterns \\
\hline
C.C.    & $ ( 0,0,0,0,0 )$ & $( s, 1, 1, 1/36, 0 )$ \\
\hline
\hline
$D_7$ &
\begin{tabular}{c} 
$D_{7}$ \\
$A_{6} \times U(1)$ \\
$D_{6} \times U(1)$ \\
$A_{1} \times D_{5} \times U(1)$ \\
$A_{2} \times D_{4} \times U(1)$ \\
$A_{3}^2 \times U(1)$ \\ 
$A_{5} \times U(1)^2$ \\
$A_{1}^2 \times A_{4} \times U(1)$ \\
\end{tabular}
&
\begin{tabular}{c} 
$D_{7}$ \\
$A_{6} \times U(1)$ \\
$D_{6} \times U(1)$ \\
$A_{1} \times D_{5} \times U(1)$ \\
$A_{2} \times D_{4} \times U(1)$ \\
$A_{3}^2 \times U(1)$ \\ 
$A_{5} \times U(1)^2$ \\
$A_{1}^2 \times A_{4} \times U(1)$ \\
\end{tabular}
\\
\hline
$E_6$ &
\begin{tabular}{c} 
$E_{6}$ \\
$A_{5} \times U(1)$ \\
$A_{2} \times A_{2} \times A_{2}$ \\
$D_{4} \times U(1)^2$ \\
$D_{5} \times U(1)$ \\
$A_{4} \times A_1 \times U(1)$ \\ 
\end{tabular}
&
\begin{tabular}{c} 
$D_{5} \times U(1)$ \\
$A_{4} \times A_1 \times U(1)$ \\
\end{tabular}
\\
\hline
$A_{8}$ 
&
\begin{tabular}{c} 
$A_{8}$ \\
$A_{6} \times U(1)^2$ \\
$A_{5} \times A_2 \times U(1)$ \\
$A_{4} \times A_1^2 \times U(1)^2$ \\
$A_{3}^2 \times U(1)^2$ \\
$A_{2}^3 \times U(1)^2$ \\
\end{tabular}
& 
\begin{tabular}{c} 
$A_{7} \times U(1)$ \\
$A_{6} \times A_{1} \times U(1)$ \\
$A_{5} \times A_{1} \times U(1)^2$ \\
$A_{4} \times A_{3} \times U(1)$ \\
$A_{4} \times A_{2} \times U(1)^2$ \\
$A_{3} \times A_{2} \times A_1 \times U(1)^2$ \\
\end{tabular}
\\
\hline
$U(1)$ &
$U(1)$
&
$U(1)$
\\
\hline
\end{tabular}
\caption[smallcaption]{Group breaking patterns for 
$D_7 \times E_6 \times A_8 \times U(1) \times \overline{E}_6$ lattice.}
\label{GroupBreakingForD7E6A8U1Lat}
\end{center}
\end{table}

Next, we construct a ${\bf Z}_3$ asymmetric orbifold model from the above lattice. We consider the ${\bf Z}_3$ twist only for the right-movers and a shift for the left-movers. The twist vector is given by $t_{{\rm R}} = ( 0, 1, 1, -2)/3$ and the shift vector is
\begin{align}   \label{eq:shiftvector1}
	V = ( 0, 0, \omega_1^{A_8} + 2\omega_4^{A_8}, 6, 0 )/3.
\end{align}
This shift vector belongs to the conjugacy class $(0, 0, 0, 0, 0)$ of $D_7 \times E_6 \times A_8 \times U(1) \times \overline{E}_6$. We can show that this shift vector leads to a consistent model since the modular invariance condition which for this setup is given by 
\begin{align}
	\frac{3V^2}{2}  \in {\bf Z}
\end{align}
is obeyed. This modular invariance condition is applicable to other ${\bf Z}_3$ models without a left-moving twist action. The shift vector $V$ breaks the original gauge symmetry to
\begin{align*}
SO(14) \times E_6 \times SU(6) \times U(1) \times SU(3) \times U(1).
\end{align*}
The massless spectrum in the untwisted sector can be read off as in symmetric orbifold models by considering the cancellation of orbifold phases. Shifting and twisting acts on the adjoint modes $\vert p \rangle$ of $D_7 \times E_6 \times A_8 \times U(1)$ and on the right-moving H-momentum modes 
\begin{align}
	\vert q \rangle \in \lbrace \vert \underline{\pm 1,0,0,0} \rangle,
	\vert \pm \frac{1}{2}, \pm \frac{1}{2}, \pm \frac{1}{2}, 
	\pm \frac{1}{2},  \rangle_{+{\rm even}} \rbrace.
\end{align}
These massless modes survive if the phases due to the orbifold action cancel, i.e.
\begin{align}
	p \cdot V - t_{{\rm R}} \cdot q \in {\bf Z}.
\end{align} 
The gauge multiplets of $SO(14) \times E_6 \times SU(6) \times U(1) \times SU(3) \times U(1)$ are coming from invariant modes in the $\mathcal{N} = 4$ vector multiplet of  $D_7 \times E_6 \times A_8 \times U(1)$. A nontrivial chiral field comes from the diagonal part of the adjoint modes of $A_8$,
\begin{align}
	p' = ( 0, 0, p_{A_5},  \frac{1}{\sqrt{2}}, p_{A_2}, 0 ),
\end{align}
of $D_7 \times E_6 \times A_5 \times U(1) \times A_2 \times U(1) $, where $p_{A_5} \in \Gamma_{A_5} + \omega_1^{A_5}$ with $p_{A_5}^2/2 = 5/12$, and $p_{A_2} \in \Gamma_{A_2} + \omega_2^{A_2}$ with $p_{A_2}^2/2 = 1/3$. We can check that these momentum modes satisfy the massless condition 
\begin{align}
	p'^2/2 -1 = 5/12 + 1/4 + 1/3 -1 = 0.
\end{align} 
The orbifold phase is evaluated as 
\begin{align}
	p' \cdot V =  \omega_1^{A_5} \cdot \omega_1^{A_5}  + \sqrt{2}/2 \cdot \sqrt{2}/2  = 5/6 +1/2 = 4/3 \sim 1/3 
\end{align} 
by using the decomposition $\omega_1^{A_8} + 2\omega_4^{A_8} = ( 3\omega_1^{A_5}, 3/\sqrt{2}, 0)$. On the other hand, the right-moving H-momentum modes $\vert q'\rangle = \vert 0, \underline{ 1,0,0 } \rangle $ or $\vert \frac{1}{2}, \underline{ \frac{1}{2}, - \frac{1}{2}, -  \frac{1}{2} } \rangle $ get phases $- t_{{\rm R}} \cdot q' \sim -1/3$. Then, as a result, we find that the following states survive orbifold projection: 
\begin{align}
	\vert p' \rangle \otimes \vert 0, \underline{ 1,0,0 } \rangle, \ 
	\vert p' \rangle \otimes \vert \frac{1}{2}, 
	\underline{ \frac{1}{2}, - \frac{1}{2}, - \frac{1}{2} } \rangle. 
\end{align}
For the fermionic states with the first component $1/2$, we define the four-dimensional chirality as left-handed. Combined with CPT conjugate states, we obtain three chiral supermultiplets in the representation $3( {\bf 1}, {\bf 1},
{\bf 6}, 1/2, \overline{{\bf 3}}, 0 )_{{\rm L}}$. Here we take the $U(1)$ normalizations as $\sqrt{2}$ and $1$. There are also four-dimensional gravity, anti-symmetric tensor and dilaton supermultiplets, but no other singlets.

Let us proceed to the ${\bf Z}_3$ twisted sector. In order to evaluate momentum modes in the twisted sector, we first consider the invariant sublattice $I_{\theta}( \Gamma )$, which is the sublattice of the original (22,6)-dimensional lattice $\Gamma$ invariant under the twist action $t_{{\rm R}}$. Since we twist all right-moving six dimensions $I_{\theta}(\Gamma )$ is a 22-dimensional lattice. In terms of conjugacy classes $I_{\theta}(\Gamma )$ is described by the generator $	( s, 1, 1, 1/36 )$ of $D_7 \times E_6 \times A_8 \times U(1)$. This invariant lattice can be spanned by the following basis
\begin{align} \label{BasisForInvSubLattice}
	\alpha_{1\ldots 7} &= ( \alpha^{D_7}_{ 1\ldots 7 }, 0, 0, 0 ), \nonumber \\ 
	\alpha_{8\ldots 13} &= ( 0, \alpha^{E_6}_{ 1\ldots 6 }, 0, 0 ), \nonumber \\
	\alpha_{14\ldots 21} &= ( 0, 0, \alpha^{A_8}_{ 1\ldots 8 }, 0 ), \nonumber \\
	\alpha_{22} &= ( \omega_1^{D_7},  \omega_1^{E_6}, \omega_1^{A_8}, 1/6 ).
\end{align}
We can also evaluate the dual lattice of the invariant sublattice $\tilde{  I_{\theta}}( \Gamma ) = \sum_{i=1}^{22} \tilde{\alpha}_i$, where the dual basis $\tilde{\alpha}_i$ satisfies $\alpha_i \cdot \tilde{\alpha}_j = \delta_{ij}$. In terms of conjugacy classes the dual lattice is generated by $( s, 0, 0, 1/4 )$, $( 0, 1, 0,  - 1/3 )$ and $( 0, 0, 1, 1/9 )$. To read off massless states in the $\alpha =1, 2$ twisted sector, we solve a masslessness condition
\begin{align}
	\frac{p^2}{2} + \Delta c_{{\rm L}} -1 = 0,
\end{align}
where $p \in \tilde{I_{\theta}}( \Gamma ) + \alpha V$. Note that, since we do not consider any left-moving twist, the change of the zero point energy in the twisted sector is trivial, i.e. $\Delta c_{{\rm L}} = 0$. In ${\bf Z}_3$ orbifolds, we do not need to consider a cancellation of phases of massless states in the twisted sector since all massless states survive the orbifold projection. We can easily read off momentum modes that satisfy the massless condition for each conjugacy class. For example, for the conjugacy class $( 0, 1, 0,  - 1/3 )$ and $\alpha=1$, \begin{align}
	p \in (\Gamma_{D_7} ,\Gamma_{E_6} + \omega_1^{E_6}, \Gamma_{A_8} + (\omega_1^{A_8} + 2\omega_4^{A_8}) / 3, 6n) \;.
\end{align}
There one finds a solution of momentum modes in the representation $( {\bf 1}, {\bf 27}, {\bf 1}, 0, {\bf 3}, 0 )$. We also find that there are H-momentum modes $\vert 0, 1/3, 1/3, 1/3 \rangle$ and $| 1/2, -1/6, -1/6, -1/6 \rangle$ in the right-movers. Combining with CPT conjugate states in the $\alpha=2$ sector we finally obtain left-handed chiral multiplets $3( {\bf 1}, {\bf 27}, {\bf 1}, 0, {\bf 3}, 0 )_{{\rm L}}$, where the degeneracy factor three comes from the number of fixed points which can be evaluated as
\begin{align}
	D = \frac{ \prod_{ \{i \vert \eta^i \neq 0 \} } 
	\left[ 2 \sin(\pi \eta^i ) \right] }{ \sqrt{ {\rm Vol} 
	( I_{\theta}( \Gamma ) ) } }.
\end{align}
Here, the $\eta^i$ are defined as $0 \leq t^i + n^i = \eta^i < 1 $ for suitable intergers $n^i$, and the volume of the invariant lattice is $ {\rm Vol} ( I_{\theta}( \Gamma ) ) = \det (g_{ij} ) = \det ( \alpha_i \cdot \alpha_j ) $, using \eqref{BasisForInvSubLattice}. A similar analysis can be performed for the other conjugacy classes. The resulting massless spectrum of this model is shown in table \ref{ExampleModel}. Although in this paper only one example is shown, we can construct other models by considering other choices of shift vectors in \eqref{ShiftVectorInRoot} and \eqref{ShiftVectorInWeight}, or by regarding other (22,6)-dimensional Narain lattices with $\overline{E}_6$ or $\overline{A}_2^3$.

There are some characteristic properties of asymmetric orbifold models that we can read off from the example model. First, the number of fixed points is three. In asymmetric orbifolds, the number of $\eta_i$ with $\eta_i \neq 0$ can be smaller than in the symmetric case. Also, the volume of the invariant lattice $ I_{\theta}( \Gamma )$ can be larger. Both effects tend to reduce the number of fixed points. Next, a non-Abelian gauge flavor symmetry can appear in the twisted sector, as the chiral superfield $( {\bf 1}, {\bf 27}, {\bf 1},0, {\bf 3}, 0 )_{{\rm L}}$ in the example model, in which ${\bf 27}$ fields are unified into the fundamental representation of some $SU(3)$ flavor group. This type of situation is impossible in the symmetric orbifold construction.\footnote{Non-Abelian discrete flavor symmetries can be realized in symmetric orbifold models \cite{Kobayashi:2006wq,Ko:2007dz}.} This is due to the asymmetric orbifold action (no twist for the left-mover), by which the zero point energy for the left-mover becomes $-1$ ($\Delta c_{{\rm L}}  = 0$). Then, matter states that are nontrivially charged under another non-Abelian gauge symmetry can be massless. Also, for the orbifolds we consider, the gauge group has rank 22, so there is a rich source of gauge symmetries in a hidden sector.

\begin{table}
\begin{center}
\begin{tabular}{|c||c|}
\hline
Gauge symmetry    & $SO(14) \times E_6 \times SU(6) \times SU(3) 
\times U(1)^2$ \\
\hline
\hline
Untwisted sector    & $3( {\bf 1}, {\bf 1}, {\bf 6}, \overline{{\bf 3}}, 1/2,  0 )_{{\rm L}}$ \\
\hline
${\bf Z}_3$ twisted sector
&
\begin{tabular}{c} 
$3( {\bf 1},{\bf  1}, {\bf 20}, {\bf 1}, -1/6,  -2/3 )_{{\rm L}}$ \\
$3( {\bf 1}, {\bf 1}, \overline{{\bf 6}}, \overline{{\bf 3}}, 1/2,  0 )_{{\rm L}}$ \\
$3( {\bf 1}, {\bf 1}, {\bf 1}, \overline{{\bf 3}}, 2/3,  2/3 )_{{\rm L}}$  \\
$3( {\bf 1}, {\bf 1}, {\bf 1}, {\bf 1}, 1/3,  4/3 )_{{\rm L}}$  \\
$3( {\bf 1}, {\bf 27}, {\bf 1}, {\bf 3}, 0,  0 )_{{\rm L}}$ \\
$3( {\bf 1}, {\bf 27}, {\bf 1}, {\bf 1}, 1/3,  -2/3 )_{{\rm L}}$ \\
$3({\bf 14}_{v}, {\bf 1}, {\bf 1}, \overline{{\bf 3}}, -1/3,  -1/3 )_{{\rm L}}$ \\
$3({\bf 14}_{v}, {\bf 1}, {\bf 1}, {\bf 1}, -2/3,  1/3 )_{{\rm L}}$ \\
$3({\bf 64}_{c}, {\bf 1}, {\bf 1}, {\bf 1}, 1/3,  -1/6 )_{{\rm L}}$ \\
\end{tabular}
\\
\hline
$U(1)$ normalization    & $\sqrt{2}, 1$ \\
\hline
\end{tabular}
\caption[smallcaption]{Massless spectrum of an example ${\bf Z}_3$ 
asymmetric orbifold model using $D_7 \times E_6 \times A_8 
\times U(1) \times \overline{E}_6$ lattice. 
The gravity and gauge supermultiplets are omitted.}
\label{ExampleModel}
\end{center}
\end{table}


\section{Narain lattices with a right-moving non-abelian group factor}

In this section, we comment on other types of (22,6)-dimensional Narain lattices that can be used for ${\bf Z}_3$ asymmetric orbifold constructions. If we impose the condition that the right-moving group factor has to be non-Abelian, Narain lattices with $\overline{D}_4 \times\overline{A}_2$ can also be starting points. These lattices can also be constructed by the lattice engineering technique as follows. From 24-dimensional Euclidean lattices we construct 
\begin{description}
\item[(vi)] (18,2)-dimensional lattices with $\overline{A}_2$ by the lattice engineering $E_6 \to \overline{A}_2$ together with $D_4 \times \overline{D}_4$ lattice,
\item[(vii)] (20,4)-dimensional lattices with $\overline{D}_4$ by the lattice engineering $D_4 \to \overline{D}_4$ together with $A_2 \times \overline{A}_2$ lattice,
\item[(viii)] (14,6)-dimensional lattices with $\overline{A}_2 \times \overline{D}_4$ by the lattice engineering $E_6 \to \overline{A}_2$ and $D_4 \to \overline{D}_4$ together with $E_8$ lattice.
\end{description}
For (vi), six types of (18,2)-dimensional lattices with $\overline{A}_2$ are listed in table \ref{Tab.182DimLatticeWithA2bar}. For ${\bf Z}_3$ orbifold models, Narain lattices must have a ${\bf Z}_3$ rotation symmetry. That condition restricts possible lattices. Actually, among the lattices which contain $\overline{D}_4$, not all lattices do have the ${\bf Z}_3$ symmetry. For (vii), eight types of (20,4)-dimensional lattices with $\overline{D}_4$ satisfy the
${\bf Z}_3$ symmetry, and we list them in tables \ref{Tab.204DimLatticeWithD4bar1}, \ref{Tab.204DimLatticeWithD4bar2}. For (viii), two (14,6)-dimensional lattices are listed in table \ref{Tab.226DimLatticeWithD4A2bar}. There are 16 types of Narain lattices with $\overline{D}_4 \times \overline{A}_2$. As we can see in table \ref{Tab.LET}, it is possible to perform the engineering $U(1)^2 \to \overline{D}_4 \times \overline{A}_2$, however, we can show that the lattices built this way are already included in the above constructions, i.e. we can neglect such a possibility. 

In the case of Narain lattices with $\overline{D}_4$, we have to be careful about the ${\bf Z}_3$ discrete symmetry. Since we twist the right-mover we need to simultaneously rotate the left-mover. We can see that the left-moving twist leads to a rank reduction \cite{Ibanez:1987xa} of gauge groups or higher level Kac-Moody models \cite{higherlevel}. For example, if we consider a ${\bf Z}_3$ orbifold action on the (4,4)-dimensional $D_4 \times \overline{D}_4$ lattice which is given by conjugacy classes
\begin{align}
	( 0_{D_4}, 0_{\overline{D}_4} ), \ ( s_{D_4}, s_{\overline{D}_4} ), 
	\ ( v_{D_4}, v_{\overline{D}_4} ), \ ( c_{D_4}, c_{\overline{D}_4} ),
\end{align}
the ${\bf Z}_3$ action acts as $0_{\overline{D}_4} \to 0_{\overline{D}_4}$ and $s_{\overline{D}_4} \to v_{\overline{D}_4} \to c_{\overline{D}_4} \to s_{\overline{D}_4}$. So at the same time we have to consider a left-moving twist which maps $0_{D_4} \to 0_{D_4}$ and $s_{D_4} \to v_{D_4} \to c_{D_4} \to s_{D_4}$ in order to retain ${\bf Z}_3$ rotation symmetry of the lattice. The left-moving orbifold action gives rise to group breaking $D_4 \to G_2$ or $D_4 \to A_2$. Also for some lattices, e.g. the third lattice in table \ref{Tab.204DimLatticeWithD4bar1}, which originally has $A_3 \times A_{17}$ gauge symmetry, we have to include a left-moving orbifold action which permutes the three $A_1$ factors as $(A_1)_1 \to (A_1)_2 \to (A_1)_3 \to (A_1)_1$. This diagonal embedding action \cite{DiagonalEmbedding} realizes the $A_1$ group with Kac-Moody level $k=3$.


\section{Conclusions}

We have studied ${\bf Z}_3$ asymmetric orbifold models of heterotic string theory systematically. Narain lattices in (22,6) dimensions that can be used as starting point are classified by utilizing the lattice engineering technique with 24-dimensional lattices. We find 90 (22,6)-dimensional Narain lattices for the ${\bf Z}_3$ asymmetric orbifold construction without left-moving twist action. Possible breaking patterns of gauge symmetry arising from the lattices are listed in tables \ref{Tab:SMGUT1} and \ref{Tab:SMGUT2}. In these tables, we find that many lattices can give rise to the SM group and other interesting gauge groups. If we allow a left-moving twist action, extra 16 types of (22,6)-dimensional Narain lattices which have $\overline{D}_4 \times \overline{A}_2 $ in the right-mover can be used as starting point for model building. In that case, the resulting orbifold models will have gauge symmetry with a higher Kac-Moody level or with lower rank than 22.

Using the group breaking patterns and the Narain lattices collected in this paper, the next step should be to search for phenomenologically realistic models in the ${\bf Z}_3$ asymmetric orbifold vacua. The model building procedure in section 4 is applicable to those 90 Narain lattices with $\overline{E}_6$ or $\overline{A}_2^3$ which are listed in this paper. Also we can extend the procedure to the case of other asymmetric orbifold constructions like ${\bf Z}_6$, ${\bf Z}_{12}$ or ${\bf Z}_3 \times {\bf Z}_3$. Furthermore, in order to calculate superpotentials of effective theories, the study of string selection rules in asymmetric orbifolds will become necessary. Also, since patterns of masses and mixings depend on flavor symmetries, it will be important to consider flavor gauge symmetries and discrete flavor symmetries \cite{Kobayashi:2006wq,Ko:2007dz} related to "fixed points" as they arise from asymmetric orbifold compactifications.


\subsection*{Acknowledgement}
F.B. was supported by the Grant-in-Aid for the Nagoya University Global COE Program, "Quest for Fundamental Principles in the Universe: from Particles to the Solar System and the Cosmos", from the Ministry of Education, Culture, Sports, Science and Technology of Japan and by the "Leadership Development Program for Space Exploration and Research" from the Japan Society for the Promotion of Science. T.K. was supported in part by the Grant-in-Aid for Scientific Research No.~25400252 and the Grant-in-Aid for the Global COE Program "The Next Generation of Physics, Spun from Universality and Emergence" from the Ministry of Education, Culture,Sports, Science and Technology of Japan. S.K. was supported by the Taiwan's National Science Council under grant NSC102-2811-M-033-002.



\begin{table}[h]
\begin{center}
\scriptsize
\begin{tabular}{|c|c|}
\hline
 {Lattice name} & 1 \\ \hline
 {24-dim lattice} & $D_{24}$ \\ \hline
 {Lattice engineering} & $A_2 \to \overline{E}_6$ for $A_2 \in D_{24}$ \\ \hline
 {(22,6)-dim lattice} & $ D_{21} \times U(1) \times \overline{E}_6 $ \\ \hline
 {$U(1)$ normalization} & $2 \sqrt{3} $ \\ \hline
 {C.C. generators of (22,6)-dim lattice} & $\left(v,\frac{1}{6},1\right),\left(s,-\frac{1}{4},0\right)$ \\ \hline
 {C.C. generators related to shift vectors} & $\left(s,-\frac{1}{4},0\right)$ \\ \hline
\hline
 {Lattice name} & 2 \\ \hline
 {24-dim lattice} & $ D_{16} \times E_8 $ \\ \hline
 {Lattice engineering} & $A_2 \to \overline{E}_6$ for $A_2 \in D_{16}$ \\ \hline
 {(22,6)-dim lattice} & $ D_{13} \times E_8 \times U(1) \times \overline{E}_6 $ \\ \hline
 {$U(1)$ normalization} & $ 2 \sqrt{3} $ \\ \hline
 {C.C. generators of (22,6)-dim lattice} & $ \left(v,0,\frac{1}{6},1\right),\left(s,0,-\frac{1}{4},0\right)$ \\ \hline
 {C.C. generators related to shift vectors} & $\left(s,0,-\frac{1}{4},0\right)$ \\ \hline
\hline
 {Lattice name} & 3 \\ \hline
 {24-dim lattice} & $ D_{16} \times E_8 $ \\ \hline
 {Lattice engineering} & $A_2 \to \overline{E}_6$ for $A_2 \in E_{8}$ \\ \hline
 {(22,6)-dim lattice} & $ D_{16} \times E_6 \times \overline{E}_6 $ \\ \hline
 {$U(1)$ normalization} & --- \\ \hline
 {C.C. generators of (22,6)-dim lattice} & $(s,1,1)$ \\ \hline
 {C.C. generators related to shift vectors} & $(s,0,0)$ \\ \hline
\hline
 {Lattice name} & 4 \\ \hline
 {24-dim lattice} & $ E_8^3$ \\ \hline
 {Lattice engineering} & $A_2 \to \overline{E}_6$ for $A_2 \in E_{8}$ \\ \hline
 {(22,6)-dim lattice} & $ E_6 \times E_8^2 \times \overline{E}_6 $ \\ \hline
 {$U(1)$ normalization} & --- \\ \hline
 {C.C. generators of (22,6)-dim lattice} & $(1,0,0,1)$ \\ \hline
 {C.C. generators related to shift vectors} & --- \\ \hline
\hline
 {Lattice name} & 5 \\ \hline
 {24-dim lattice} & $ A_{24} $ \\ \hline
 {Lattice engineering} & $A_2 \to \overline{E}_6$ for $A_2 \in A_{24}$ \\ \hline
 {(22,6)-dim lattice} & $ A_{21} \times U(1) \times \overline{E}_6 $ \\ \hline
 {$U(1)$ normalization} & $5 \sqrt{66} $ \\ \hline
 {C.C. generators of (22,6)-dim lattice} & $ \left(1,\frac{1}{66},1\right),\left(5,\frac{1}{110},0\right) $ \\ \hline
 {C.C. generators related to shift vectors} & $ \left(5,\frac{1}{110},0\right) $ \\ \hline
\hline
 {Lattice name} & 6 \\ \hline
 {24-dim lattice} & $ D_{12}^2 $ \\ \hline
 {Lattice engineering} & $A_2 \to \overline{E}_6$ for $A_2 \in D_{12}$ \\ \hline
 {(22,6)-dim lattice} & $ D_{12} \times D_9 \times U(1) \times \overline{E}_6 $ \\ \hline
 {$U(1)$ normalization} & $ 2 \sqrt{3} $ \\ \hline
 {C.C. generators of (22,6)-dim lattice} & $ \left(0,v,\frac{1}{6},1\right),\left(v,s,-\frac{1}{4},0\right),(s,v,0,0) $ \\ \hline
 {C.C. generators related to shift vectors} & $ \left(v,s,-\frac{1}{4},0\right),(s,v,0,0) $ \\ \hline
\hline
 {Lattice name} & 7 \\ \hline
 {24-dim lattice} & $ A_{17} \times E_7 $ \\ \hline
 {Lattice engineering} & $A_2 \to \overline{E}_6$ for $A_2 \in A_{17}$ \\ \hline
 {(22,6)-dim lattice} & $ A_{14} \times E_7 \times U(1) \times \overline{E}_6 $ \\ \hline
 {$U(1)$ normalization} & $ 3 \sqrt{10} $ \\ \hline
 {C.C. generators of (22,6)-dim lattice} & $ \left(1,0,\frac{1}{15},1\right),\left(3,1,\frac{1}{30},0\right) $ \\ \hline
 {C.C. generators related to shift vectors} & $ \left(3,1,\frac{1}{30},0\right) $ \\ \hline
\end{tabular}
\caption[smallcaption]{(22,6)-dimensional lattices with $\overline{E}_6$. These lattices correspond to (22,6)-dimensional Narain lattices which are numbered \#1 $\ldots$ \#31.}
\label{Tab.226DimLatticeWithE6bar1}
\end{center}
\end{table}

\begin{table}[h]
\begin{center}
\scriptsize
\begin{tabular}{|c|c|}
\hline
 {Lattice name} & 8 \\ \hline
 {24-dim lattice} & $ A_{17} \times E_7 $ \\ \hline
 {Lattice engineering} & $A_2 \to \overline{E}_6$ for $A_2 \in E_{7}$ \\ \hline
 {(22,6)-dim lattice} & $ A_{17} \times A_5 \times \overline{E}_6 $ \\ \hline
 {$U(1)$ normalization} & --- \\ \hline
 {C.C. generators of (22,6)-dim lattice} & $(0,2,1),(3,3,0)$ \\ \hline
 {C.C. generators related to shift vectors} & $(3,3,0)$ \\ \hline
\hline
 {Lattice name} & 9 \\ \hline
 {24-dim lattice} & $ D_{10} \times E_7^2 $ \\ \hline
 {Lattice engineering} & $A_2 \to \overline{E}_6$ for $A_2 \in D_{10}$ \\ \hline
 {(22,6)-dim lattice} & $ D_7 \times E_7^2 \times U(1) \times \overline{E}_6 $ \\ \hline
 {$U(1)$ normalization} & $ 2 \sqrt{3} $ \\ \hline
 {C.C. generators of (22,6)-dim lattice} & $ \left(v,0,0,\frac{1}{6},1\right),\left(s,1,0,-\frac{1}{4},0\right),\left(c,0,1,-\frac{1}{4},0\right) $ \\ \hline
 {C.C. generators related to shift vectors} & $ \left(s,1,0,-\frac{1}{4},0\right),\left(c,0,1,-\frac{1}{4},0\right) $ \\ \hline
\hline
 {Lattice name} & 10 \\ \hline
 {24-dim lattice} & $ D_{10} \times E_7^2 $ \\ \hline
 {Lattice engineering} & $A_2 \to \overline{E}_6$ for $A_2 \in E_{7}$ \\ \hline
 {(22,6)-dim lattice} & $ A_5 \times D_{10} \times E_7 \times \overline{E}_6 $ \\ \hline
 {$U(1)$ normalization} & --- \\ \hline
 {C.C. generators of (22,6)-dim lattice} & $(3,s,0,0),(2,c,1,1)$ \\ \hline
 {C.C. generators related to shift vectors} & $(3,s,0,0),(0,c,1,0)$ \\ \hline
\hline
 {Lattice name} & 11 \\ \hline
 {24-dim lattice} & $ A_{15} \times D_9 $ \\ \hline
 {Lattice engineering} & $A_2 \to \overline{E}_6$ for $A_2 \in A_{15}$ \\ \hline
 {(22,6)-dim lattice} & $ A_{12} \times D_9 \times U(1) \times \overline{E}_6 $ \\ \hline
 {$U(1)$ normalization} & $ 4 \sqrt{39} $ \\ \hline
 {C.C. generators of (22,6)-dim lattice} & $ \left(1,0,\frac{1}{39},1\right),\left(2,s,\frac{1}{104},0\right) $ \\ \hline
 {C.C. generators related to shift vectors} & $ \left(2,s,\frac{1}{104},0\right) $ \\ \hline
\hline
 {Lattice name} & 12 \\ \hline
 {24-dim lattice} & $ A_{15} \times D_9 $ \\ \hline
 {Lattice engineering} & $A_2 \to \overline{E}_6$ for $A_2 \in D_{9}$ \\ \hline
 {(22,6)-dim lattice} & $ A_{15} \times D_6 \times U(1) \times \overline{E}_6 $ \\ \hline
 {$U(1)$ normalization} & $ 2 \sqrt{3} $ \\ \hline
 {C.C. generators of (22,6)-dim lattice} & $ \left(0,v,\frac{1}{6},1\right),\left(2,s,-\frac{1}{4},0\right) $ \\ \hline
 {C.C. generators related to shift vectors} & $ \left(2,s,-\frac{1}{4},0\right),\left(0,v,\frac{1}{2},0\right) $ \\ \hline
\hline
 {Lattice name} & 13 \\ \hline
 {24-dim lattice} & $ D_8^3 $ \\ \hline
 {Lattice engineering} & $A_2 \to \overline{E}_6$ for $A_2 \in D_{8}$ \\ \hline
 {(22,6)-dim lattice} & $ D_5 \times D_8^2 \times U(1) \times \overline{E}_6 $ \\ \hline
 {$U(1)$ normalization} & $ 2 \sqrt{3} $ \\ \hline
 {C.C. generators of (22,6)-dim lattice} & \begin{tabular}{c} $\left(v,0,0,\frac{1}{6},1\right), \left(s,v,v,-\frac{1}{4},0\right),$ \\ $(v,s,v,0,0),(v,v,s,0,0)$ \\ \end{tabular} \\ \hline
 {C.C. generators related to shift vectors} & $ \left(s,v,v,-\frac{1}{4},0\right),(v,s,v,0,0),(v,v,s,0,0) $ \\ \hline
\hline
 {Lattice name} & 14 \\ \hline
 {24-dim lattice} & $ A_{12}^2 $ \\ \hline
 {Lattice engineering} & $A_2 \to \overline{E}_6$ for $A_2 \in A_{12}$ \\ \hline
 {(22,6)-dim lattice} & $ A_{12} \times A_9 \times U(1) \times \overline{E}_6 $ \\ \hline
 {$U(1)$ normalization} & $ \sqrt{390} $ \\ \hline
 {C.C. generators of (22,6)-dim lattice} & $ \left(0,1,\frac{1}{30},1\right),\left(5,1,\frac{1}{130},0\right) $ \\ \hline
 {C.C. generators related to shift vectors} & $ \left(5,1,\frac{1}{130},0\right) $ \\ \hline
\end{tabular}
\caption[smallcaption]{(22,6)-dimensional lattices with $\overline{E}_6$ (continued). These lattices correspond to (22,6)-dimensional Narain lattices which are numbered \#1 $\ldots$ \#31.}
\label{Tab.226DimLatticeWithE6bar2}
\end{center}
\end{table}

\begin{table}[h]
\begin{center}
\scriptsize
\begin{tabular}{|c|c|}
\hline
 {Lattice name} & 15 \\ \hline
 {24-dim lattice} & $ A_{11} \times D_7 \times E_6 $ \\ \hline
 {Lattice engineering} & $A_2 \to \overline{E}_6$ for $A_2 \in A_{11}$ \\ \hline
 {(22,6)-dim lattice} & $ A_8 \times D_7 \times E_6 \times U(1) \times \overline{E}_6 $ \\ \hline
 {$U(1)$ normalization} & 6 \\ \hline
 {C.C. generators of (22,6)-dim lattice} & $\left(1,0,0,\frac{1}{9},1\right),\left(1,s,1,\frac{1}{36},0\right)$ \\ \hline
 {C.C. generators related to shift vectors} & $\left(1,s,1,\frac{1}{36},0\right)$ \\ \hline
\hline
 {Lattice name} & 16 \\ \hline
 {24-dim lattice} & $ A_{11} \times D_7 \times E_6 $ \\ \hline
 {Lattice engineering} & $A_2 \to \overline{E}_6$ for $A_2 \in D_{7}$ \\ \hline
 {(22,6)-dim lattice} & $ A_{11} \times D_4 \times E_6 \times U(1) \times \overline{E}_6 $ \\ \hline
 {$U(1)$ normalization} & $ 2 \sqrt{3} $ \\ \hline
 {C.C. generators of (22,6)-dim lattice} & $ \left(0,v,0,\frac{1}{6},1\right),\left(1,s,1,-\frac{1}{4},0\right) $ \\ \hline
 {C.C. generators related to shift vectors} & $ \left(1,s,1,-\frac{1}{4},0\right),\left(0,v,0,\frac{1}{2},0\right) $ \\ \hline
\hline
 {Lattice name} & 17 \\ \hline
 {24-dim lattice} & $ A_{11} \times D_7 \times E_6 $ \\ \hline
 {Lattice engineering} & $A_2 \to \overline{E}_6$ for $A_2 \in E_{6}$ \\ \hline
 {(22,6)-dim lattice} & $ A_{11} \times A_2^2 \times D_7 \times \overline{E}_6 $ \\ \hline
 {$U(1)$ normalization} & --- \\ \hline
 {C.C. generators of (22,6)-dim lattice} & $(0,1,2,0,1),(1,1,1,s,0)$ \\ \hline
 {C.C. generators related to shift vectors} & $(1,1,1,s,0)$ \\ \hline
\hline
 {Lattice name} & 18 \\ \hline
 {24-dim lattice} & $E_6^4 $ \\ \hline
 {Lattice engineering} & $A_2 \to \overline{E}_6$ for $A_2 \in E_{6}$ \\ \hline
 {(22,6)-dim lattice} & $ A_2^2 \times E_6^3 \times \overline{E}_6 $ \\ \hline
 {$U(1)$ normalization} & --- \\ \hline
 {C.C. generators of (22,6)-dim lattice} & $(1,2,0,0,0,1),(1,1,1,2,0,0),(1,1,2,0,1,0)$ \\ \hline
 {C.C. generators related to shift vectors} & $(1,1,1,2,0,0),(1,1,2,0,1,0)$ \\ \hline
\hline
 {Lattice name} & 19 \\ \hline
 {24-dim lattice} & $ A_9^2 \times D_6 $ \\ \hline
 {Lattice engineering} & $A_2 \to \overline{E}_6$ for $A_2 \in A_{9}$ \\ \hline
 {(22,6)-dim lattice} & $ A_6 \times A_9 \times D_6 \times U(1) \times \overline{E}_6 $ \\ \hline
 {$U(1)$ normalization} & $ \sqrt{210} $ \\ \hline
 {C.C. generators of (22,6)-dim lattice} & $ \left(2,4,0,\frac{1}{35},0\right),\left(2,0,s,-\frac{1}{14},0\right),\left(1,5,c,\frac{1}{21},1\right) $ \\ \hline
 {C.C. generators related to shift vectors} & $ \left(2,4,0,\frac{1}{35},0\right),\left(2,0,s,-\frac{1}{14},0\right),\left(3,5,c,\frac{1}{7},0\right) $ \\ \hline
\hline
 {Lattice name} & 20 \\ \hline
 {24-dim lattice} & $A_9^2 \times D_6 $ \\ \hline
 {Lattice engineering} & $A_2 \to \overline{E}_6$ for $A_2 \in D_{6}$ \\ \hline
 {(22,6)-dim lattice} & $ A_3 \times A_9^2 \times U(1) \times \overline{E}_6 $ \\ \hline
 {$U(1)$ normalization} & $ 2 \sqrt{3}$ \\ \hline
 {C.C. generators of (22,6)-dim lattice} & $ \left(2,2,4,\frac{1}{6},1\right),\left(1,5,0,-\frac{1}{4},0\right),\left(3,0,5,-\frac{1}{4},0\right) $ \\ \hline
 {C.C. generators related to shift vectors} & $ \left(1,5,0,-\frac{1}{4},0\right),\left(3,0,5,-\frac{1}{4},0\right),\left(2,2,4,\frac{1}{2},0\right) $ \\ \hline
\hline
 {Lattice name} & 21 \\ \hline
 {24-dim lattice} & $ D_6^4 $ \\ \hline
 {Lattice engineering} & $A_2 \to \overline{E}_6$ for $A_2 \in D_{6}$ \\ \hline
 {(22,6)-dim lattice} & $ A_3 \times D_6^3 \times U(1) \times \overline{E}_6 $ \\ \hline
 {$U(1)$ normalization} & $ 2 \sqrt{3} $ \\ \hline
 {C.C. generators of (22,6)-dim lattice} & \begin{tabular}{c} $\left(2,v,c,s,\frac{1}{6},1\right),\left(3,c,c,c,-\frac{1}{4},0\right), $ \\ $\left(3,v,s,0,-\frac{1}{4},0\right),(2,v,v,v,0,0) $  \\ \end{tabular} \\ \hline
 {C.C. generators related to shift vectors} & \begin{tabular}{c} $\left(3,v,s,0,-\frac{1}{4},0\right),(2,v,v,v,0,0),\left(2,c,s,v,\frac{1}{2},0\right),\left(2,s,v,c,\frac{1}{2},0\right)$ \\ \end{tabular} \\ \hline
\hline
 {Lattice name} & 22 \\ \hline
 {24-dim lattice} & $ A_8^3 $ \\ \hline
 {Lattice engineering} & $A_2 \to \overline{E}_6$ for $A_2 \in A_{8}$ \\ \hline
 {(22,6)-dim lattice} & $ A_5 \times A_8^2 \times U(1) \times \overline{E}_6 $ \\ \hline
 {$U(1)$ normalization} & $ 3 \sqrt{2} $ \\ \hline
 {C.C. generators of (22,6)-dim lattice} & $ \left(1,0,0,\frac{1}{6},1\right),\left(1,1,1,-\frac{5}{18},0\right),\left(1,4,1,\frac{1}{18},0\right) $ \\ \hline
 {C.C. generators related to shift vectors} & $ \left(1,1,1,-\frac{5}{18},0\right),\left(1,4,1,\frac{1}{18},0\right) $ \\ \hline
\end{tabular}
\caption[smallcaption]{(22,6)-dimensional lattices with $\overline{E}_6$ (continued). These lattices correspond to (22,6)-dimensional Narain lattices which are numbered \#1 $\ldots$ \#31.}
\label{Tab.226DimLatticeWithE6bar3}
\end{center}
\end{table}

\begin{table}[h]
\begin{center}
\scriptsize
\begin{tabular}{|c|c|}
\hline
 {Lattice name} & 23 \\ \hline
 {24-dim lattice} & $ A_7^2 \times D_5^2 $ \\ \hline
 {Lattice engineering} & $A_2 \to \overline{E}_6$ for $A_2 \in A_{7}$ \\ \hline
 {(22,6)-dim lattice} & $ A_4 \times A_7 \times D_5^2 \times U(1) \times \overline{E}_6 $ \\ \hline
 {$U(1)$ normalization} & $ 2 \sqrt{30} $ \\ \hline
 {C.C. generators of (22,6)-dim lattice} & $ \left(1,0,0,0,\frac{1}{15},1\right),\left(1,1,s,v,\frac{1}{40},0\right),\left(1,7,v,s,\frac{1}{40},0\right) $ \\ \hline
 {C.C. generators related to shift vectors} & $ \left(1,1,s,v,\frac{1}{40},0\right),\left(1,7,v,s,\frac{1}{40},0\right) $ \\ \hline
\hline
 {Lattice name} & 24 \\ \hline
 {24-dim lattice} & $ A_7^2 \times D_5^2 $ \\ \hline
 {Lattice engineering} & $A_2 \to \overline{E}_6$ for $A_2 \in D_{5}$ \\ \hline
 {(22,6)-dim lattice} & $ A_1^2 \times A_7^2 \times D_5 \times U(1) \times \overline{E}_6 $ \\ \hline
 {$U(1)$ normalization} & $ 2 \sqrt{3} $ \\ \hline
 {C.C. generators of (22,6)-dim lattice} & $ \left(1,1,0,0,0,\frac{1}{6},1\right),\left(0,1,1,1,v,-\frac{1}{4},0\right),(1,1,1,7,s,0,0) $ \\ \hline
 {C.C. generators related to shift vectors} & $ \left(0,1,1,1,v,-\frac{1}{4},0\right),(1,1,1,7,s,0,0),\left(1,1,0,0,0,\frac{1}{2},0\right) $ \\ \hline
\hline
 {Lattice name} & 25 \\ \hline
 {24-dim lattice} & $ A_6^4 $ \\ \hline
 {Lattice engineering} & $A_2 \to \overline{E}_6$ for $A_2 \in A_{6}$ \\ \hline
 {(22,6)-dim lattice} & $ A_3 \times A_6^3 \times U(1) \times \overline{E}_6 $ \\ \hline
 {$U(1)$ normalization} & $ 2 \sqrt{21} $ \\ \hline
 {C.C. generators of (22,6)-dim lattice} & $ \left(1,0,0,0,\frac{1}{12},1\right),\left(1,6,2,1,\frac{1}{28},0\right),\left(1,1,6,2,\frac{1}{28},0\right) $ \\ \hline
 {C.C. generators related to shift vectors} & $ \left(1,6,2,1,\frac{1}{28},0\right),\left(1,1,6,2,\frac{1}{28},0\right) $ \\ \hline
\hline
 {Lattice name} & 26 \\ \hline
 {24-dim lattice} & $ A_5^4 \times D_4 $ \\ \hline
 {Lattice engineering} & $A_2 \to \overline{E}_6$ for $A_2 \in A_{5}$ \\ \hline
 {(22,6)-dim lattice} & $ A_2 \times A_5^3 \times D_4 \times U(1) \times \overline{E}_6 $ \\ \hline
 {$U(1)$ normalization} & $ \sqrt{6} $ \\ \hline
 {C.C. generators of (22,6)-dim lattice} & \begin{tabular}{c} $ \left(1,0,0,0,0,\frac{1}{3},1\right),\left(2,4,0,2,0,\frac{1}{3},0\right),\left(2,2,4,0,0,\frac{1}{3},0\right),$ \\
$\left(0,3,0,0,s,\frac{1}{2},0\right),\left(0,0,3,0,v,\frac{1}{2},0\right),\left(0,0,0,3,c,\frac{1}{2},0\right) $ \\ \end{tabular} \\ \hline
 {C.C. generators related to shift vectors} & \begin{tabular}{c} $ \left(2,4,0,2,0,\frac{1}{3},0\right),\left(2,2,4,0,0,\frac{1}{3},0\right),\left(0,3,0,0,s,\frac{1}{2},0\right),$ \\
$\left(0,0,3,0,v,\frac{1}{2},0\right),\left(0,0,0,3,c,\frac{1}{2},0\right) $ \\ \end{tabular} \\ \hline
\hline
 {Lattice name} & 27 \\ \hline
 {24-dim lattice} & $ A_5^4 \times D_4 $ \\ \hline
 {Lattice engineering} & $A_2 \to \overline{E}_6$ for $A_2 \in D_{4}$ \\ \hline
 {(22,6)-dim lattice} & $ A_5^4 \times U(1)^2 \times \overline{E}_6 $ \\ \hline
 {$U(1)$ normalization} & $ 2,2 \sqrt{3} $ \\ \hline
 {C.C. generators of (22,6)-dim lattice} & \begin{tabular}{c} $ \left(2,4,0,2,\frac{1}{2},\frac{1}{6},1\right),(2,2,4,0,0,0,0), $ \\ $\left(2,2,4,0,\frac{1}{2},\frac{1}{6},1\right), \left(3,3,0,0,-\frac{1}{4},-\frac{1}{4},0\right), $ \\ $\left(3,0,3,0,\frac{1}{4},-\frac{1}{4},0\right),\left(3,0,0,3,\frac{1}{2},0,0\right) $ \\ \end{tabular} \\ \hline
 {C.C. generators related to shift vectors} & \begin{tabular}{c} $ \left(3,3,0,0,-\frac{1}{4},-\frac{1}{4},0\right),\left(3,0,3,0,\frac{1}{4},-\frac{1}{4},0\right), $ \\ $\left(3,0,0,3,\frac{1}{2},0,0\right), \left(2,4,0,2,\frac{1}{2},\frac{1}{2},0\right), $ \\ $\left(2,0,2,4,\frac{1}{2},\frac{1}{2},0\right)$ \\ \end{tabular} \\ \hline
\hline
 {Lattice name} & 28 \\ \hline
 {24-dim lattice} & $ D_4^6 $ \\ \hline
 {Lattice engineering} & $A_2 \to \overline{E}_6$ for $A_2 \in D_{4}$ \\ \hline
 {(22,6)-dim lattice} & $ D_4^5 \times U(1)^2 \times \overline{E}_6 $ \\ \hline
 {$U(1)$ normalization} & $ 2,2 \sqrt{3} $ \\ \hline
 {C.C. generators of (22,6)-dim lattice} & \begin{tabular}{c} $ \left(0,0,0,0,0,\frac{1}{2},\frac{1}{6},1\right),\left(s,s,s,s,s,-\frac{1}{4},-\frac{1}{4},0\right), $ \\ $\left(v,0,c,c,0,\frac{1}{4},-\frac{1}{4},0\right), \left(0,v,0,c,c,\frac{1}{4},-\frac{1}{4},0\right), $ \\ $\left(c,0,v,0,c,\frac{1}{4},-\frac{1}{4},0\right),\left(c,c,0,v,0,\frac{1}{4},-\frac{1}{4},0\right),$ \\ $\left(0,c,c,0,v,\frac{1}{4},-\frac{1}{4},0\right) $ \\ \end{tabular} \\ \hline
 {C.C. generators related to shift vectors} & \begin{tabular}{c} $ \left(s,s,s,s,s,-\frac{1}{4},-\frac{1}{4},0\right),\left(v,0,c,c,0,\frac{1}{4},-\frac{1}{4},0\right),$ \\ $\left(0,v,0,c,c,\frac{1}{4},-\frac{1}{4},0\right), \left(c,0,v,0,c,\frac{1}{4},-\frac{1}{4},0\right), $ \\ $ \left(c,c,0,v,0,\frac{1}{4},-\frac{1}{4},0\right),\left(0,c,c,0,v,\frac{1}{4},-\frac{1}{4},0\right) $ \\ \end{tabular} \\ \hline
\end{tabular}
\caption[smallcaption]{(22,6)-dimensional lattices with $\overline{E}_6$ (continued). These lattices correspond to (22,6)-dimensional Narain lattices which are numbered \#1 $\ldots$ \#31.}
\label{Tab.226DimLatticeWithE6bar4}
\end{center}
\end{table}

\begin{table}[h]
\begin{center}
\scriptsize
\begin{tabular}{|c|c|}
\hline
 {Lattice name} & 29 \\ \hline
 {24-dim lattice} & $ A_4^6 $ \\ \hline
 {Lattice engineering} & $A_2 \to \overline{E}_6$ for $A_2 \in A_{4}$ \\ \hline
 {(22,6)-dim lattice} & $ A_1 \times A_4^5 \times U(1) \times \overline{E}_6 $ \\ \hline
 {$U(1)$ normalization} & $ \sqrt{30} $ \\ \hline
 {C.C. generators of (22,6)-dim lattice} & \begin{tabular}{c} $ \left(1,0,0,0,0,0,\frac{1}{6},2\right),\left(0,4,1,0,1,4,\frac{1}{15},2\right),$ \\ $\left(0,4,4,1,0,1,\frac{1}{15},2\right),\left(0,1,4,4,1,0,\frac{1}{15},2\right) $ \\ \end{tabular} \\ \hline
 {C.C. generators related to shift vectors} & \begin{tabular}{c} $ \left(0,4,1,0,1,4,\frac{2}{5},0\right),\left(0,1,0,1,4,4,\frac{2}{5},0\right), $ \\ $\left(0,0,1,4,4,1,\frac{2}{5},0\right),\left(1,0,0,0,0,0,\frac{1}{2},0\right) $ \\ \end{tabular} \\ \hline
\hline
 {Lattice name} & 30 \\ \hline
 {24-dim lattice} & $ A_3^8 $ \\ \hline
 {Lattice engineering} & $A_2 \to \overline{E}_6$ for $A_2 \in A_{3}$ \\ \hline
 {(22,6)-dim lattice} & $ A_3^7 \times U(1) \times \overline{E}_6 $ \\ \hline
 {$U(1)$ normalization} & $ 2 \sqrt{3} $ \\ \hline
 {C.C. generators of (22,6)-dim lattice} & \begin{tabular}{c} $ \left(0,0,0,0,0,0,0,\frac{1}{3},2\right),\left(0,1,1,2,0,0,1,\frac{1}{4},0\right), $ \\ $\left(1,0,1,1,2,0,0,\frac{1}{4},0\right),\left(0,1,0,1,1,2,0,\frac{1}{4},0\right), $ \\ $\left(0,0,1,0,1,1,2,\frac{1}{4},0\right) $ \\ \end{tabular} \\ \hline
 {C.C. generators related to shift vectors} & \begin{tabular}{c} $ \left(0,1,1,2,0,0,1,\frac{1}{4},0\right),\left(1,0,1,1,2,0,0,\frac{1}{4},0\right), $ \\ $\left(0,1,0,1,1,2,0,\frac{1}{4},0\right),\left(0,0,1,0,1,1,2,\frac{1}{4},0\right) $ \\ \end{tabular} \\ \hline
\hline
 {Lattice name} & 31 \\ \hline
 {24-dim lattice} & $ A_2^{12} $ \\ \hline
 {Lattice engineering} & $A_2 \to \overline{E}_6$ for $A_2$ \\ \hline
 {(22,6)-dim lattice} & $ A_2^{11} \times \overline{E}_6 $ \\ \hline
 {$U(1)$ normalization} & --- \\ \hline
 {C.C. generators of (22,6)-dim lattice} & \begin{tabular}{c} $(2,2,2,1,2,1,1,2,1,1,1,2),(1,2,2,2,1,2,1,1,2,1,1,2), $ \\ $(1,1,2,2,2,1,2,1,1,2,1,2),(1,1,1,2,2,2,1,2,1,1,2,2), $ \\ $(2,1,1,1,2,2,2,1,2,1,1,2),(1,2,1,1,1,2,2,2,1,2,1,2)$ \\ \end{tabular} \\ \hline
 {C.C. generators related to shift vectors} & \begin{tabular}{c} $(1,1,0,0,1,0,2,0,2,0,2,0),(1,1,2,1,0,0,0,2,2,0,0,0), $ \\ $(1,0,0,0,0,1,2,2,2,1,0,0),(0,1,2,0,1,0,2,2,0,1,0,0), $ \\ $(1,0,2,1,0,0,2,0,0,1,2,0) $ \\ \end{tabular} \\ \hline
\end{tabular}
\caption[smallcaption]{(22,6)-dimensional lattices with $\overline{E}_6$ (continued). These lattices correspond to (22,6)-dimensional Narain lattices which are numbered \#1 $\ldots$ \#31.}
\label{Tab.226DimLatticeWithE6bar5}
\end{center}
\end{table}

\begin{table}[h]
\begin{center}
\scriptsize
\begin{tabular}{|c|c|}
\hline
 {Lattice name} & 1 \\ \hline
 {24-dim lattice} & $ D_{16} \times E_8 $ \\ \hline
 {Lattice engineering} & $ E_6 \to \overline{A}_2 $ for $E_6 \in E_{8}$ \\ \hline
 {(18,2)-dim lattice} & $ A_2 \times D_{16} \times \overline{A}_2 $ \\ \hline
 {$U(1)$ normalization} & --- \\ \hline
 {C.C. generators of (18,2)-dim lattice} & $(1,s,1)$ \\ \hline
 {C.C. generators related to shift vectors} & $(0,s,0)$ \\ \hline
\hline
 {Lattice name} & 2 \\ \hline
 {24-dim lattice} & $ E_8^3 $ \\ \hline
 {Lattice engineering} & $ E_6 \to \overline{A}_2 $ for $E_6 \in E_{8}$ \\ \hline
 {(18,2)-dim lattice} & $ A_2 \times E_8^2 \times \overline{A}_2 $ \\ \hline
 {$U(1)$ normalization} & --- \\ \hline
 {C.C. generators of (18,2)-dim lattice} & $(1,0,0,1)$ \\ \hline
 {C.C. generators related to shift vectors} & --- \\ \hline
\hline
 {Lattice name} & 3 \\ \hline
 {24-dim lattice} & $ A_{17} \times E_7 $ \\ \hline
 {Lattice engineering} & $ E_6 \to \overline{A}_2 $ for $E_6 \in E_{7}$ \\ \hline
 {(18,2)-dim lattice} & $ A_{17} \times U(1) \times \overline{A}_2 $ \\ \hline
 {$U(1)$ normalization} & $ \sqrt{6} $ \\ \hline
 {C.C. generators of (18,2)-dim lattice} & $ \left(0,\frac{1}{3},1\right),\left(3,-\frac{1}{6},1\right) $ \\ \hline
 {C.C. generators related to shift vectors} & $ \left(3,\frac{1}{2},0\right) $ \\ \hline
\hline
 {Lattice name} & 4 \\ \hline
 {24-dim lattice} & $ D_{10} \times E_7^2 $ \\ \hline
 {Lattice engineering} & $ E_6 \to \overline{A}_2 $ for $E_6 \in E_{7}$ \\ \hline
 {(18,2)-dim lattice} & $ D_{10} \times E_7 \times U(1) \times \overline{A}_2 $ \\ \hline
 {$U(1)$ normalization} & $ \sqrt{6} $ \\ \hline
 {C.C. generators of (18,2)-dim lattice} & $ \left(s,0,-\frac{1}{6},1\right),(c,1,0,0) $ \\ \hline
 {C.C. generators related to shift vectors} & $ \left(s,0,\frac{1}{2},0\right),\left(v,1,\frac{1}{2},0\right) $ \\ \hline
\hline
 {Lattice name} & 5 \\ \hline
 {24-dim lattice} & $ A_{11} \times D_7 \times E_6 $ \\ \hline
 {Lattice engineering} & $ E_6 \to \overline{A}_2 $ for $E_6 $ \\ \hline
 {(18,2)-dim lattice} & $ A_{11} \times D_7 \times \overline{A}_2 $ \\ \hline
 {$U(1)$ normalization} & --- \\ \hline
 {C.C. generators of (18,2)-dim lattice} & $(1,s,1)$ \\ \hline
 {C.C. generators related to shift vectors} & $(3,c,0)$ \\ \hline
\hline
 {Lattice name} & 6 \\ \hline
 {24-dim lattice} & $ E_6^4 $ \\ \hline
 {Lattice engineering} & $ E_6 \to \overline{A}_2 $ for $E_6 $ \\ \hline
 {(18,2)-dim lattice} & $ E_6^3 \times \overline{A}_2 $ \\ \hline
 {$U(1)$ normalization} & --- \\ \hline
 {C.C. generators of (18,2)-dim lattice} & $(1,2,0,1),(2,0,1,1)$ \\ \hline
 {C.C. generators related to shift vectors} & $(1,1,1,0)$ \\ \hline
\end{tabular}
\caption[smallcaption]{(18,2)-dimensional lattices with $\overline{A}_2$. These lattices combined with two $A_2 \times \overline{A}_2 $ lattices correspond to (22,6)-dimensional Narain lattices which are numbered \#32 $\ldots$ \#37. Also, these lattices combined with the $D_4 \times \overline{D}_4 $ lattice lead to (22,6)-dimensional Narain lattices.}
\label{Tab.182DimLatticeWithA2bar}
\end{center}
\end{table}

\clearpage

\begin{table}[h]
\begin{center}
\scriptsize
\begin{tabular}{|c|c|}
\hline
 {Lattice name} & 1 \\ \hline
 {24-dim lattice} & $ D_{24} $ \\ \hline
 {Lattice engineering} & $ A_2^2 \to \overline{A}_2^2 $ for $A_2^2 \in D_{24}$ \\ \hline
 {(20,4)-dim lattice} & $ D_{18} \times U(1)^2 \times \overline{A}_2^2 $ \\ \hline
 {$U(1)$ normalization} & $ 2 \sqrt{3},2 \sqrt{3} $ \\ \hline
 {C.C. generators of (20,4)-dim lattice} & $ \left(v,\frac{1}{6},0,2,1\right),\left(v,0,\frac{1}{6},1,1\right),\left(s,-\frac{1}{4},-\frac{1}{4},0,0\right) $ \\ \hline
 {C.C. generators related to shift vectors} & $ \left(s,-\frac{1}{4},-\frac{1}{4},0,0\right),\left(v,\frac{1}{2},0,0,0\right) $ \\ \hline
\hline
 {Lattice name} & 2 \\ \hline
 {24-dim lattice} & $ D_{16} \times E_8 $ \\ \hline
 {Lattice engineering} & $ A_2^2 \to \overline{A}_2^2 $ for $A_2^2 \in D_{16}$ \\ \hline
 {(20,4)-dim lattice} & $ D_{10} \times E_8 \times U(1)^2 \times \overline{A}_2^2 $ \\ \hline
 {$U(1)$ normalization} &$ 2 \sqrt{3},2 \sqrt{3} $ \\ \hline
 {C.C. generators of (20,4)-dim lattice} & $ \left(v,0,\frac{1}{6},0,2,1\right),\left(v,0,0,\frac{1}{6},1,1\right),\left(s,0,-\frac{1}{4},-\frac{1}{4},0,0\right) $ \\ \hline
 {C.C. generators related to shift vectors} & $ \left(s,0,-\frac{1}{4},-\frac{1}{4},0,0\right),\left(v,0,\frac{1}{2},0,0,0\right) $ \\ \hline
\hline
 {Lattice name} & 3 \\ \hline
 {24-dim lattice} & $ D_{16} \times E_8 $ \\ \hline
 {Lattice engineering} & $ A_2^2 \to \overline{A}_2^2 $ for $A_2 \in D_{16}$ and $ A_2 \in E_8 $ \\ \hline
 {(20,4)-dim lattice} & $ D_{13} \times E_6 \times U(1) \times \overline{A}_2^2 $ \\ \hline
 {$U(1)$ normalization} & $ 2 \sqrt{3} $ \\ \hline
 {C.C. generators of (20,4)-dim lattice} & $ \left(v,0,\frac{1}{6},2,1\right),\left(s,1,-\frac{1}{4},1,1\right) $ \\ \hline
 {C.C. generators related to shift vectors} & $ \left(s,0,\frac{3}{4},0,0\right) $ \\ \hline
\hline
 {Lattice name} & 4 \\ \hline
 {24-dim lattice} & $E_8^3 $ \\ \hline
 {Lattice engineering} & $ A_2^2 \to \overline{A}_2^2 $ for $A_2 \in E_{8} $ and $ A_2 \in E_8'$ \\ \hline
 {(20,4)-dim lattice} & $ E_6^2 \times E_8 \times \overline{A}_2^2 $ \\ \hline
 {$U(1)$ normalization} & --- \\ \hline
 {C.C. generators of (20,4)-dim lattice} & $(1,0,0,2,1),(0,1,0,1,1)$ \\ \hline
 {C.C. generators related to shift vectors} & --- \\ \hline
\hline
 {Lattice name} & 5 \\ \hline
 {24-dim lattice} & $ A_{24} $ \\ \hline
 {Lattice engineering} & $ A_2^2 \to \overline{A}_2^2 $ for $A_2^2 \in A_{24}$ \\ \hline
 {(20,4)-dim lattice} & $ A_{18} \times U(1)^2 \times \overline{A}_2^2 $ \\ \hline
 {$U(1)$ normalization} & $ \sqrt{1254},5 \sqrt{66} $ \\ \hline
 {C.C. generators of (20,4)-dim lattice} & $ \left(1,\frac{1}{418},\frac{1}{66},2,1\right),\left(1,\frac{1}{57},0,1,1\right),\left(5,\frac{5}{418},\frac{1}{110},0,0\right) $ \\ \hline
 {C.C. generators related to shift vectors} & $ \left(5,\frac{5}{418},\frac{1}{110},0,0\right) $ \\ \hline
\hline
 {Lattice name} & 6 \\ \hline
 {24-dim lattice} & $ D_{12}^2 $ \\ \hline
 {Lattice engineering} & $ A_2^2 \to \overline{A}_2^2 $ for $A_2^2 \in D_{12}$ \\ \hline
 {(20,4)-dim lattice} & $ D_{12} \times D_6 \times U(1)^2 \times \overline{A}_2^2 $ \\ \hline
 {$U(1)$ normalization} & $ 2 \sqrt{3},2 \sqrt{3} $ \\ \hline
 {C.C. generators of (20,4)-dim lattice} & $ \left(0,v,\frac{1}{6},0,2,1\right),\left(0,v,0,\frac{1}{6},1,1\right),\left(v,s,-\frac{1}{4},-\frac{1}{4},0,0\right),(s,v,0,0,0,0) $ \\ \hline
 {C.C. generators related to shift vectors} & $ \left(v,s,-\frac{1}{4},-\frac{1}{4},0,0\right),(s,v,0,0,0,0),\left(0,v,\frac{1}{2},0,0,0\right) $ \\ \hline
\hline
 {Lattice name} & 7 \\ \hline
 {24-dim lattice} & $ D_{12}^2 $ \\ \hline
 {Lattice engineering} & $ A_2^2 \to \overline{A}_2^2 $ for $A_2 \in D_{12}$ and $ A_2 \in D_{12}' $ \\ \hline
 {(20,4)-dim lattice} & $ D_9^2 \times U(1)^2 \times \overline{A}_2^2 $ \\ \hline
 {$U(1)$ normalization} & $ 2 \sqrt{3},2 \sqrt{3} $ \\ \hline
 {C.C. generators of (20,4)-dim lattice} & $ \left(v,0,\frac{1}{6},0,2,1\right),\left(0,v,0,\frac{1}{6},1,1\right),\left(s,v,-\frac{1}{4},0,0,0\right),\left(v,s,0,-\frac{1}{4},0,0\right) $ \\ \hline
 {C.C. generators related to shift vectors} & $ \left(s,v,-\frac{1}{4},0,0,0\right),\left(v,s,0,-\frac{1}{4},0,0\right) $ \\ \hline
\end{tabular}
\caption[smallcaption]{(20,4)-dimensional lattices with $\overline{A}_2^2$. These lattices combined with the $A_2 \times \overline{A}_2 $ lattice correspond to (22,6)-dimensional Narain lattices which are numbered \#38 $\ldots$ \#83.}
\label{Tab.204DimLatticeWithA2A2bar1}
\end{center}
\end{table}

\begin{table}[h]
\begin{center}
\scriptsize
\begin{tabular}{|c|c|}
\hline
 {Lattice name} & 8 \\ \hline
 {24-dim lattice} & $ A_{17} \times E_7 $ \\ \hline
 {Lattice engineering} & $ A_2^2 \to \overline{A}_2^2 $ for $A_2^2 \in A_{17}$ \\ \hline
 {(20,4)-dim lattice} & $ A_{11} \times E_7 \times U(1)^2 \times \overline{A}_2^2 $ \\ \hline
 {$U(1)$ normalization} & $ 2 \sqrt{15},3 \sqrt{10} $ \\ \hline
 {C.C. generators of (20,4)-dim lattice} & $ \left(1,0,\frac{1}{60},\frac{1}{15},2,1\right),\left(1,0,\frac{1}{12},0,1,1\right),\left(3,1,\frac{1}{20},\frac{1}{30},0,0\right) $ \\ \hline
 {C.C. generators related to shift vectors} & $ \left(3,1,\frac{1}{20},\frac{1}{30},0,0\right),\left(3,0,\frac{1}{20},\frac{1}{5},0,0\right) $ \\ \hline
\hline
 {Lattice name} & 9 \\ \hline
 {24-dim lattice} & $ A_{17} \times E_7 $ \\ \hline
 {Lattice engineering} & $ A_2^2 \to \overline{A}_2^2 $ for $A_2 \in A_{17}$ and $ A_2 \in E_7 $ \\ \hline
 {(20,4)-dim lattice} & $ A_{14} \times A_5 \times U(1) \times \overline{A}_2^2 $ \\ \hline
 {$U(1)$ normalization} & $ 3 \sqrt{10} $ \\ \hline
 {C.C. generators of (20,4)-dim lattice} & $ \left(1,0,\frac{1}{15},2,1\right),(0,2,0,1,1),\left(3,3,\frac{1}{30},0,0\right) $ \\ \hline
 {C.C. generators related to shift vectors} & $ \left(3,3,\frac{1}{30},0,0\right) $ \\ \hline
\hline
 {Lattice name} & 10 \\ \hline
 {24-dim lattice} & $ D_{10} \times E_7^2 $ \\ \hline
 {Lattice engineering} & $ A_2^2 \to \overline{A}_2^2 $ for $A_2^2 \in D_{10}$ \\ \hline
 {(20,4)-dim lattice} & $ D_4 \times E_7^2 \times U(1)^2 \times \overline{A}_2^2 $ \\ \hline
 {$U(1)$ normalization} & $ 2 \sqrt{3},2 \sqrt{3} $ \\ \hline
 {C.C. generators of (20,4)-dim lattice} &  \begin{tabular}{c} $ \left(v,0,0,\frac{1}{6},0,2,1\right),\left(v,0,0,0,\frac{1}{6},1,1\right), $ \\ $ 
\left(s,1,0,-\frac{1}{4},-\frac{1}{4},0,0\right),\left(c,0,1,-\frac{1}{4},-\frac{1}{4},0,0\right) $ \\  \end{tabular} \\ \hline
 {C.C. generators related to shift vectors} & $ \left(s,1,0,-\frac{1}{4},-\frac{1}{4},0,0\right),\left(c,0,1,-\frac{1}{4},-\frac{1}{4},0,0\right),\left(v,0,0,\frac{1}{2},0,0,0\right) $ \\ \hline
\hline
 {Lattice name} & 11 \\ \hline
 {24-dim lattice} & $ D_{10} \times E_7^2 $ \\ \hline
 {Lattice engineering} & $ A_2^2 \to \overline{A}_2^2 $ for $A_2 \in D_{10}$ and $ A_2 \in E_7 $ \\ \hline
 {(20,4)-dim lattice} & $ A_5 \times D_7 \times E_7 \times U(1) \times \overline{A}_2^2 $ \\ \hline
 {$U(1)$ normalization} & $ 2 \sqrt{3} $ \\ \hline
 {C.C. generators of (20,4)-dim lattice} & $ \left(0,v,0,\frac{1}{6},2,1\right),\left(3,s,0,-\frac{1}{4},0,0\right),\left(2,c,1,-\frac{1}{4},1,1\right) $ \\ \hline
 {C.C. generators related to shift vectors} & $ \left(3,s,0,-\frac{1}{4},0,0\right),\left(0,c,1,\frac{3}{4},0,0\right) $ \\ \hline
\hline
 {Lattice name} & 12 \\ \hline
 {24-dim lattice} & $ D_{10} \times E_7^2 $ \\ \hline
 {Lattice engineering} & $ A_2^2 \to \overline{A}_2^2 $ for $A_2 \in E_{7}$ and $ A_2 \in E_7' $ \\ \hline
 {(20,4)-dim lattice} & $ A_5^2 \times D_{10} \times \overline{A}_2^2 $ \\ \hline
 {$U(1)$ normalization} & --- \\ \hline
 {C.C. generators of (20,4)-dim lattice} & $(3,2,s,1,1),(2,3,c,2,1)$ \\ \hline
 {C.C. generators related to shift vectors} & $(0,3,c,0,0),(3,3,v,0,0)$ \\ \hline
\hline
 {Lattice name} & 13 \\ \hline
 {24-dim lattice} & $ A_{15} \times D_9 $ \\ \hline
 {Lattice engineering} & $ A_2^2 \to \overline{A}_2^2 $ for $A_2^2 \in A_{15}$ \\ \hline
 {(20,4)-dim lattice} & $ A_9 \times D_9 \times U(1)^2 \times \overline{A}_2^2 $ \\ \hline
 {$U(1)$ normalization} & $ \sqrt{390},4 \sqrt{39} $ \\ \hline
 {C.C. generators of (20,4)-dim lattice} & $ \left(1,0,\frac{1}{130},\frac{1}{39},2,1\right),\left(1,0,\frac{1}{30},0,1,1\right),\left(2,s,\frac{1}{65},\frac{1}{104},0,0\right) $ \\ \hline
 {C.C. generators related to shift vectors} & $ \left(2,s,\frac{1}{65},\frac{1}{104},0,0\right),\left(3,0,\frac{3}{130},\frac{1}{13},0,0\right) $ \\ \hline
\hline
 {Lattice name} & 14 \\ \hline
 {24-dim lattice} & $ A_{15} \times D_9 $ \\ \hline
 {Lattice engineering} & $ A_2^2 \to \overline{A}_2^2 $ for $A_2^2 \in D_{9}$ \\ \hline
 {(20,4)-dim lattice} & $ A_{15} \times A_3 \times U(1)^2 \times \overline{A}_2^2 $ \\ \hline
 {$U(1)$ normalization} & $ 2 \sqrt{3},2 \sqrt{3} $ \\ \hline
 {C.C. generators of (20,4)-dim lattice} & $ \left(0,2,\frac{1}{6},0,2,1\right),\left(0,2,0,\frac{1}{6},1,1\right),\left(2,1,-\frac{1}{4},-\frac{1}{4},0,0\right) $ \\ \hline
 {C.C. generators related to shift vectors} & $ \left(2,1,-\frac{1}{4},-\frac{1}{4},0,0\right),\left(0,2,\frac{1}{2},0,0,0\right),\left(0,2,0,\frac{1}{2},0,0\right) $ \\ \hline
\end{tabular}
\caption[smallcaption]{(20,4)-dimensional lattices with $\overline{A}_2^2$ (continued). These lattices combined with the $A_2 \times \overline{A}_2 $ lattice correspond to (22,6)-dimensional Narain lattices which are numbered \#38 $\ldots$ \#83.}
\label{Tab.204DimLatticeWithA2A2bar2}
\end{center}
\end{table}

\begin{table}[h]
\begin{center}
\scriptsize
\begin{tabular}{|c|c|}
\hline
 {Lattice name} & 15 \\ \hline
 {24-dim lattice} & $ A_{15} \times D_9 $ \\ \hline
 {Lattice engineering} & $ A_2^2 \to \overline{A}_2^2 $ for $A_2 \in A_{15}$ and $ A_2 \in D_9 $ \\ \hline
 {(20,4)-dim lattice} & $ A_{12} \times D_6 \times U(1)^2 \times \overline{A}_2^2 $ \\ \hline
 {$U(1)$ normalization} & $ 2 \sqrt{3},4 \sqrt{39} $ \\ \hline
 {C.C. generators of (20,4)-dim lattice} & $ \left(1,0,0,\frac{1}{39},2,1\right),\left(0,v,\frac{1}{6},0,1,1\right),\left(2,s,-\frac{1}{4},\frac{1}{104},0,0\right) $ \\ \hline
 {C.C. generators related to shift vectors} & $ \left(2,s,-\frac{1}{4},\frac{1}{104},0,0\right),\left(3,v,\frac{1}{2},\frac{1}{13},0,0\right) $ \\ \hline
\hline
 {Lattice name} & 16 \\ \hline
 {24-dim lattice} & $ D_8^3 $ \\ \hline
 {Lattice engineering} & $ A_2^2 \to \overline{A}_2^2 $ for $A_2^2 \in D_{8}$ \\ \hline
 {(20,4)-dim lattice} & $ A_1^2 \times D_8^2 \times U(1)^2 \times \overline{A}_2^2 $ \\ \hline
 {$U(1)$ normalization} & $ 2 \sqrt{3},2 \sqrt{3} $ \\ \hline
 {C.C. generators of (20,4)-dim lattice} & \begin{tabular}{c} $ \left(1,1,0,0,\frac{1}{6},0,2,1\right),\left(1,1,0,0,0,\frac{1}{6},1,1\right),\left(0,1,v,v,-\frac{1}{4},-\frac{1}{4},0,0\right), $ \\ $ (1,1,s,v,0,0,0,0),(1,1,v,s,0,0,0,0)$ \\ \end{tabular} \\ \hline
 {C.C. generators related to shift vectors} & \begin{tabular}{c} $ \left(0,1,v,v,-\frac{1}{4},-\frac{1}{4},0,0\right),(1,1,s,v,0,0,0,0), $ \\ $ 
(1,1,v,s,0,0,0,0),\left(1,1,0,0,\frac{1}{2},0,0,0\right) $ \\ \end{tabular} \\ \hline
\hline
 {Lattice name} & 17 \\ \hline
 {24-dim lattice} & $ D_8^3 $ \\ \hline
 {Lattice engineering} & $ A_2^2 \to \overline{A}_2^2 $ for $A_2 \in D_{8}$ and $ A_2 \in D_8' $ \\ \hline
 {(20,4)-dim lattice} & $ D_5^2 \times D_8 \times U(1)^2 \times \overline{A}_2^2 $ \\ \hline
 {$U(1)$ normalization} & $ 2 \sqrt{3},2 \sqrt{3} $ \\ \hline
 {C.C. generators of (20,4)-dim lattice} & \begin{tabular}{c} $\left(v,0,0,\frac{1}{6},0,2,1\right),\left(0,v,0,0,\frac{1}{6},1,1\right),\left(s,v,v,-\frac{1}{4},0,0,0\right), $ \\ $ \left(v,s,v,0,-\frac{1}{4},0,0\right),(v,v,s,0,0,0,0)$ \\ \end{tabular} \\ \hline
 {C.C. generators related to shift vectors} & $ \left(s,v,v,-\frac{1}{4},0,0,0\right),\left(v,s,v,0,-\frac{1}{4},0,0\right),(v,v,s,0,0,0,0) $ \\ \hline
\hline
 {Lattice name} & 18 \\ \hline
 {24-dim lattice} & $ A_{12}^2 $ \\ \hline
 {Lattice engineering} & $ A_2^2 \to \overline{A}_2^2 $ for $A_2^2 \in A_{12}$  \\ \hline
 {(20,4)-dim lattice} & $ A_{12} \times A_6 \times U(1)^2 \times \overline{A}_2^2 $ \\ \hline
 {$U(1)$ normalization} & $ \sqrt{210},\sqrt{390} $ \\ \hline
 {C.C. generators of (20,4)-dim lattice} & $ \left(0,1,\frac{1}{70},\frac{1}{30},2,1\right),\left(0,1,\frac{1}{21},0,1,1\right),\left(5,1,\frac{1}{70},\frac{1}{130},0,0\right) $ \\ \hline
 {C.C. generators related to shift vectors} & $ \left(5,1,\frac{1}{70},\frac{1}{130},0,0\right) $ \\ \hline
\hline
 {Lattice name} & 19 \\ \hline
 {24-dim lattice} & $ A_{12}^2 $ \\ \hline
 {Lattice engineering} & $ A_2^2 \to \overline{A}_2^2 $ for $A_2 \in A_{12}$ and $ A_2 \in A_{12}' $ \\ \hline
 {(20,4)-dim lattice} & $ A_9^2 \times U(1)^2 \times \overline{A}_2^2 $ \\ \hline
 {$U(1)$ normalization} & $ \sqrt{390},\sqrt{390} $ \\ \hline
 {C.C. generators of (20,4)-dim lattice} & $ \left(1,0,\frac{1}{30},0,2,1\right),\left(0,1,0,\frac{1}{30},1,1\right),\left(1,5,\frac{1}{130},\frac{1}{26},0,0\right) $ \\ \hline
 {C.C. generators related to shift vectors} & $ \left(1,5,\frac{1}{130},\frac{1}{26},0,0\right),\left(3,3,\frac{1}{10},\frac{1}{10},0,0\right), \left(3,0,\frac{1}{10},0,0,0\right) $ \\ \hline
\hline
 {Lattice name} & 20 \\ \hline
 {24-dim lattice} & $ A_{11} \times D_7 \times E_6 $ \\ \hline
 {Lattice engineering} & $ A_2^2 \to \overline{A}_2^2 $ for $A_2^2 \in A_{11}$ \\ \hline
 {(20,4)-dim lattice} & $ A_5 \times D_7 \times E_6 \times U(1)^2 \times \overline{A}_2^2 $ \\ \hline
 {$U(1)$ normalization} & $ 3 \sqrt{2},6 $ \\ \hline
 {C.C. generators of (20,4)-dim lattice} & $ \left(1,0,0,\frac{1}{18},\frac{1}{9},2,1\right),\left(1,0,0,\frac{1}{6},0,1,1\right),\left(1,s,1,\frac{1}{18},\frac{1}{36},0,0\right) $ \\ \hline
 {C.C. generators related to shift vectors} & $ \left(1,s,1,\frac{1}{18},\frac{1}{36},0,0\right),\left(3,0,0,\frac{1}{6},\frac{1}{3},0,0\right) $ \\ \hline
\hline
 {Lattice name} & 21 \\ \hline
 {24-dim lattice} & $ A_{11} \times D_7 \times E_6 $ \\ \hline
 {Lattice engineering} & $ A_2^2 \to \overline{A}_2^2 $ for $A_2^2 \in D_{7}$ \\ \hline
 {(20,4)-dim lattice} & $ A_{11} \times E_6 \times U(1)^3 \times \overline{A}_2^2 $ \\ \hline
 {$U(1)$ normalization} & $ 2,2 \sqrt{3},2 \sqrt{3} $ \\ \hline
 {C.C. generators of (20,4)-dim lattice} & $ \left(0,0,\frac{1}{4},\frac{1}{6},-\frac{1}{4},2,1\right),\left(0,0,\frac{1}{2},0,\frac{1}{6},1,1\right),\left(1,1,-\frac{1}{4},-\frac{1}{4},-\frac{1}{4},0,0\right) $ \\ \hline
 {C.C. generators related to shift vectors} & $ \left(1,1,-\frac{1}{4},-\frac{1}{4},-\frac{1}{4},0,0\right),\left(0,0,\frac{3}{4},\frac{1}{2},\frac{1}{4},0,0\right) $ \\ \hline
\end{tabular}
\caption[smallcaption]{(20,4)-dimensional lattices with $\overline{A}_2^2$ (continued). These lattices combined with the $A_2 \times \overline{A}_2 $ lattice correspond to (22,6)-dimensional Narain lattices which are numbered \#38 $\ldots$ \#83.}
\label{Tab.204DimLatticeWithA2A2bar3}
\end{center}
\end{table}

\begin{table}[h]
\begin{center}
\scriptsize
\begin{tabular}{|c|c|}
\hline
 {Lattice name} & 22 \\ \hline
 {24-dim lattice} & $ A_{11} \times D_7 \times E_6 $ \\ \hline
 {Lattice engineering} & $ A_2^2 \to \overline{A}_2^2 $ for $A_2 \in A_{11}$ and $ A_2 \in D_7$ \\ \hline
 {(20,4)-dim lattice} & $ A_8 \times D_4 \times E_6 \times U(1)^2 \times \overline{A}_2^2 $ \\ \hline
 {$U(1)$ normalization} & $ 2 \sqrt{3},6 $ \\ \hline
 {C.C. generators of (20,4)-dim lattice} & $ \left(1,0,0,0,\frac{1}{9},2,1\right),\left(0,v,0,\frac{1}{6},0,1,1\right),\left(1,s,1,-\frac{1}{4},\frac{1}{36},0,0\right) $ \\ \hline
 {C.C. generators related to shift vectors} & $ \left(1,s,1,-\frac{1}{4},\frac{1}{36},0,0\right),\left(3,v,0,\frac{1}{2},\frac{1}{3},0,0\right) $ \\ \hline
\hline
 {Lattice name} & 23 \\ \hline
 {24-dim lattice} & $ A_{11} \times D_7 \times E_6 $ \\ \hline
 {Lattice engineering} & $ A_2^2 \to \overline{A}_2^2 $ for $A_2 \in D_{7}$ and $ A_2 \in E_6 $ \\ \hline
 {(20,4)-dim lattice} & $ A_{11} \times A_2^2 \times D_4 \times U(1) \times \overline{A}_2^2 $ \\ \hline
 {$U(1)$ normalization} & $ 2 \sqrt{3} $ \\ \hline
 {C.C. generators of (20,4)-dim lattice} & $ \left(0,0,0,v,\frac{1}{6},2,1\right),(0,1,2,0,0,1,1),\left(1,1,1,s,-\frac{1}{4},0,0\right) $ \\ \hline
 {C.C. generators related to shift vectors} & $ \left(1,1,1,s,-\frac{1}{4},0,0\right),\left(0,0,0,v,\frac{1}{2},0,0\right) $ \\ \hline
\hline
 {Lattice name} & 24 \\ \hline
 {24-dim lattice} & $ A_{11} \times D_7 \times E_6 $ \\ \hline
 {Lattice engineering} & $ A_2^2 \to \overline{A}_2^2 $ for $A_2 \in A_{11}$ and $ A_2 \in E_6 $ \\ \hline
 {(20,4)-dim lattice} & $ A_2^2 \times A_8 \times D_7 \times U(1) \times \overline{A}_2^2 $ \\ \hline
 {$U(1)$ normalization} & $6$ \\ \hline
 {C.C. generators of (20,4)-dim lattice} & $\left(0,0,1,0,\frac{1}{9},2,1\right),(1,2,0,0,0,1,1),\left(1,1,1,s,\frac{1}{36},0,0\right) $ \\ \hline
 {C.C. generators related to shift vectors} & $\left(1,1,1,s,\frac{1}{36},0,0\right) $ \\ \hline
\hline
 {Lattice name} & 25 \\ \hline
 {24-dim lattice} & $ E_6^4 $ \\ \hline
 {Lattice engineering} & $ A_2^2 \to \overline{A}_2^2 $ for $A_2 \in E_{6}$ and $ A_2 \in E_6' $ \\ \hline
 {(20,4)-dim lattice} & $ A_2^4 \times E_6^2 \times \overline{A}_2^2 $ \\ \hline
 {$U(1)$ normalization} & --- \\ \hline
 {C.C. generators of (20,4)-dim lattice} & $(1,2,0,0,0,0,2,1),(1,1,1,2,1,2,1,1),(1,1,1,1,2,0,0,0),(1,1,2,2,0,1,0,0)$ \\ \hline
 {C.C. generators related to shift vectors} & $(1,1,2,2,0,1,0,0),(1,1,0,0,1,2,0,0)$ \\ \hline
\hline
 {Lattice name} & 26 \\ \hline
 {24-dim lattice} & $ A_9^2 \times D_6 $ \\ \hline
 {Lattice engineering} & $ A_2^2 \to \overline{A}_2^2 $ for $A_2^2 \in A_{9}$ \\ \hline
 {(20,4)-dim lattice} & $ A_3 \times A_9 \times D_6 \times U(1)^2 \times \overline{A}_2^2 $ \\ \hline
 {$U(1)$ normalization} & $ 2 \sqrt{21},\sqrt{210} $ \\ \hline
 {C.C. generators of (20,4)-dim lattice} & \begin{tabular}{c} $\left(2,4,0,\frac{1}{14},\frac{1}{35},0,0\right),\left(2,0,s,\frac{1}{14},-\frac{1}{14},0,0\right),\left(1,5,c,\frac{1}{28},\frac{1}{21},2,1\right), $ \\ $ \left(1,5,c,\frac{1}{12},0,1,1\right),(0,5,c,0,0,0,0) $ \end{tabular} \\ \hline
 {C.C. generators related to shift vectors} & $ \left(2,4,0,\frac{1}{14},\frac{1}{35},0,0\right),\left(2,0,s,\frac{1}{14},-\frac{1}{14},0,0\right),\left(3,5,c,\frac{3}{28},\frac{1}{7},0,0\right),\left(3,0,0,\frac{3}{28},\frac{1}{7},0,0\right) $ \\ \hline
\hline
 {Lattice name} & 27 \\ \hline
 {24-dim lattice} & $ A_9^2 \times D_6 $ \\ \hline
 {Lattice engineering} & $ A_2^2 \to \overline{A}_2^2 $ for $A_2^2 \in D_{6}$ \\ \hline
 {(20,4)-dim lattice} & $ A_9^2 \times U(1)^2 \times \overline{A}_2^2 $ \\ \hline
 {$U(1)$ normalization} & $ 2 \sqrt{3},2 \sqrt{3} $ \\ \hline
 {C.C. generators of (20,4)-dim lattice} & $ \left(2,4,\frac{1}{6},-\frac{1}{6},1,0\right),\left(2,4,0,\frac{1}{3},2,2\right),\left(5,0,-\frac{1}{4},-\frac{1}{4},0,0\right),\left(0,5,-\frac{1}{4},\frac{1}{4},0,0\right) $ \\ \hline
 {C.C. generators related to shift vectors} & $ \left(5,0,-\frac{1}{4},-\frac{1}{4},0,0\right),\left(0,5,-\frac{1}{4},\frac{1}{4},0,0\right),\left(8,6,0,0,0,0\right) $ \\ \hline
\hline
 {Lattice name} & 28 \\ \hline
 {24-dim lattice} & $ A_9^2 \times D_6 $ \\ \hline
 {Lattice engineering} & $ A_2^2 \to \overline{A}_2^2 $ for $A_2 \in A_{9}$ and $ A_2 \in A_9' $ \\ \hline
 {(20,4)-dim lattice} & $ A_6^2 \times D_6 \times U(1)^2 \times \overline{A}_2^2 $ \\ \hline
 {$U(1)$ normalization} & $ \sqrt{210},\sqrt{210} $ \\ \hline
 {C.C. generators of (20,4)-dim lattice} & $ \left(2,1,0,\frac{1}{35},-\frac{3}{35},0,0\right),\left(2,1,s,-\frac{1}{14},\frac{1}{21},1,1\right),\left(1,2,c,\frac{1}{21},-\frac{1}{14},2,1\right) $ \\ \hline
 {C.C. generators related to shift vectors} & $ \left(2,1,0,\frac{1}{35},-\frac{3}{35},0,0\right),\left(5,3,s,\frac{1}{14},\frac{1}{7},0,0\right),\left(3,2,c,\frac{1}{7},\frac{13}{14},0,0\right) $ \\ \hline
\end{tabular}
\caption[smallcaption]{(20,4)-dimensional lattices with $\overline{A}_2^2$ (continued). These lattices combined with the $A_2 \times \overline{A}_2 $ lattice correspond to (22,6)-dimensional Narain lattices which are numbered \#38 $\ldots$ \#83.}
\label{Tab.204DimLatticeWithA2A2bar4}
\end{center}
\end{table}

\begin{table}[h]
\begin{center}
\scriptsize
\begin{tabular}{|c|c|}
\hline
 {Lattice name} & 29 \\ \hline
 {24-dim lattice} & $A_9^2 \times D_6 $ \\ \hline
 {Lattice engineering} & $ A_2^2 \to \overline{A}_2^2 $ for $A_2 \in A_{9}$ and $ A_2 \in D_6 $ \\ \hline
 {(20,4)-dim lattice} & $ A_3 \times A_6 \times A_9 \times U(1)^2 \times \overline{A}_2^2 $ \\ \hline
 {$U(1)$ normalization} & $ 2 \sqrt{3},\sqrt{210} $ \\ \hline
 {C.C. generators of (20,4)-dim lattice} & $ \left(2,2,4,\frac{1}{6},\frac{1}{35},1,1\right),\left(1,2,0,-\frac{1}{4},-\frac{1}{14},0,0\right),\left(3,1,5,-\frac{1}{4},\frac{1}{21},2,1\right) $ \\ \hline
 {C.C. generators related to shift vectors} & $ \left(1,2,0,-\frac{1}{4},-\frac{1}{14},0,0\right),\left(2,5,4,\frac{1}{2},\frac{6}{35},0,0\right),\left(3,3,5,\frac{3}{4},\frac{1}{7},0,0\right) $ \\ \hline
\hline
 {Lattice name} & 30 \\ \hline
 {24-dim lattice} & $ D_6^4 $ \\ \hline
 {Lattice engineering} & $ A_2^2 \to \overline{A}_2^2 $ for $A_2^2 \in D_{6}$ \\ \hline
 {(20,4)-dim lattice} & $ D_6^3 \times U(1)^2 \times \overline{A}_2^2 $ \\ \hline
 {$U(1)$ normalization} & $ 2 \sqrt{3},2 \sqrt{3} $ \\ \hline
 {C.C. generators of (20,4)-dim lattice} & $ \left(v,c,s,\frac{1}{6},-\frac{1}{6},1,0\right),\left(c,c,c,-\frac{1}{4},\frac{1}{4},0,0\right),\left(v,s,0,-\frac{1}{4},\frac{1}{4},0,0\right),\left(v,v,v,0,-\frac{1}{6},2,2\right) $ \\ \hline
 {C.C. generators related to shift vectors} & $ \left(v,s,0,-\frac{1}{4},\frac{1}{4},0,0\right),\left(0,s,c,0,\frac{1}{2},0,0\right),\left(c,s,v,0,0,0,0\right),\left(s,v,c,0,0,0,0\right) $ \\ \hline
\hline
 {Lattice name} & 31 \\ \hline
 {24-dim lattice} & $ D_6^4 $ \\ \hline
 {Lattice engineering} & $ A_2^2 \to \overline{A}_2^2 $ for $A_2 \in D_{6}$ and $A_2 \in D_6' $ \\ \hline
 {(20,4)-dim lattice} & $ A_3^2 \times D_6^2 \times U(1)^2 \times \overline{A}_2^2 $ \\ \hline
 {$U(1)$ normalization} & $ 2 \sqrt{3},2 \sqrt{3} $ \\ \hline
 {C.C. generators of (20,4)-dim lattice} &  \begin{tabular}{c} $ \left(2,2,c,s,\frac{1}{6},0,2,1\right),\left(3,3,c,c,-\frac{1}{4},-\frac{1}{4},0,0\right), $ \\ $ 
\left(1,2,c,v,-\frac{1}{4},\frac{1}{6},1,1\right),(2,2,v,v,0,0,0,0) $ \end{tabular} \\ \hline
 {C.C. generators related to shift vectors} & \begin{tabular}{c} $ \left(3,0,0,c,\frac{1}{4},\frac{1}{2},0,0\right),\left(0,0,v,v,\frac{1}{2},\frac{1}{2},0,0\right), $ \\ $ 
\left(2,3,s,v,\frac{1}{2},\frac{3}{4},0,0\right),\left(2,1,v,c,\frac{1}{2},\frac{3}{4},0,0\right) $ \end{tabular} \\ \hline
\hline
 {Lattice name} & 32 \\ \hline
 {24-dim lattice} & $ A_8^3 $ \\ \hline
 {Lattice engineering} & $ A_2^2 \to \overline{A}_2^2 $ for $A_2^2 \in A_{8}$ \\ \hline
 {(20,4)-dim lattice} & $ A_2 \times A_8^2 \times U(1)^2 \times \overline{A}_2^2 $ \\ \hline
 {$U(1)$ normalization} & $ \sqrt{6},3 \sqrt{2} $ \\ \hline
 {C.C. generators of (20,4)-dim lattice} & $ \left(1,0,0,\frac{1}{6},\frac{1}{6},2,1\right),\left(1,0,0,\frac{1}{3},0,1,1\right),\left(1,1,1,\frac{1}{6},-\frac{5}{18},0,0\right),\left(1,4,1,\frac{1}{6},\frac{1}{18},0,0\right) $ \\ \hline
 {C.C. generators related to shift vectors} & $\ \left(1,1,1,\frac{1}{6},-\frac{5}{18},0,0\right),\left(1,4,1,\frac{1}{6},\frac{1}{18},0,0\right) $ \\ \hline
\hline
 {Lattice name} & 33 \\ \hline
 {24-dim lattice} & $ A_8^3 $ \\ \hline
 {Lattice engineering} & $ A_2^2 \to \overline{A}_2^2 $ for $A_2 \in A_{8}$ and $ A_2 \in A_8' $  \\ \hline
 {(20,4)-dim lattice} & $ A_5^2 \times A_8 \times U(1)^2 \times \overline{A}_2^2 $ \\ \hline
 {$U(1)$ normalization} & $ 3 \sqrt{2},3 \sqrt{2} $ \\ \hline
 {C.C. generators of (20,4)-dim lattice} & $ \left(1,0,0,\frac{1}{6},0,2,1\right),\left(0,1,0,0,\frac{1}{6},1,1\right),\left(1,1,1,-\frac{5}{18},\frac{1}{18},0,0\right),\left(1,1,1,\frac{1}{18},-\frac{5}{18},0,0\right) $ \\ \hline
 {C.C. generators related to shift vectors} & $ \left(1,1,1,-\frac{5}{18},\frac{1}{18},0,0\right),\left(1,1,1,\frac{1}{18},-\frac{5}{18},0,0\right),\left(3,0,0,\frac{1}{2},0,0,0\right) $ \\ \hline
\hline
 {Lattice name} & 34 \\ \hline
 {24-dim lattice} & $ A_7^2 \times D_5^2 $ \\ \hline
 {Lattice engineering} & $ A_2^2 \to \overline{A}_2^2 $ for $A_2^2 \in A_{7}$ \\ \hline
 {(20,4)-dim lattice} & $ A_1 \times A_7 \times D_5^2 \times U(1)^2 \times \overline{A}_2^2 $ \\ \hline
 {$U(1)$ normalization} & $ \sqrt{30},2 \sqrt{30} $ \\ \hline
 {C.C. generators of (20,4)-dim lattice} & \begin{tabular}{c} $ \left(0,0,0,0,\frac{1}{15},\frac{1}{15},1,0\right),\left(1,0,0,0,\frac{1}{6},0,2,2\right), $ \\ $ 
\left(0,1,s,v,\frac{1}{15},\frac{1}{40},2,2\right),\left(0,7,v,s,\frac{1}{15},\frac{1}{40},2,2\right) $ \end{tabular} \\ \hline
 {C.C. generators related to shift vectors} & $ \left(0,7,v,s,\frac{2}{5},\frac{1}{40},0,0\right),\left(0,1,s,v,\frac{2}{5},\frac{1}{40},0,0\right),\left(1,0,0,0,\frac{1}{2},0,0,0\right) $ \\ \hline
\hline
 {Lattice name} & 35 \\ \hline
 {24-dim lattice} & $ A_7^2 \times D_5^2 $ \\ \hline
 {Lattice engineering} & $ A_2^2 \to \overline{A}_2^2 $ for $A_2 \in A_{7}$ and $ A_2 \in A_7' $ \\ \hline
 {(20,4)-dim lattice} & $ A_4^2 \times D_5^2 \times U(1)^2 \times \overline{A}_2^2 $ \\ \hline
 {$U(1)$ normalization} & $ 2 \sqrt{30},2 \sqrt{30} $ \\ \hline
 {C.C. generators of (20,4)-dim lattice} & \begin{tabular}{c} $ \left(1,0,0,0,\frac{1}{15},0,2,1\right),\left(0,1,0,0,0,\frac{1}{15},1,1\right), $ \\ $ 
\left(1,1,s,v,\frac{1}{40},\frac{1}{40},0,0\right),\left(1,4,v,s,\frac{1}{40},-\frac{1}{40},0,0\right) $ \end{tabular} \\ \hline
 {C.C. generators related to shift vectors} & $ \left(1,1,s,v,\frac{1}{40},\frac{1}{40},0,0\right),\left(1,4,v,s,\frac{1}{40},-\frac{1}{40},0,0\right) $ \\ \hline

\end{tabular}
\caption[smallcaption]{(20,4)-dimensional lattices with $\overline{A}_2^2$ (continued). These lattices combined with the $A_2 \times \overline{A}_2 $ lattice correspond to (22,6)-dimensional Narain lattices which are numbered \#38 $\ldots$ \#83.}
\label{Tab.204DimLatticeWithA2A2bar5}
\end{center}
\end{table}

\begin{table}[h]
\begin{center}
\scriptsize
\begin{tabular}{|c|c|}
\hline
 {Lattice name} & 36 \\ \hline
 {24-dim lattice} & $ A_7^2 \times D_5^2 $ \\ \hline
 {Lattice engineering} & $ A_2^2 \to \overline{A}_2^2 $ for $A_2 \in D_{5}$ and $ A_2 \in D_5' $ \\ \hline
 {(20,4)-dim lattice} & $ A_1^4 \times A_7^2 \times U(1)^2 \times \overline{A}_2^2 $ \\ \hline
 {$U(1)$ normalization} & $ 2 \sqrt{3},2 \sqrt{3} $ \\ \hline
 {C.C. generators of (20,4)-dim lattice} & \begin{tabular}{c} $ \left(1,1,0,0,0,0,\frac{1}{6},0,2,1\right),\left(1,1,1,1,0,0,\frac{1}{6},\frac{1}{6},0,2\right), $ \\ $ \left(0,1,1,1,1,1,-\frac{1}{4},0,0,0\right),  \left(1,1,0,1,1,7,0,-\frac{1}{4},0,0\right)$ \end{tabular} \\ \hline
 {C.C. generators related to shift vectors} & \begin{tabular}{c} $ \left(0,1,1,1,1,1,-\frac{1}{4},0,0,0\right),\left(1,1,0,1,1,7,0,-\frac{1}{4},0,0\right), $ \\ $ \left(1,1,1,1,0,0,\frac{1}{2},\frac{1}{2},0,0\right),\left(1,1,0,0,0,0,\frac{1}{2},0,0,0\right)$ \end{tabular} \\ \hline
\hline
 {Lattice name} & 37 \\ \hline
 {24-dim lattice} & $ A_7^2 \times D_5^2 $ \\ \hline
 {Lattice engineering} & $ A_2^2 \to \overline{A}_2^2 $ for $A_2 \in A_{7}$ and $ A_2 \in D_5 $ \\ \hline
 {(20,4)-dim lattice} & $ A_1^2 \times A_4 \times A_7 \times D_5 \times U(1)^2 \times \overline{A}_2^2 $ \\ \hline
 {$U(1)$ normalization} & $ 2 \sqrt{3},2 \sqrt{30} $ \\ \hline
 {C.C. generators of (20,4)-dim lattice} & \begin{tabular}{c} $ \left(1,1,1,0,0,\frac{1}{6},\frac{1}{15},0,2\right),\left(1,1,0,0,0,\frac{1}{6},0,1,1\right), $ \\ $ \left(0,1,1,1,v,-\frac{1}{4},\frac{1}{40},0,0\right),\left(1,1,4,1,s,0,-\frac{1}{40},0,0\right)$ \end{tabular} \\ \hline
 {C.C. generators related to shift vectors} & $ \left(0,1,1,1,v,-\frac{1}{4},\frac{1}{40},0,0\right),\left(1,1,4,1,s,0,-\frac{1}{40},0,0\right),\left(1,1,3,0,0,\frac{1}{2},\frac{1}{5},0,0\right) $ \\ \hline
\hline
 {Lattice name} & 38 \\ \hline
 {24-dim lattice} & $ A_6^4 $ \\ \hline
 {Lattice engineering} & $ A_2^2 \to \overline{A}_2^2 $ for $A_2^2 \in A_{6}$ \\ \hline
 {(20,4)-dim lattice} & $ A_6^3 \times U(1)^2 \times \overline{A}_2^2 $ \\ \hline
 {$U(1)$ normalization} & $ 2 \sqrt{3},2 \sqrt{21} $ \\ \hline
 {C.C. generators of (20,4)-dim lattice} & \begin{tabular}{c} 
$ \left(0,0,0,-\frac{1}{4},\frac{1}{12},2,1\right),\left(0,0,0,\frac{1}{3},0,2,2\right), $ \\ $ 
\left(6,2,1,-\frac{1}{4},\frac{1}{28},0,0\right),\left(1,6,2,-\frac{1}{4},\frac{1}{28},0,0\right) $ \end{tabular} \\ \hline
 {C.C. generators related to shift vectors} & $ \left(6,2,1,-\frac{1}{4},\frac{1}{28},0,0\right),\left(1,6,2,-\frac{1}{4},\frac{1}{28},0,0\right) $ \\ \hline
\hline
 {Lattice name} & 39 \\ \hline
 {24-dim lattice} & $ A_6^4 $ \\ \hline
 {Lattice engineering} & $ A_2^2 \to \overline{A}_2^2 $ for $A_2 \in A_{6}$ and $ A_2 \in A_6' $ \\ \hline
 {(20,4)-dim lattice} & $ A_3^2 \times A_6^2 \times U(1)^2 \times \overline{A}_2^2 $ \\ \hline
 {$U(1)$ normalization} & $ 2 \sqrt{21},2 \sqrt{21} $ \\ \hline
 {C.C. generators of (20,4)-dim lattice} & \begin{tabular}{c} 
$ \left(1,0,0,0,\frac{1}{12},0,2,1\right),\left(0,1,0,0,0,\frac{1}{12},1,1\right), $ \\ $
\left(1,3,2,1,\frac{1}{28},-\frac{1}{28},0,0\right),\left(1,1,6,2,\frac{1}{28},\frac{1}{28},0,0\right) $ \end{tabular} \\ \hline
 {C.C. generators related to shift vectors} & $ \left(1,2,1,6,\frac{1}{28},\frac{1}{14},0,0\right),\left(1,1,6,2,\frac{1}{28},\frac{1}{28},0,0\right) $ \\ \hline
\hline
 {Lattice name} & 40 \\ \hline
 {24-dim lattice} & $ A_5^4 \times D_4 $ \\ \hline
 {Lattice engineering} & $ A_2^2 \to \overline{A}_2^2 $ for $A_2^2 \in A_{5}$ \\ \hline
 {(20,4)-dim lattice} & $ A_5^3 \times D_4 \times U(1) \times \overline{A}_2^2 $ \\ \hline
 {$U(1)$ normalization} & $ \sqrt{6} $ \\ \hline
 {C.C. generators of (20,4)-dim lattice} & \begin{tabular}{c} $ \left(0,0,0,0,\frac{1}{3},1,0\right),\left(4,0,2,0,\frac{1}{3},1,1\right),\left(2,4,0,0,\frac{1}{3},1,1\right), $ \\ $ \left(3,0,0,s,\frac{1}{2},0,0\right),\left(0,3,0,v,\frac{1}{2},0,0\right),\left(0,0,3,c,\frac{1}{2},0,0\right)$ \end{tabular} \\ \hline
 {C.C. generators related to shift vectors} & $ \left(3,0,0,s,\frac{1}{2},0,0\right),\left(0,3,0,v,\frac{1}{2},0,0\right),\left(0,0,3,c,\frac{1}{2},0,0\right),(4,4,4,0,0,0,0) $ \\ \hline
\hline
 {Lattice name} & 41 \\ \hline
 {24-dim lattice} & $ A_5^4 \times D_4 $ \\ \hline
 {Lattice engineering} & $ A_2^2 \to \overline{A}_2^2 $ for $A_2 \in A_{5}$ and $ A_2 \in A_5' $ \\ \hline
 {(20,4)-dim lattice} & $ A_2^2 \times A_5^2 \times D_4 \times U(1)^2 \times \overline{A}_2^2 $ \\ \hline
 {$U(1)$ normalization} & $ \sqrt{6},\sqrt{6} $ \\ \hline
 {C.C. generators of (20,4)-dim lattice} & \begin{tabular}{c} $ \left(1,0,0,0,0,\frac{1}{3},0,2,1\right),\left(2,1,0,2,0,\frac{1}{3},-\frac{1}{3},0,0\right),\left(2,2,4,0,0,\frac{1}{3},\frac{1}{3},0,0\right), $ \\ $ \left(0,0,0,0,s,\frac{1}{2},\frac{1}{2},0,0\right),\left(0,0,3,0,v,\frac{1}{2},0,0,0\right),\left(0,1,0,3,c,\frac{1}{2},\frac{1}{3},1,1\right)$ \end{tabular} \\ \hline
 {C.C. generators related to shift vectors} & \begin{tabular}{c} 
$ \left(2,2,4,0,0,\frac{1}{3},\frac{1}{3},0,0\right),\left(0,0,0,0,s,\frac{1}{2},\frac{1}{2},0,0\right), $ \\ $
\left(2,0,5,4,v,\frac{5}{6},0,0,0\right),\left(1,0,1,5,s,\frac{2}{3},0,0,0\right) $ \end{tabular} \\ \hline
\end{tabular}
\caption[smallcaption]{(20,4)-dimensional lattices with $\overline{A}_2^2$ (continued). These lattices combined with the $A_2 \times \overline{A}_2 $ lattice correspond to (22,6)-dimensional Narain lattices which are numbered \#38 $\ldots$ \#83.}
\label{Tab.204DimLatticeWithA2A2bar6}
\end{center}
\end{table}

\begin{table}[h]
\begin{center}
\scriptsize
\begin{tabular}{|c|c|}
\hline
 {Lattice name} & 42 \\ \hline
 {24-dim lattice} & $ A_5^4 \times D_4 $ \\ \hline
 {Lattice engineering} & $ A_2^2 \to \overline{A}_2^2 $ for $A_2 \in A_{5}$ and $ A_2 \in D_4 $ \\ \hline
 {(20,4)-dim lattice} & $ A_2 \times A_5^3 \times U(1)^3 \times \overline{A}_2^2 $ \\ \hline
 {$U(1)$ normalization} & $ 2,\sqrt{6},2 \sqrt{3} $ \\ \hline
 {C.C. generators of (20,4)-dim lattice} & \begin{tabular}{c} $ \left(1,0,0,0,\frac{1}{2},\frac{1}{3},\frac{1}{6},0,2\right),\left(2,4,0,2,\frac{1}{2},\frac{1}{3},\frac{1}{6},1,1\right),\left(2,2,4,0,0,\frac{1}{3},0,0,0\right), $ \\ $  \left(2,2,4,0,\frac{1}{2},\frac{1}{3},\frac{1}{6},1,1\right),\left(0,3,0,0,-\frac{1}{4},\frac{1}{2},-\frac{1}{4},0,0\right), $ \\ $ \left(0,0,3,0,\frac{1}{4},\frac{1}{2},-\frac{1}{4},0,0\right),\left(0,0,0,3,\frac{1}{2},\frac{1}{2},0,0,0\right)$ \end{tabular} \\ \hline
 {C.C. generators related to shift vectors} & \begin{tabular}{c} $ \left(2,2,4,0,0,\frac{1}{3},0,0,0\right),\left(0,3,0,0,-\frac{1}{4},\frac{1}{2},-\frac{1}{4},0,0\right),\left(0,0,3,0,\frac{1}{4},\frac{1}{2},-\frac{1}{4},0,0\right),$ \\ $ \left(0,0,0,3,\frac{1}{2},\frac{1}{2},0,0,0\right),\left(2,0,2,4,\frac{1}{2},\frac{1}{3},\frac{1}{2},0,0\right)$ \end{tabular} \\ \hline
\hline
 {Lattice name} & 43 \\ \hline
 {24-dim lattice} & $ D_4^6 $ \\ \hline
 {Lattice engineering} & $ A_2^2 \to \overline{A}_2^2 $ for $A_2 \in D_{4}$ and $ A_2 \in D_4' $  \\ \hline
 {(20,4)-dim lattice} & $ D_4^4 \times U(1)^4 \times \overline{A}_2^2 $ \\ \hline
 {$U(1)$ normalization} & $ 2,2,2 \sqrt{3},2 \sqrt{3} $ \\ \hline
 {C.C. generators of (20,4)-dim lattice} & \begin{tabular}{c} 
$ \left(0,0,0,0,\frac{1}{2},\frac{1}{2},\frac{1}{6},\frac{1}{6},0,2\right),\left(s,s,s,s,-\frac{1}{4},-\frac{1}{4},-\frac{1}{4},-\frac{1}{4},0,0\right), $ \\ $
\left(0,c,c,0,\frac{1}{4},\frac{1}{4},-\frac{1}{4},-\frac{1}{4},0,0\right)  ,\left(v,0,c,c,\frac{1}{4},\frac{1}{2},-\frac{1}{4},\frac{1}{6},1,1\right), $ \\ $
\left(0,v,0,c,\frac{1}{4},\frac{1}{2},-\frac{1}{4},0,0,0\right),\left(c,0,v,0,\frac{1}{4},\frac{1}{2},-\frac{1}{4},0,0,0\right),  $ \\ $ 
\left(c,c,0,v,\frac{1}{4},\frac{1}{2},-\frac{1}{4},\frac{1}{6},1,1\right)$ \end{tabular} \\ \hline
 {C.C. generators related to shift vectors} & \begin{tabular}{c} 
$ \left(s,s,s,s,-\frac{1}{4},-\frac{1}{4},-\frac{1}{4},-\frac{1}{4},0,0\right),\left(0,c,c,0,\frac{1}{4},\frac{1}{4},-\frac{1}{4},-\frac{1}{4},0,0\right), $ \\ $
\left(0,v,0,c,\frac{1}{4},\frac{1}{2},-\frac{1}{4},0,0,0\right),  \left(c,0,v,0,\frac{1}{4},\frac{1}{2},-\frac{1}{4},0,0,0\right), $ \\ $
\left(c,c,0,v,\frac{3}{4},0,\frac{1}{4},0,0,0\right),\left(v,0,c,c,\frac{3}{4},0,\frac{1}{4},0,0,0\right)$ \end{tabular} \\ \hline
\hline
 {Lattice name} & 44 \\ \hline
 {24-dim lattice} & $ A_4^6 $ \\ \hline
 {Lattice engineering} & $ A_2^2 \to \overline{A}_2^2 $ for $A_2 \in A_{4}$ and $ A_2 \in A_4' $ \\ \hline
 {(20,4)-dim lattice} & $ A_1^2 \times A_4^4 \times U(1)^2 \times \overline{A}_2^2 $ \\ \hline
 {$U(1)$ normalization} & $ \sqrt{30},\sqrt{30} $ \\ \hline
 {C.C. generators of (20,4)-dim lattice} & \begin{tabular}{c} $\left(1,0,0,0,0,0,\frac{1}{6},0,1,2\right),\left(0,1,1,4,4,1,\frac{1}{15},\frac{1}{6},0,1\right), $ \\ $ \left(0,0,4,1,0,1,\frac{1}{15},-\frac{1}{15},2,0\right),\left(0,0,4,4,1,0,\frac{1}{15},\frac{1}{15},0,1\right)$ \end{tabular} \\ \hline
 {C.C. generators related to shift vectors} & $\left(0,1,0,2,0,4,\frac{1}{5},\frac{1}{10},0,0\right),\left(1,0,0,1,1,0,\frac{3}{10},\frac{1}{5},0,0\right),\left(1,0,2,0,2,3,\frac{3}{10},\frac{1}{5},0,0\right) $ \\ \hline
\hline
 {Lattice name} & 45 \\ \hline
 {24-dim lattice} & $ A_3^8 $ \\ \hline
 {Lattice engineering} & $ A_2^2 \to \overline{A}_2^2 $ for $A_2 \in A_{3}$ and $ A_2 \in A_3' $ \\ \hline
 {(20,4)-dim lattice} & $ A_3^6 \times U(1)^2 \times \overline{A}_2^2 $ \\ \hline
 {$U(1)$ normalization} & $ 2 \sqrt{3},2 \sqrt{3} $ \\ \hline
 {C.C. generators of (20,4)-dim lattice} & \begin{tabular}{c} $\left(0,0,0,0,0,0,\frac{1}{3},0,1,2\right),\left(1,1,2,0,0,1,\frac{1}{4},0,0,0\right),\left(0,1,1,2,0,0,\frac{1}{4},-\frac{1}{4},0,0\right), $ \\ $ \left(1,0,1,1,2,0,\frac{1}{4},0,0,0\right),\left(0,1,0,1,1,2,\frac{1}{4},\frac{1}{3},2,2\right)$ \end{tabular} \\ \hline
 {C.C. generators related to shift vectors} & \begin{tabular}{c} $\left(0,1,1,2,0,0,\frac{1}{4},-\frac{1}{4},0,0\right),\left(1,0,1,1,2,0,\frac{1}{4},0,0,0\right), $ \\ $ \left(0,1,0,1,1,2,\frac{1}{4},0,0,0\right),\left(0,0,1,0,1,1,\frac{1}{4},\frac{1}{2},0,0\right) $ \end{tabular} \\ \hline
\hline
 {Lattice name} & 46 \\ \hline
 {24-dim lattice} & $ A_2^{12} $ \\ \hline
 {Lattice engineering} & $ A_2^2 \to \overline{A}_2^2 $ for $A_2 \times A_2' $ \\ \hline
 {(20,4)-dim lattice} & $ A_2^{10} \times \overline{A}_2^2 $ \\ \hline
 {$U(1)$ normalization} & --- \\ \hline
 {C.C. generators of (20,4)-dim lattice} & \begin{tabular}{c} 
$(2,2,1,2,1,1,2,1,1,1,0,2),(2,2,2,1,2,1,1,2,1,1,1,0), $ \\ $
(1,2,2,2,1,2,1,1,2,1,1,0),  (1,1,2,2,2,1,2,1,1,2,1,0), $ \\ $
(1,1,1,2,2,2,1,2,1,1,0,2),(2,1,1,1,2,2,2,1,2,1,1,0)$  \end{tabular} \\ \hline
 {C.C. generators related to shift vectors} & \begin{tabular}{c} 
$(1,0,1,1,0,1,1,1,0,0,0,0),(0,2,0,2,2,0,2,2,2,0,0,0), $ \\ $(1,1,0,1,1,1,0,0,0,1,0,0),(2,2,2,0,0,0,2,0,2,2,0,0)$ \end{tabular} \\ \hline
\end{tabular}
\caption[smallcaption]{(20,4)-dimensional lattices with $\overline{A}_2^2$ (continued). These lattices combined with the $A_2 \times \overline{A}_2 $ lattice correspond to (22,6)-dimensional Narain lattices which are numbered \#38 $\ldots$ \#83.}
\label{Tab.204DimLatticeWithA2A2bar7}
\end{center}
\end{table}

\begin{table}[h]
\begin{center}
\scriptsize
\begin{tabular}{|c|c|}
\hline
 {Lattice name} & 1 \\ \hline
 {24-dim lattice} & $ A_{17} \times E_7 $ \\ \hline
 {Lattice engineering} & $ A_2^2 \to \overline{A}_2^2 $ for $A_2^2 \in A_{17}$, $ E_6 \to \overline{A}_2 $ for $E_6 \in E_7$ \\ \hline
 {(14,6)-dim lattice} & $ A_{11} \times U(1)^3 \times \overline{A}_2^3 $ \\ \hline
 {$U(1)$ normalization} & $ \sqrt{6},2 \sqrt{15},3 \sqrt{10} $ \\ \hline
 {C.C. generators of (14,6)-dim lattice} & \begin{tabular}{c}$\left(1,-\frac{1}{3},\frac{1}{60},\frac{1}{15},2,2,1\right),\left(1,-\frac{1}{3},\frac{1}{12},0,2,1,1\right), $ \\ $\left(3,-\frac{1}{6},\frac{1}{20},\frac{1}{30},1,0,0\right), \left(3,-\frac{1}{2},\frac{1}{20},\frac{1}{30},0,0,0\right) $ \end{tabular} \\ \hline
 {C.C. generators related to shift vectors} & $ \left(3,\frac{1}{2},\frac{1}{4},\frac{1}{6},0,0,0\right),\left(0,0,\frac{1}{5},\frac{2}{15},0,0,0\right),\left(3,0,\frac{1}{4},0,0,0,0\right) $ \\ \hline
\hline
 {Lattice name} & 2 \\ \hline
 {24-dim lattice} & $ D_{10} \times E_7^2 $ \\ \hline
 {Lattice engineering} & $ A_2^2 \to \overline{A}_2^2 $ for $A_2^2 \in D_{10}$, $ E_6 \to \overline{A}_2 $ for $E_6 \in E_7$ \\ \hline
 {(14,6)-dim lattice} & $ D_4 \times E_7 \times U(1)^3 \times \overline{A}_2^3 $ \\ \hline
 {$U(1)$ normalization} & $ \sqrt{6},2 \sqrt{3},2 \sqrt{3} $ \\ \hline
 {C.C. generators of (14,6)-dim lattice} & \begin{tabular}{c}$\left(v,0,-\frac{1}{3},\frac{1}{6},0,2,2,1\right),\left(v,0,-\frac{1}{3},0,\frac{1}{6},2,1,1\right), $ \\ $\left(s,0,-\frac{1}{2},-\frac{1}{4},-\frac{1}{4},0,0,0\right) ,\left(c,1,-\frac{1}{3},-\frac{1}{4},-\frac{1}{4},2,0,0\right)$ \end{tabular} \\ \hline
 {C.C. generators related to shift vectors} & $ \left(s,0,\frac{1}{2},\frac{1}{4},\frac{1}{4},0,0,0\right),\left(c,1,0,\frac{3}{4},\frac{3}{4},0,0,0\right),\left(v,0,0,\frac{1}{2},0,0,0,0\right) $ \\ \hline
\hline
 {Lattice name} & 3 \\ \hline
 {24-dim lattice} & $ D_{10} \times E_7^2 $ \\ \hline
 {Lattice engineering} & $ A_2^2 \to \overline{A}_2^2 $ for $A_2 \in D_{7}$ and $ A_2 \in E_7 $, $ E_6 \to \overline{A}_2 $ for $E_6 \in E_7'$ \\ \hline
 {(14,6)-dim lattice} & $ A_5 \times D_7 \times U(1)^2 \times \overline{A}_2^3 $ \\ \hline
 {$U(1)$ normalization} & $ \sqrt{6},2 \sqrt{3} $ \\ \hline
 {C.C. generators of (14,6)-dim lattice} & $ \left(0,v,-\frac{1}{3},\frac{1}{6},2,2,1\right),\left(3,s,-\frac{1}{3},-\frac{1}{4},2,0,0\right),\left(2,c,-\frac{1}{2},-\frac{1}{4},0,1,1\right) $ \\ \hline
 {C.C. generators related to shift vectors} & $ \left(3,c,0,\frac{1}{4},0,0,0\right),\left(3,0,\frac{1}{2},\frac{1}{2},0,0,0\right) $ \\ \hline
\hline
 {Lattice name} & 4 \\ \hline
 {24-dim lattice} & $ A_{11} \times D_7 \times E_6 $ \\ \hline
 {Lattice engineering} & $ A_2^2 \to \overline{A}_2^2 $ for $A_2^2 \in A_{11}$, $ E_6 \to \overline{A}_2 $ for $E_6 $ \\ \hline
 {(14,6)-dim lattice} & $ A_5 \times D_7 \times U(1)^2 \times \overline{A}_2^3 $ \\ \hline
 {$U(1)$ normalization} & $ 3 \sqrt{2},6 $ \\ \hline
 {C.C. generators of (14,6)-dim lattice} & $ \left(1,0,\frac{1}{18},\frac{1}{9},0,2,1\right),\left(1,0,\frac{1}{6},0,0,1,1\right),\left(1,s,\frac{1}{18},\frac{1}{36},1,0,0\right) $ \\ \hline
 {C.C. generators related to shift vectors} & $ \left(0,c,\frac{2}{3},\frac{1}{12},0,0,0\right),\left(3,c,\frac{1}{6},\frac{1}{12},0,0,0\right) $ \\ \hline
\hline
 {Lattice name} & 5 \\ \hline
 {24-dim lattice} & $ A_{11} \times D_7 \times E_6 $ \\ \hline
 {Lattice engineering} & $ A_2^2 \to \overline{A}_2^2 $ for $A_2^2 \in D_{7}$, $ E_6 \to \overline{A}_2 $ for $E_6 $ \\ \hline
 {(14,6)-dim lattice} & $ A_{11} \times U(1)^3 \times \overline{A}_2^3 $ \\ \hline
 {$U(1)$ normalization} & $ 2,2 \sqrt{3},2 \sqrt{3} $ \\ \hline
 {C.C. generators of (14,6)-dim lattice} & $ \left(0,\frac{1}{4},\frac{1}{6},-\frac{1}{4},0,2,1\right),\left(0,\frac{1}{2},0,\frac{1}{6},0,1,1\right),\left(1,-\frac{1}{4},-\frac{1}{4},-\frac{1}{4},1,0,0\right) $ \\ \hline
 {C.C. generators related to shift vectors} & $ \left(3,0,\frac{3}{4},\frac{1}{2},0,0,0\right),\left(3,\frac{3}{4},\frac{1}{4},\frac{3}{4},0,0,0\right) $ \\ \hline
\hline
 {Lattice name} & 6 \\ \hline
 {24-dim lattice} & $ A_{11} \times D_7 \times E_6 $ \\ \hline
 {Lattice engineering} & $ A_2^2 \to \overline{A}_2^2 $ for $A_2 \in A_{11}$ and $ A_2 \in D_7 $, $ E_6 \to \overline{A}_2 $ for $E_6 $ \\ \hline
 {(14,6)-dim lattice} & $ A_8 \times D_4 \times U(1)^2 \times \overline{A}_2^3 $ \\ \hline
 {$U(1)$ normalization} & $ 2 \sqrt{3},6 $ \\ \hline
 {C.C. generators of (14,6)-dim lattice} & $ \left(1,0,0,\frac{1}{9},0,2,1\right),\left(0,v,\frac{1}{6},0,0,1,1\right),\left(1,s,-\frac{1}{4},\frac{1}{36},1,0,0\right) $ \\ \hline
 {C.C. generators related to shift vectors} & $ \left(3,s,\frac{1}{4},\frac{1}{12},0,0,0\right),\left(3,c,\frac{3}{4},\frac{1}{12},0,0,0\right) $ \\ \hline
\hline
 {Lattice name} & 7 \\ \hline
 {24-dim lattice} & $ E_6^4 $ \\ \hline
 {Lattice engineering} & $ A_2^2 \to \overline{A}_2^2 $ for $A_2 \in E_{6}$ and $ A_2 \in E_6' $, $ E_6 \to \overline{A}_2 $ for $E_6'' $ \\ \hline
 {(14,6)-dim lattice} & $ A_2^4 \times E_6 \times \overline{A}_2^3 $ \\ \hline
 {$U(1)$ normalization} & --- \\ \hline
 {C.C. generators of (14,6)-dim lattice} & \begin{tabular}{c} $(1,2,0,0,0,0,2,1),(1,1,1,2,2,1,1,1), $ \\ $(1,1,1,1,0,2,0,0),(1,1,2,2,1,0,0,0)$ \end{tabular} \\ \hline
 {C.C. generators related to shift vectors} & $(1,1,2,2,1,0,0,0)$ \\ \hline
\end{tabular}
\caption[smallcaption]{(14,6)-dimensional lattices with $\overline{A}_2^3$. These lattices combined with the $E_8 $ lattice correspond to (22,6)-dimensional Narain lattices which are numbered \#84 $\ldots$ \#90.}
\label{Tab.146DimLatticeWithA2A2A2bar1}
\end{center}
\end{table}

\clearpage

\begin{table}[h]
\begin{center}
\scriptsize
\begin{tabular}{|c|c|}
\hline
 {Lattice name} & 1 \\ \hline
 {24-dim lattice} & $ D_{16} \times E_8 $ \\ \hline
 {Lattice engineering} & $ D_4 \to \overline{D}_4 $ for $D_4 \in E_{8}$ \\ \hline
 {(20,4)-dim lattice} & $ D_{16} \times D_4 \times \overline{D}_4 $ \\ \hline
 {$U(1)$ normalization} & --- \\ \hline
 {C.C. generators of (20,4)-dim lattice} & $(s,0,0),(s,v,v),(s,s,s)$ \\ \hline
 {C.C. generators related to shift vectors} & $(s,0,0)$ \\ \hline
\hline
 {Lattice name} & 2 \\ \hline
 {24-dim lattice} & $ E_8^3 $ \\ \hline
 {Lattice engineering} & $ D_4 \to \overline{D}_4 $ for $D_4 \in E_{8}$ \\ \hline
 {(20,4)-dim lattice} & $ D_4 \times E_8^2 \times \overline{D}_4 $ \\ \hline
 {$U(1)$ normalization} & --- \\ \hline
 {C.C. generators of (20,4)-dim lattice} & $(v,0,0,v),(s,0,0,s)$ \\ \hline
 {C.C. generators related to shift vectors} & --- \\ \hline
\hline
 {Lattice name} & 3 \\ \hline
 {24-dim lattice} & $ A_{17} \times E_7 $ \\ \hline
 {Lattice engineering} & $ D_4 \to \overline{D}_4 $ for $D_4 \in E_{7}$ \\ \hline
 {(20,4)-dim lattice} & $ A_1^3 \times A_{17} \times \overline{D}_4 $ \\ \hline
 {$U(1)$ normalization} & --- \\ \hline
 {C.C. generators of (20,4)-dim lattice} & $(1,0,1,0,s),(1,1,0,0,v),(0,1,0,3,s)$ \\ \hline
 {C.C. generators related to shift vectors} & $(1,1,1,3,0)$ \\ \hline
 \hline
{Lattice name} & 4 \\ \hline
 {24-dim lattice} & $ D_{10} \times E_7^2 $ \\ \hline
 {Lattice engineering} & $ D_4 \to \overline{D}_4 $ for $D_4 \in E_{7}$ \\ \hline
 {(20,4)-dim lattice} & $ A_1^3 \times D_{10} \times E_7 \times \overline{D}_4 $ \\ \hline
 {$U(1)$ normalization} & --- \\ \hline
 {C.C. generators of (20,4)-dim lattice} & \begin{tabular}{c} $ (0,1,0,s,0,s),(1,0,1,c,1,s),$ \\ $(0,0,0,c,1,0),(1,1,0,c,1,v)$ \\ \end{tabular} \\ \hline
 {C.C. generators related to shift vectors} & $(0,0,0,c,1,0),(1,1,1,s,0,0)$ \\ \hline
\hline
 {Lattice name} & 5 \\ \hline
 {24-dim lattice} & $ A_{11} \times D_7 \times E_6 $ \\ \hline
 {Lattice engineering} & $ D_4 \to \overline{D}_4 $ for $D_4 \in E_{6}$ \\ \hline
 {(20,4)-dim lattice} & $ A_{11} \times D_7 \times U(1)^2 \times \overline{D}_4 $ \\ \hline
 {$U(1)$ normalization} & $ 2,2 \sqrt{3} $ \\ \hline
 {C.C. generators of (20,4)-dim lattice} & $ \left(0,0,\frac{1}{4},\frac{1}{4},s\right),\left(0,0,\frac{1}{2},0,v\right),\left(1,s,0,-\frac{1}{3},0\right) $ \\ \hline
 {C.C. generators related to shift vectors} & $ \left(3,c,0,0,0 \right) $ \\ \hline
\hline
 {Lattice name} & 6 \\ \hline
 {24-dim lattice} & $ E_6^4 $ \\ \hline
 {Lattice engineering} & $ D_4 \to \overline{D}_4 $ for $D_4 \in E_{6}$ \\ \hline
 {(20,4)-dim lattice} & $ E_6^3 \times U(1)^2 \times \overline{D}_4 $ \\ \hline
 {$U(1)$ normalization} & $ 2,2 \sqrt{3} $ \\ \hline
 {C.C. generators of (20,4)-dim lattice} & \begin{tabular}{c} $ \left(0,0,0,\frac{1}{4},\frac{1}{4},s\right),\left(0,0,0,\frac{1}{2},0,v\right), $ \\ $\left(1,2,0,0,-\frac{1}{3},0\right),\left(2,0,1,0,-\frac{1}{3},0\right) $ \\ \end{tabular} \\ \hline
 {C.C. generators related to shift vectors} & ---  \\ \hline
\hline
 {Lattice name} & 7 \\ \hline
 {24-dim lattice} & $ A_5^4 \times D_4 $ \\ \hline
 {Lattice engineering} & $ D_4 \to \overline{D}_4 $ for $D_4 $ \\ \hline
 {(20,4)-dim lattice} & $ A_5^4 \times \overline{D}_4 $ \\ \hline
 {$U(1)$ normalization} & --- \\ \hline
 {C.C. generators of (20,4)-dim lattice} & \begin{tabular}{c} $ (2,4,0,2,0),(2,2,4,0,0),(3,3,0,0,s), $ \\ $(3,0,3,0,v),(3,0,0,3,c)$ \\ \end{tabular} \\ \hline
 {C.C. generators related to shift vectors} & $(3,3,3,3,0)$ \\ \hline
\end{tabular}
\caption[smallcaption]{(20,4)-dimensional lattices with $\overline{D}_4$. These lattices combined with the $A_2 \times \overline{A}_2 $ lattice lead to (22,6)-dimensional Narain lattices.}
\label{Tab.204DimLatticeWithD4bar1}
\end{center}
\end{table}

\begin{table}[h]
\begin{center}
\scriptsize
\begin{tabular}{|c|c|}
\hline
 {Lattice name} & 8 \\ \hline
 {24-dim lattice} & $ D_4^6 $ \\ \hline
 {Lattice engineering} & $ D_4 \to \overline{D}_4 $ for $D_4 $ \\ \hline
 {(20,4)-dim lattice} & $ D_4^5 \times \overline{D}_4 $ \\ \hline
 {$U(1)$ normalization} & --- \\ \hline
 {C.C. generators of (20,4)-dim lattice} & \begin{tabular}{c} $ (s,s,s,s,s,s),(v,0,c,c,0,v),(0,v,0,c,c,v), $ \\ $ (c,0,v,0,c,v),(c,c,0,v,0,v),(0,c,c,0,v,v) $ \\ \end{tabular} \\ \hline
 {C.C. generators related to shift vectors} & --- \\ \hline
\end{tabular}
\caption[smallcaption]{(20,4)-dimensional lattices with $\overline{D}_4$ (continued). These lattices combined with the $A_2 \times \overline{A}_2 $ lattice lead to (22,6)-dimensional Narain lattices.}
\label{Tab.204DimLatticeWithD4bar2}
\end{center}
\end{table}

\begin{table}[h]
\begin{center}
\scriptsize
\begin{tabular}{|c|c|}
\hline
 {Lattice name} & 1 \\ \hline
 {24-dim lattice} & $ D_{10} \times E_7^2 $ \\ \hline
 {Lattice engineering} & $ D_4 \to \overline{D}_4 $ for $D_4 \in E_{7}$, $ E_6 \to \overline{A}_2 $ for $E_6 \in E_{7}'$ \\ \hline
 {(14,6)-dim lattice} & $ A_1^3 \times D_{10} \times U(1) \times \overline{D}_4 \times \overline{A}_2 $ \\ \hline
 {$U(1)$ normalization} & $ \sqrt{6} $ \\ \hline
 {C.C. generators of (14,6)-dim lattice} & \begin{tabular}{c} $\left(0,1,0,s,-\frac{1}{3},s,2\right),\left(1,0,1,c,-\frac{1}{2},s,0\right),\left(0,0,0,c,-\frac{1}{2},0,0\right), $ \\  $\left(1,1,0,c,-\frac{1}{2},v,0\right)$ \\ \end{tabular} \\ \hline
 {C.C. generators related to shift vectors} & $(1,1,1,s,0,0,0),\left(1,1,1,v,\frac{1}{2},0,0\right) $ \\ \hline
\hline
 {Lattice name} & 2 \\ \hline
 {24-dim lattice} & $ E_6^4 $ \\ \hline
 {Lattice engineering} & $ D_4 \to \overline{D}_4 $ for $D_4 \in E_{6}$, $ E_6 \to \overline{A}_2 $ for $E_6'$ \\ \hline
 {(14,6)-dim lattice} & $ E_6^2 \times U(1)^2 \times \overline{D}_4 \times \overline{A}_2 $ \\ \hline
 {$U(1)$ normalization} & $ 2,2 \sqrt{3} $ \\ \hline
 {C.C. generators of (14,6)-dim lattice} & \begin{tabular}{c} $\left(0,0,\frac{1}{4},\frac{1}{4},s,0\right),\left(0,0,\frac{1}{2},0,v,0\right),\left(2,0,0,-\frac{1}{3},0,1\right), $ \\ $\left(0,1,0,-\frac{1}{3},0,2 \right)$ \\ \end{tabular} \\ \hline
 {C.C. generators related to shift vectors} & --- \\ \hline
\hline
\end{tabular}
\caption[smallcaption]{(14,6)-dimensional lattices with $\overline{D}_4 \times \overline{A}_2$. These lattices combined with the $E_8 $ lattice lead to (22,6)-dimensional Narain lattices.}
\label{Tab.226DimLatticeWithD4A2bar}
\end{center}
\end{table}


\begin{table}[h]
\begin{center}
\scriptsize
\begin{tabular}{|c|c|c|}
\hline
Group & C.C & Breaking patterns \\
\hline
\hline
 $A_1$ & All & $ A_1, U(1) $ \\
\hline
 $A_2$ & $0 \bmod 3$ & $A_2,U(1)^2 $ \\
\hline
 $A_2$ & \begin{tabular}{c} $1 \bmod 3$ \\ $2 \bmod 3$ \end{tabular} & $ A_1 \times U(1)$ \\
\hline
 $A_3$ & All & $ A_3, A_2 \times U(1), A_1^2 \times U(1), A_1 \times U(1)^2 $ \\
\hline
 $A_4$ & All & $ A_4, A_3 \times U(1), A_1 \times A_2 \times U(1), A_1^2 \times U(1)^2, A_2 \times U(1)^2 $ \\
\hline
 $A_5$ & $0 \bmod 3$ & $ A_5, A_3 \times U(1)^2, A_2^2 \times U(1), A_1^3 \times U(1)^2 $ \\
\hline
 $A_5$ & \begin{tabular}{c} $1 \bmod 3$ \\ $2 \bmod 3$ \end{tabular} & $ A_4 \times U(1), A_1 \times A_3 \times U(1), A_1 \times A_2 \times U(1)^2 $ \\
\hline
 $A_6$ & All & 
\begin{tabular}{c} $ A_6, A_5 \times U(1), A_1 \times A_4 \times U(1), A_1 \times A_3 \times U(1)^2, A_2^2 \times U(1)^2,   $ \\ $
A_2 \times A_3 \times U(1), A_1^2 \times A_2 \times U(1)^2, A_4 \times U(1)^2 $ 
\end{tabular} \\
\hline
 $A_7$ & All & \begin{tabular}{c} $
 A_7, A_6 \times U(1), A_1 \times A_5 \times U(1) , A_1 \times A_4 \times U(1)^2, A_3^2 \times U(1), A_2 \times A_3 \times U(1)^2,  $ \\ 
$A_1 \times A_2^2 \times U(1)^2, A_2 \times A_4 \times U(1), A_1^2 \times A_3 \times U(1)^2, A_5 \times U(1)^2 $
\end{tabular} \\
\hline
 $A_8$ & $0 \bmod 3$ & $A_8, A_6 \times U(1)^2, A_2 \times A_5 \times U(1), A_1^2 \times A_4 \times U(1)^2, A_3^2 \times U(1)^2, A_2^3 \times U(1)^2 $ \\
\hline
 $A_8$ & \begin{tabular}{c} $1 \bmod 3$ \\ $2 \bmod 3$ \end{tabular} &  
\begin{tabular}{c} $
A_7 \times U(1), A_1 \times A_6 \times U(1), A_1 \times A_5 \times U(1)^2, A_3 \times A_4 \times U(1),  $ \\ $
A_2 \times A_4 \times U(1)^2, A_1 \times A_2 \times A_3 \times U(1)^2 $ 
\end{tabular} \\
\hline
 $A_9$ & All & \begin{tabular}{c} $
A_9, A_8 \times U(1), A_1 \times A_7 \times U(1), A_1 \times A_6 \times U(1)^2, A_2 \times A_5 \times U(1)^2, A_4^2 \times U(1),  $ \\ $
A_3 \times A_5 \times U(1), A_1 \times A_2 \times A_4 \times U(1)^2, A_1 \times A_3^2 \times U(1)^2, A_3 \times A_4 \times U(1)^2,  $ \\ $
A_2^2 \times A_3 \times U(1)^2, A_2 \times A_6 \times U(1), A_1^2 \times A_5 \times U(1)^2, A_7 \times U(1)^2 $
\end{tabular} \\
\hline
 $A_{10}$ & All & \begin{tabular}{c} $
A_{10}, A_9 \times U(1), A_1 \times A_8 \times U(1), A_1 \times A_7 \times U(1)^2, A_3 \times A_6 \times U(1), A_2 \times A_6 \times U(1)^2,  $ \\ $
A_1 \times A_2 \times A_5 \times U(1)^2, A_4^2 \times U(1)^2, A_4 \times A_5 \times U(1), A_1 \times A_3 \times A_4 \times U(1)^2,  $ \\ $
A_2 \times A_3^2 \times U(1)^2, A_3 \times A_5 \times U(1)^2, A_2^2 \times A_4 \times U(1)^2, A_2 \times A_7 \times U(1),  $ \\ $
A_1^2 \times A_6 \times U(1)^2, A_8 \times U(1)^2 $
\end{tabular} \\
\hline
 $A_{11}$ & $0 \bmod 3$ & \begin{tabular}{c} $
A_{11}, A_9 \times U(1)^2, A_2 \times A_8 \times U(1), A_1^2 \times A_7 \times U(1)^2, A_3 \times A_6 \times U(1)^2, A_5^2 \times U(1),  $ \\ $
A_2^2 \times A_5 \times U(1)^2, A_1 \times A_4^2 \times U(1)^2, A_3^3 \times U(1)^2 $
\end{tabular} \\
\hline
 $A_{11}$ & \begin{tabular}{c} $1 \bmod 3$ \\ $2 \bmod 3$ \end{tabular} & \begin{tabular}{c} $
A_{10} \times U(1), A_1 \times A_9 \times U(1), A_1 \times A_8 \times U(1)^2, A_3 \times A_7 \times U(1), A_2 \times A_7 \times U(1)^2,  $ \\ $
A_1 \times A_2 \times A_6 \times U(1)^2, A_4 \times A_5 \times U(1)^2, A_4 \times A_6 \times U(1), A_1 \times A_3 \times A_5 \times U(1)^2,  $ \\ $
A_2 \times A_3 \times A_4 \times U(1)^2 $
\end{tabular} \\
\hline
 $A_{12}$ & All & \begin{tabular}{c} $
A_{12}, A_{11} \times U(1), A_1 \times A_{10} \times U(1), A_1 \times A_9 \times U(1)^2, A_2 \times A_8 \times U(1)2^, A_4 \times A_7 \times U(1),  $ \\ $
A_3 \times A_8 \times U(1), A_1 \times A_2 \times A_7 \times U(1)^2, A_1 \times A_3 \times A_6 \times U(1)^2, A_5^2 \times U(1)^2,  $ \\ $
A_4 \times A_6 \times U(1)^2, A_2 \times A_3 \times A_5 \times U(1)^2, A_2vA_4^2 \times U(1)^2, A_5 \times A_6 \times U(1),  $ \\ $
A_1 \times A_4 \times A_5 \times U(1)^2, A_3^2 \times A_4 \times U(1)^2, A_3 \times A_7 \times U(1)^2, A_2^2 \times A_6 \times U(1)^2,  $ \\ $
A_2 \times A_9 \times U(1), A_1^2 \times A_8 \times U(1)^2,  $ \\ $
A_{10} \times U(1)^2 $
\end{tabular} \\
\hline
 $A_{13}$ & All & \begin{tabular}{c} $
A_{13}, A_{12} \times U(1), A_1 \times A_{11} \times U(1), A_1 \times A_{10} \times U(1)^2, A_3 \times A_9 \times U(1), A_2 \times A_9 \times U(1)^2,  $ \\ $
A_1 \times A_2 \times A_8 \times U(1)^2, A_4 \times A_7 \times U(1)^2, A_6^2 \times U(1), A_4 \times A_8 \times U(1),  $ \\ $
A_1 \times A_3 \times A_7 \times U(1)^2, A_2 \times A_3 \times A_6 \times U(1)^2, A_1 \times A_5^2 \times U(1)^2, A_5 \times A_6 \times U(1)^2,  $ \\ $
A_2 \times A_4 \times A_5 \times U(1)^2, A_3 \times A_4^2 \times U(1)^2, A_5 \times A_7 \times U(1), A_1 \times A_4 \times A_6 \times U(1)^2,  $ \\ $
A_3^2 \times A_5 \times U(1)^2, A_3 \times A_8 \times U(1)^2, A_2^2 \times A_7 \times U(1)^2, A_{10} \times A_2 \times U(1),  $ \\ $
A_1^2 \times A_9 \times U(1)^2,  A_{11} \times U(1)^2 $
\end{tabular} \\
\hline
 $A_{14}$ & $0 \bmod 3$ & \begin{tabular}{c} $
A_{14}, A_{12} \times U(1)^2, A_{11} \times A_2 \times U(1), A_1^2 \times A_{10} \times U(1)^2, A_3 \times A_9 \times U(1)^2, A_5 \times A_8 \times U(1),  $ \\ $
A_2^2 \times A_8 \times U(1)^2, A_1 \times A_4 \times A_7 \times U(1)^2, A_6^2 \times U(1)^2, A_3 \times A_6 \times U(1)^2,  $ \\ $
A_2 \times A_5^2 \times U(1)^2, A_4^3 \times U(1)^2 $
\end{tabular} \\
\hline
 $A_{14}$ & \begin{tabular}{c} $1 \bmod 3$ \\ $2 \bmod 3$ \end{tabular} &  \begin{tabular}{c} $
A_{13} \times U(1), A_1 \times A_{12} \times U(1), A_1 \times A_{11} \times U(1)^2, A_{10} \times A_3 \times U(1), A_{10} \times A_2 \times U(1)^2,  $ \\ $
A_1 \times A_2 \times A_9 \times U(1)^2, A_4 \times A_8 \times U(1)^2, A_6 \times A_7 \times U(1), A_4 \times A_9 \times U(1),  $ \\ $
A_1 \times A_3 \times A_8 \times U(1)^2, A_2 \times A_3 \times A_7 \times U(1)^2, A_1 \times A_5 \times A_6 \times U(1)^2,  $ \\ $
A_5 \times A_7 \times U(1)^2, A_2 \times A_4 \times A_6 \times U(1)^2, A_3 \times A_4 \times A_5 \times U(1)^2 $
\end{tabular} \\
\hline
 $A_{15}$ & All & \begin{tabular}{c} $
A_{15}, A_{14} \times U(1), A_1 \times A_{13} \times U(1), A_1 \times A_{12} \times U(1)^2, A_{11} \times A_2 \times U(1)^2, A_{10} \times A_4 \times U(1),  $ \\ $
A_{11} \times A_3 \times U(1), A_1 \times A_{10} \times A_2 \times U(1)^2, A_1 \times A_3 \times A_9 \times U(1)^2, A_5 \times A_8 \times U(1)^2,  $ \\ $
A_7^2 \times U(1), A_4 \times A_9 \times U(1)^2, A_2 \times A_3 \times A_8 \times U(1)^2, A_2 \times A_4 \times A_7 \times U(1)^2,  $ \\ $
A_1 \times A_6^2 \times U(1)^2, A_6 \times A_8 \times U(1), A_1 \times A_5 \times A_7 \times U(1)^2, A_3 \times A_4 \times A_6 \times U(1)^2,  $ \\ $
A_3 \times A_5^2 \times U(1)^2, A_6 \times A_7 \times U(1)^2, A_2 \times A_5 \times A_6 \times U(1)^2, A_4^2 \times A_5 \times U(1)^2,  $ \\ $
A_5 \times A_9 \times U(1), A_1 \times A_4 \times A_8 \times U(1)^2, A_3^2 \times A_7 \times U(1)^2, A_{10} \times A_3 \times U(1)^2,  $ \\ $
 A_2^2 \times A_9 \times U(1)^2, A_{12} \times A_2 \times U(1), A_1^2 \times A_{11} \times U(1)^2, A_{13} \times U(1)^2  $
\end{tabular} \\
\hline
\end{tabular}
\caption[smallcaption]{$A_n (n=1 \ldots 22)$ group breaking by a ${\bf Z}_3$ shift action.}
\label{Tab.AnGroupBreaking1}
\end{center}
\end{table}

\begin{table}[h]
\begin{center}
\scriptsize
\begin{tabular}{|c|c|c|}
\hline
Group & C.C & Breaking patterns \\
\hline
\hline
 $A_{16}$ & All & \begin{tabular}{c} $
A_{16}, A_{15} \times U(1), A_1 \times A_{14} \times U(1), A_1 \times A_{13} \times U(1)^2, A_{12} \times A_3 \times U(1), A_{12} \times A_2 \times U(1)^2,  $ \\ $
A_1 \times A_{11} \times A_2 \times U(1)^2, A_{10} \times A_4 \times U(1)^2, A_6 \times A_9 \times U(1), A_{11} \times A_4 \times U(1),  $ \\ $
A_1 \times A_{10} \times A_3 \times U(1)^2, A_2 \times A_3 \times A_9 \times U(1)^2, A_1 \times A_5 \times A_8 \times U(1)^2,  $ \\ $
A_7^2 \times U(1)^2, A_5 \times A_9 \times U(1)^2, A_2 \times A_4 \times A_8 \times U(1)^2, A_3 \times A_4 \times A_7 \times U(1)^2,  $ \\ $
A_2 \times A_6^2 \times U(1)^2, A_7 \times A_8 \times U(1), A_1 \times A_6 \times A_7 \times U(1)^2, A_3 \times A_5 \times A_6 \times U(1)^2,  $ \\ $
A_4 \times A_5^2 \times U(1)^2, A_6 \times A_8 \times U(1)^2, A_2 \times A_5 \times A_7 \times U(1)^2, A_4^2 \times A_6 \times U(1)^2,  $ \\ $
A_{10} \times A_5 \times U(1), A_1 \times A_4 \times A_9 \times U(1)^2,  \times A_3^2 \times A_8 \times U(1)^2, A_{11} \times A_3 \times U(1)^2,  $ \\ $
A_{10} \times A_2^2 \times U(1)^2, A_{13} \times A_2 \times U(1), A_1^2 \times A_{12} \times U(1)^2, A_{14} \times U(1)^2 $
\end{tabular} \\
\hline
 $A_{17}$ & $0 \bmod 3$ & \begin{tabular}{c} $ 
A_{17}, A_{15} \times U(1)^2, A_{14} \times A_2 \times U(1), A_1^2 \times A_{13} \times U(1)^2, A_{12} \times A_3 \times U(1)^2,  $ \\ $
A_{11} \times A_5 \times U(1), A_{11} \times A_2^2 \times U(1)^2,  A_1 \times A_{10} \times A_4 \times U(1)^2, A_6 \times A_9 \times U(1)^2,  $ \\ $
A_8^2 \times U(1), A_3^2 \times A_9 \times U(1)^2, A_2 \times A_5 \times A_8 \times U(1)^2, A_1 \times A_7^2 \times U(1)^2,  $ \\ $
A_4^2 \times A_7 \times U(1)^2, A_3 \times A_6^2 \times U(1)^2, A_5^3 \times U(1)^2  $
\end{tabular} \\
\hline
 $A_{17}$ & \begin{tabular}{c} $1 \bmod 3$ \\ $2 \bmod 3$ \end{tabular} & \begin{tabular}{c} $ 
A_{16} \times U(1), A_1 \times A_{15} \times U(1), A_1 \times A_{14} \times U(1)^2, A_{13} \times A_3 \times U(1), A_{13} \times A_2 \times U(1)^2,  $ \\ $
A_1 \times A_{12} \times A_2 \times U(1)^2, A_{11} \times A_4 \times U(1)^2, A_{10} \times A_6 \times U(1), A_{12} \times A_4 \times U(1),  $ \\ $
A_1 \times A_{11} \times A_3 \times U(1)^2, A_{10} \times A_2 \times A_3 \times U(1)^2, A_1 \times A_5 \times A_9 \times U(1)^2 ,  $ \\ $
A_7 \times A_8 \times U(1)^2, A_{10} \times A_5 \times U(1)^2, A_2 \times A_4 \times A_9 \times U(1)^2, A_3 \times A_4 \times A_8 \times U(1)^2,  $ \\ $
A_2 \times A_6 \times A_7 \times U(1)^2, A_7 \times A_9 \times U(1), A_1 \times A_6 \times A_8 \times U(1)^2, A_3 \times A_5 \times A_7 \times U(1)^2,  $ \\ $
A_4 \times A_5 \times A_6 \times U(1)^2 $
\end{tabular} \\
\hline
 $A_{18}$ & All & \begin{tabular}{c} $ 
A_{18}, A_{17} \times U(1), A_1 \times A_{16} \times U(1), A_1 \times A_{15} \times U(1)^2, A_{14} \times A_2 \times U(1)^2, A_{13} \times A_4 \times U(1),  $ \\ $
A_{14} \times A_3 \times U(1), A_1 \times A_{13} \times A_2 \times U(1)^2, A_1 \times A_{12} \times A_3 \times U(1)^2, A_{11} \times A_5 \times U(1)^2,  $ \\ $
A_{10} \times A_7 \times U(1), A_{12} \times A_4 \times U(1)^2, A_{11} \times A_2 \times A_3 \times U(1)^2, A_{10} \times A_2 \times A_4 \times U(1)^2,  $ \\ $
A_1 \times A_6 \times A_9 \times U(1)^2, A_8^2 \times U(1)^2, A_{11} \times A_6 \times U(1), A_1 \times A_{10} \times A_5 \times U(1)^2 ,  $ \\ $
A_3 \times A_4 \times A_9 \times U(1)^2, A_3 \times A_5 \times A_8 \times U(1)^2, A_2 \times A_7^2 \times U(1)^2, A_7 \times A_9 \times U(1)^2,  $ \\ $
A_2 \times A_6 \times A_8 \times U(1)^2, A_4 \times A_5 \times A_7 \times U(1)^2, A_4 \times A_6^2 \times U(1)^2, A_8 \times A_9 \times U(1),  $ \\ $
A_1 \times A_7 \times A_8 \times U(1)^2, A_3 \times A_6 \times A_7 \times U(1)^2, A_5^2 \times A_6 \times U(1)^2, A_{10} \times A_6 \times U(1)^2,  $ \\ $
A_2 \times A_5 \times A_9 \times U(1)^2, A_4^2 \times A_8 \times U(1)^2, A_{12} \times A_5 \times U(1), A_1 \times A_{11} \times A_4 \times U(1)^2,  $ \\ $
A_{10} \times A_3^2 \times U(1)^2, A_{13} \times A_3 \times U(1)^2, A_{12} \times A_2^2 \times U(1)^2, A_{15} \times A_2 \times U(1),  $ \\ $
A_1^2 \times A_{14} \times U(1)^2, A_{16} \times U(1)^2 $
\end{tabular} \\
\hline
 $A_{19}$ & All & \begin{tabular}{c} $ 
A_{19}, A_{18} \times U(1), A_1 \times A_{17} \times U(1), A_1 \times A_{16} \times U(1)^2, A_{15} \times A_3 \times U(1), A_{15} \times A_2 \times U(1)^2,  $ \\ $
A_1 \times A_{14} \times A_2 \times U(1)^2, A_{13} \times A_4 \times U(1)^2, A_{12} \times A_6 \times U(1), A_{14} \times A_4 \times U(1),  $ \\ $
A_1 \times A_{13} \times A_3 \times U(1)^2, A_{12} \times A_2 \times A_3 \times U(1)^2, A_1 \times A_{11} \times A_5 \times U(1)^2,  $ \\ $
A_{10} \times A_7 \times U(1)^2, A_9^2 \times U(1), A_{12} \times A_5 \times U(1)^2, A_{11} \times A_2 \times A_4 \times U(1)^2,  $ \\ $
A_{10} \times A_3 \times A_4 \times U(1)^2, A_2 \times A_6 \times A_9 \times U(1)^2, A_1 \times A_8^2 \times U(1)^2, A_{11} \times A_7 \times U(1),  $ \\ $
A_1 \times A_{10} \times A_6 \times U(1)^2, A_3 \times A_5 \times A_9 \times U(1)^2, A_4 \times A_5 \times A_8 \times U(1)^2,  $ \\ $
A_3 \times A_7^2 \times U(1)^2, A_8 \times A_9 \times U(1)^2, A_2 \times A_7 \times A_8 \times U(1)^2, A_4 \times A_6 \times A_7 \times U(1)^2,  $ \\ $
A_5 \times A_6^2 \times U(1)^2, A_{10} \times A_8 \times U(1), A_1 \times A_7 \times A_9 \times U(1)^2, A_3 \times A_6 \times A_8 \times U(1)^2,  $ \\ $
A_5^2 \times A_7 \times U(1)^2, A_{11} \times A_6 \times U(1)^2, A_{10} \times A_2 \times A_5 \times U(1)^2, A_4^2 \times A_9 \times U(1)^2,  $ \\ $
A_{13} \times A_5 \times U(1), A_1 \times A_{12} \times A_4 \times U(1)^2, A_{11} \times A_3^2 \times U(1)^2, A_{14} \times A_3 \times U(1)^2,  $ \\ $
A_{13} \times A_2^2 \times U(1)^2, A_{16} \times A_2 \times U(1), A_1^2 \times A_{15} \times U(1)^2, A_{17} \times U(1)^2 $ 
\end{tabular} \\
\hline
 $A_{20}$ & $0 \bmod 3$ & \begin{tabular}{c} $ 
A_{20}, A_{18} \times U(1)^2, A_{17} \times A_2 \times U(1), A_1^2 \times A_{16} \times U(1)^2, A_{15} \times A_3 \times U(1)^2, A_{14} \times A_5 \times U(1),  $ \\ $
A_{14} \times A_2^2 \times U(1)^2, A_1 \times A_{13} \times A_4 \times U(1)^2, A_{12} \times A_6 \times U(1)^2, A_{11} \times A_8 \times U(1),  $ \\ $
A_{12} \times A_3^2 \times U(1)^2, A_{11} \times A_2 \times A_5 \times U(1)^2, A_1 \times A_{10} \times A_7 \times U(1)^2, A_9^2 \times U(1)^2,  $ \\ $
A_{10} \times A_4^2 \times U(1)^2, A_3 \times A_6 \times A_9 \times U(1)^2, A_2 \times A_8^2 \times U(1)^2, A_5^2 \times A_8 \times U(1)^2,  $ \\ $
A_4 \times A_7^2 \times U(1)^2, A_6^3 \times U(1)^2 $
\end{tabular} \\
\hline
 $A_{20}$ & \begin{tabular}{c} $1 \bmod 3$ \\ $2 \bmod 3$ \end{tabular} & \begin{tabular}{c} $ 
A_{19} \times U(1), A_1 \times A_{18} \times U(1), A_1 \times A_{17} \times U(1)^2, A_{16} \times A_3 \times U(1), A_{16} \times A_2 \times U(1)^2,  $ \\ $
A_1 \times A_{15} \times A_2 \times U(1)^2, A_{14} \times A_4 \times U(1)^2, A_{13} \times A_6 \times U(1), A_{15} \times A_4 \times U(1),  $ \\ $
A_1 \times A_{14} \times A_3 \times U(1)^2,  \times A_{13} \times A_2 \times A_3 \times U(1)^2, A_1 \times A_{12} \times A_5 \times U(1)^2,  $ \\ $
A_{11} \times A_7 \times U(1)^2, A_{10} \times A_9 \times U(1), A_{13} \times A_5 \times U(1)^2, A_{12} \times A_2 \times A_4 \times U(1)^2,  $ \\ $
A_{11} \times A_3 \times A_4 \times U(1)^2, A_{10} \times A_2 \times A_6 \times U(1)^2, A_1 \times A_8 \times A_9 \times U(1)^2, $ \\ $
A_{12} \times A_7 \times U(1), A_1 \times A_{11} \times A_6 \times U(1)^2, A_{10} \times A_3 \times A_5 \times U(1)^2,  $ \\ $
A_4 \times A_5 \times A_9 \times U(1)^2, A_3 \times A_7 \times A_8 \times U(1)^2, A_{10} \times A_8 \times U(1)^2,  $ \\ $
A_2 \times A_7 \times A_9 \times U(1)^2, A_4 \times A_6 \times A_8 \times U(1)^2, A_5 \times A_6 \times A_7 \times U(1)^2 $
\end{tabular} \\
\hline
\end{tabular}
\caption[smallcaption]{$A_n (n=1 \ldots 22)$ group breaking by a ${\bf Z}_3$ shift action (continued).}
\label{Tab.AnGroupBreaking2}
\end{center}
\end{table}

\begin{table}[h]
\begin{center}
\scriptsize
\begin{tabular}{|c|c|c|}
\hline
Group & C.C & Breaking patterns \\
\hline
\hline
 $A_{21}$ & All & \begin{tabular}{c} $ 
A_{21}, A_{20} \times U(1), A_1 \times A_{19} \times U(1), A_1 \times A_{18} \times U(1)^2, A_{17} \times A_2 \times U(1)^2, A_{16} \times A_4 \times U(1),  $ \\ $
A_{17} \times A_3 \times U(1), A_1 \times A_{16} \times A_2 \times U(1)^2, A_1 \times A_{15} \times A_3 \times U(1)^2, A_{14} \times A_5 \times U(1)^2,  $ \\ $
A_{13} \times A_7 \times U(1), A_{15} \times A_4 \times U(1)^2, A_{14} \times A_2 \times A_3 \times U(1)^2, A_{13} \times A_2 \times A_4 \times U(1)^2,  $ \\ $
A_1 \times A_{12} \times A_6 \times U(1)^2, A_{11} \times A_8 \times U(1)^2, A_{10}^2 \times U(1), A_{14} \times A_6 \times U(1),  $ \\ $
A_1 \times A_{13} \times A_5 \times U(1)^2, A_{12} \times A_3 \times A_4 \times U(1)^2, A_{11} \times A_3 \times A_5 \times U(1)^2,  $ \\ $
A_{10} \times A_2 \times A_7 \times U(1)^2, A_1 \times A_9^2 \times U(1)^2, A_{12} \times A_7 \times U(1)^2, A_{11} \times A_2 \times A_6 \times U(1)^2,  $ \\ $
A_{10} \times A_4 \times A_5 \times U(1)^2, A_4 \times A_6 \times A_9 \times U(1)^2, A_3 \times A_8^2 \times U(1)^2, A_{11} \times A_9 \times U(1),  $ \\ $
A_1 \times A_{10} \times A_8 \times U(1)^2, A_3 \times A_7 \times A_9 \times U(1)^2, A_5 \times A_6 \times A_8 \times U(1)^2,  $ \\ $
A_5 \times A_7^2 \times U(1)^2, A_{10} \times A_9 \times U(1)^2, A_2 \times A_8 \times A_9 \times U(1)^2, A_4 \times A_7 \times A_8 \times U(1)^2, $ \\ $
A_6^2 \times A_7 \times U(1)^2, A_{12} \times A_8 \times U(1), A_1 \times A_{11} \times A_7 \times U(1)^2, A_{10} \times A_3 \times A_6 \times U(1)^2,  $ \\ $
A_5^2 \times A_9 \times U(1)^2, A_{13} \times A_6 \times U(1)^2, A_{12} \times A_2 \times A_5 \times U(1)^2, A_{11} \times A_4^2 \times U(1)^2,  $ \\ $
A_{15} \times A_5 \times U(1), A_1 \times A_{14} \times A_4 \times U(1)^2, A_{13} \times A_3^2 \times U(1)^2, A_{16} \times A_3 \times U(1)^2, $ \\ $
A_{15} \times A_2^2 \times U(1)^2, A_{18} \times A_2 \times U(1), A_1^2 \times A_{17} \times U(1)^2, A_{19} \times U(1)^2 $
\end{tabular} \\
\hline
 $A_{22}$ & All & \begin{tabular}{c} $ 
A_{22}, A_{21} \times U(1), A_1 \times A_{20} \times U(1), A_1 \times A_{19} \times U(1)^2, A_{18} \times A_3 \times U(1), A_{18} \times A_2 \times U(1)^2,  $ \\ $
A_1 \times A_{17} \times A_2 \times U(1)^2, A_{16} \times A_4 \times U(1)^2, A_{15} \times A_6 \times U(1), A_{17} \times A_4 \times U(1),  $ \\ $
A_1 \times A_{16} \times A_3 \times U(1)^2, A_{15} \times A_2 \times A_3 \times U(1)^2, A_1 \times A_{14} \times A_5 \times U(1)^2, $ \\ $
A_{13} \times A_7 \times U(1)^2, A_{12} \times A_9 \times U(1), A_{15} \times A_5 \times U(1)^2, A_{14} \times A_2 \times A_4 \times U(1)^2,  $ \\ $
A_{13} \times A_3 \times A_4 \times U(1)^2, A_{12} \times A_2 \times A_6 \times U(1)^2, A_1 \times A_{11} \times A_8 \times U(1)^2,  $ \\ $
A_{10}^2 \times U(1)^2, A_{14} \times A_7 \times U(1), A_1 \times A_{13} \times A_6 \times U(1)^2, A_{12} \times A_3 \times A_5 \times U(1)^2,  $ \\ $
A_{11} \times A_4 \times A_5 \times U(1)^2, A_{10} \times A_3 \times A_7 \times U(1)^2, A_2 \times A_9^2 \times U(1)^2, A_{12} \times A_8 \times U(1)^2,  $ \\ $
A_{11} \times A_2 \times A_7 \times U(1)^2, A_{10} \times A_4 \times A_6 \times U(1)^2, A_5 \times A_6 \times A_9 \times U(1)^2,  $ \\ $
A_4 \times A_8^2 \times U(1)^2, A_{10} \times A_{11} \times U(1), A_1 \times A_{10} \times A_9 \times U(1)^2, A_3 \times A_8 \times A_9 \times U(1)^2,  $ \\ $
A_5 \times A_7 \times A_8 \times U(1)^2, A_6 \times A_7^2 \times U(1)^2, A_{11} \times A_9 \times U(1)^2, A_{10} \times A_2 \times A_8 \times U(1)^2, $ \\ $
 A_4 \times A_7 \times A_9 \times U(1)^2, A_6^2 \times A_8 \times U(1)^2, A_{13} \times A_8 \times U(1), A_1 \times A_{12} \times A_7 \times U(1)^2,  $ \\ $
A_{11} \times A_3 \times A_6 \times U(1)^2, A_{10} \times A_5^2 \times U(1)^2, A_{14} \times A_6 \times U(1)^2, A_{13} \times A_2 \times A_5 \times U(1)^2,  $ \\ $
 A_{12} \times A_4^2 \times U(1)^2, A_{16} \times A_5 \times U(1), A_1 \times A_{15} \times A_4 \times U(1)^2, A_{14} \times A_3^2 \times U(1)^2 ,  $ \\ $
A_{17} \times A_3 \times U(1)^2, A_{16} \times A_2^2 \times U(1)^2, A_{19} \times A_2 \times U(1), A_1^2 \times A_{18} \times U(1)^2, A_{20} \times U(1)^2 $  
\end{tabular} \\
\hline
\end{tabular}
\caption[smallcaption]{$A_n (n=1 \ldots 22)$ group breaking by a ${\bf Z}_3$ shift action (continued).}
\label{Tab.AnGroupBreaking3}
\end{center}
\end{table}

\begin{table}[h]
\begin{center}
\scriptsize
\begin{tabular}{|c|c|c|}
\hline
Group & C.C & Breaking patterns \\
\hline
\hline
 $D_4$ & All & $ D_4, A_3 \times U(1), A_1^3 \times U(1), A_2 \times U(1)^2 $ \\
\hline
 $D_5$ & All & $ D_5, A_3 \times U(1)^2, A_1 \times A_3 \times U(1), A_4 \times U(1), A_1^2 \times A_2 \times U(1), D_4 \times U(1) $ \\
\hline
 $D_6$ & All & $ D_6, A_5 \times U(1), A_1^2 \times A_3 \times U(1), A_1 \times D_4 \times U(1), A_4 \times U(1)^2, A_2 \times A_3 \times U(1), D_5 \times U(1) $ \\
\hline
 $D_7$ & All & $ D_7, A_5 \times U(1)^2, A_3^2 \times U(1), A_1 \times D_5 \times U(1), A_6 \times U(1), A_1^2 \times A_4 \times U(1), A_2 \times D_4 \times U(1), D_6 \times U(1) $ \\
\hline
 $D_8$ & All & \begin{tabular}{c} $ D_8, A_7 \times U(1), A_1^2 \times A_5 \times U(1), A_3 \times D_4 \times U(1), A_1 \times D_6 \times U(1), $ \\ $ A_6 \times U(1)^2, A_3 \times A_4 \times U(1), A_2 \times D_5 \times U(1), D_7 \times U(1) $ \end{tabular} \\
\hline
 $D_9$ & All & \begin{tabular}{c} $ D_9, A_7 \times U(1)^2, A_3 \times A_5 \times U(1), A_3 \times D_5 \times U(1), A_1 \times D_7 \times U(1), $ \\ $ A_8 \times U(1), A_1^2 \times A_6 \times U(1), A_4 \times D_4 \times U(1), A_2 \times D_6 \times U(1), D_8 \times U(1)  $ \end{tabular} \\
\hline
 $D_{10}$ & All & \begin{tabular}{c} $ D_{10}, A_9 \times U(1), A_1^2 \times A_7 \times U(1), A_5 \times D_4 \times U(1), A_3 \times D_6 \times U(1), A_1 \times D_8 \times U(1), $ \\ $ A_8 \times U(1)^2, A_3 \times A_6 \times U(1), A_4 \times D_5 \times U(1), A_2 \times D_7 \times U(1), D_9 \times U(1) $ \end{tabular} \\
\hline
 $D_{11}$ & All & \begin{tabular}{c} $ D_{11}, A_9 \times U(1)^2, A_3 \times A_7 \times U(1), A_5 \times D_5 \times U(1), A_3 \times D_7 \times U(1), A_1 \times D_9 \times U(1), $ \\ $ A_{10} \times U(1), A_1^2 \times A_8 \times U(1), A_6 \times D_4 \times U(1), A_4 \times D_6 \times U(1), A_2 \times D_8 \times U(1), D_{10} \times U(1) $ \end{tabular} \\
\hline
 $D_{12}$ & All & \begin{tabular}{c} $ D_{12}, A_{11} \times U(1), A_1^2 \times A_9 \times U(1), A_7 \times D_4 \times U(1), A_5 \times D_6 \times U(1),  $ \\ $
 A_3 \times D_8 \times U(1), A_1 \times D_{10} \times U(1), A_{10} \times U(1)^2, A_3 \times A_8 \times U(1), A_6 \times D_5 \times U(1), $ \\ $
A_4 \times D_7 \times U(1), A_2 \times D_9 \times U(1), D_{11} \times U(1) $ \end{tabular} \\
\hline
 $D_{13}$ & All & \begin{tabular}{c} $ D_{13}, A_{11} \times U(1)^2, A_3 \times A_9 \times U(1), A_7 \times D_5 \times U(1), A_5 \times D_7 \times U(1), A_3 \times D_9 \times U(1), $ \\ $ A_1 \times D_{11} \times U(1), A_{12} \times U(1), A_1^2 \times A_{10} \times U(1), A_8 \times D_4 \times U(1), A_6 \times D_6 \times U(1), A_4 \times D_8 \times U(1), $ \\ $ A_2 \times D_{10} \times U(1), D_{12} \times U(1) $ \end{tabular} \\
\hline
 $D_{14}$ & All & \begin{tabular}{c} $ D_{14}, A_{13} \times U(1), A_1^2 \times A_{11} \times U(1), A_9 \times D_4 \times U(1), A_7 \times D_6 \times U(1), A_5 \times D_8 \times U(1), $ \\ $ A_3 \times D_{10} \times U(1), A_1 \times D_{12} \times U(1), A_{12} \times U(1)^2, A_{10} \times A_3 \times U(1), A_8 \times D_5 \times U(1), A_6 \times D_7 \times U(1), $ \\ $ A_4 \times D_9 \times U(1), A_2 \times D_{11} \times U(1), D_{13} \times U(1) $ \end{tabular} \\
\hline
 $D_{15}$ & All & \begin{tabular}{c} $ D_{15}, A_{13} \times U(1)^2, A_{11} \times A_3 \times U(1), A_9 \times D_5 \times U(1), A_7 \times D_7 \times U(1), A_5 \times D_9 \times U(1), $ \\ $ A_3 \times D_{11} \times U(1), A_1 \times D_{13} \times U(1), A_{14} \times U(1), A_1^2 \times A_{12} \times U(1), A_{10} \times D_4 \times U(1), A_8 \times D_6 \times U(1), $ \\ $ A_6 \times D_8 \times U(1), A_4 \times D_{10} \times U(1), A_2 \times D_{12} \times U(1), D_{14} \times U(1) $ \end{tabular} \\
\hline
 $D_{16}$ & All & \begin{tabular}{c} $ D_{16}, A_{15} \times U(1), A_1^2 \times A_{13} \times U(1), A_{11} \times D_4 \times U(1), A_9 \times D_6 \times U(1), A_7 \times D_8 \times U(1), $ \\ $ A_5 \times D_{10} \times U(1), A_3 \times D_{12} \times U(1), A_1 \times D_{14} \times U(1), A_{14} \times U(1)^2, A_{12} \times A_3 \times U(1), A_{10} \times D_5 \times U(1), $ \\ $ A_8 \times D_7 \times U(1),  A_6 \times D_9 \times U(1), A_4 \times D_{11} \times U(1), A_2 \times D_{13} \times U(1), D_{15} \times U(1) $ \end{tabular} \\
\hline
 $D_{17}$ & All & \begin{tabular}{c}  $ D_{17}, A_{15} \times U(1)^2, A_{13} \times A_3 \times U(1), A_{11} \times D_5 \times U(1), A_9 \times D_7 \times U(1), A_7 \times D_9 \times U(1), $ \\ $ A_5 \times D_{11} \times U(1), A_3 \times D_{13} \times U(1), A_1 \times D_{15} \times U(1), A_{16} \times U(1), A_1^2 \times A_{14} \times U(1), A_{12} \times D_4 \times U(1), $ \\ $ A_{10} \times D_6 \times U(1), A_8 \times D_8 \times U(1), A_6 \times D_{10} \times U(1), A_4 \times D_{12} \times U(1), A_2 \times D_{14} \times U(1), D_{16} \times U(1) $ \end{tabular} \\
\hline
 $D_{18}$ & All & \begin{tabular}{c} $ D_{18}, A_{17} \times U(1), A_1^2 \times A_{15} \times U(1), A_{13} \times D_4 \times U(1), A_{11} \times D_6 \times U(1), A_9 \times D_8 \times U(1), $ \\ $ A_7 \times D_{10} \times U(1), A_5 \times D_{12} \times U(1), A_3 \times D_{14} \times U(1), A_1 \times D_{16} \times U(1), A_{16} \times U(1)^2, $ \\ $ A_{14} \times A_3 \times U(1), A_{12} \times D_5 \times U(1),  A_{10} \times D_7 \times U(1), A_8 \times D_9 \times U(1), A_6 \times D_{11} \times U(1), $ \\ $ A_4 \times D_{13} \times U(1), A_2 \times D_{15} \times U(1), D_{17} \times U(1) $ \end{tabular} \\
\hline
 $D_{19}$ & All & \begin{tabular}{c} $ 
D_{19}, A_{17} \times U(1)^2, A_{15} \times A_3 \times U(1), A_{13} \times D_5 \times U(1), A_{11} \times D_7 \times U(1), $ \\ $
A_9 \times D_9 \times U(1),  A_7 \times D_{11} \times U(1), A_5 \times D_{13} \times U(1), A_3 \times D_{15} \times U(1), $ \\ $
A_1 \times D_{17} \times U(1), A_{18} \times U(1), A_1^2 \times A_{16} \times U(1),  A_{14} \times D_4 \times U(1), $ \\ $
A_{12} \times D_6 \times U(1), A_{10} \times D_8 \times U(1), A_8 \times D_{10} \times U(1), A_6 \times D_{12} \times U(1), $ \\ $
A_4 \times D_{14} \times U(1), $ \\ $ A_2 \times D_{16} \times U(1), D_{18} \times U(1) $ \end{tabular} \\
\hline
 $D_{20}$ & All & \begin{tabular}{c} $ D_{20}, A_{19} \times U(1), A_1^2 \times A_{17} \times U(1), A_{15} \times D_4 \times U(1), A_{13} \times D_6 \times U(1), A_{11} \times D_8 \times U(1), $ \\ $ A_9 \times D_{10} \times U(1), A_7 \times D_{12} \times U(1), A_5 \times D_{14} \times U(1), A_3 \times D_{16} \times U(1), A_1 \times D_{18} \times U(1), $ \\ $ A_{18} \times U(1)^2, A_{16} \times A_3 \times U(1), A_{14} \times D_5 \times U(1), A_{12} \times D_7 \times U(1), A_{10} \times D_9 \times U(1), A_8 \times D_{11} \times U(1), $ \\ $ A_6 \times D_{13} \times U(1), A_4 \times D_{15} \times U(1), A_2 \times D_{17} \times U(1), D_{19} \times U(1) $ \end{tabular} \\
\hline
 $D_{21}$ & All & \begin{tabular}{c} $ D_{21}, A_{19} \times U(1)^2, A_{17} \times A_3 \times U(1), A_{15} \times D_5 \times U(1), A_{13} \times D_7 \times U(1), A_{11} \times D_9 \times U(1), $ \\ $ A_9 \times D_{11} \times U(1), A_7 \times D_{13} \times U(1), A_5 \times D_{15} \times U(1), A_3 \times D_{17} \times U(1), A_1 \times D_{19} \times U(1), $ \\ $ A_{20} \times U(1), A_1^2 \times A_{18} \times U(1), A_{16} \times D_4 \times U(1), A_{14} \times D_6 \times U(1), A_{12} \times D_8 \times U(1), A_{10} \times D_{10} \times U(1), $ \\ $ A_8 \times D_{12} \times U(1), A_6 \times D_{14} \times U(1), A_4 \times D_{16} \times U(1), A_2 \times D_{18} \times U(1), D_{20} \times U(1) $ \end{tabular} \\
\hline
 $D_{22}$ & All & \begin{tabular}{c} $ 
D_{22}, A_{21} \times U(1), A_1^2 \times A_{19} \times U(1), A_{17} \times D_4 \times U(1), A_{15} \times D_6 \times U(1), $ \\ $
A_{13} \times D_8 \times U(1),  A_{11} \times D_{10} \times U(1), A_9 \times D_{12} \times U(1), A_7 \times D_{14} \times U(1), $ \\ $
A_5 \times D_{16} \times U(1), A_3 \times D_{18} \times U(1), A_1 \times D_{20} \times U(1),  A_{20} \times U(1)^2, A_{18} \times A_3 \times U(1), $ \\ $
A_{16} \times D_5 \times U(1), A_{14} \times D_7 \times U(1), A_{12} \times D_9 \times U(1), A_{10} \times D_{11} \times U(1), $ \\ $ 
A_8 \times D_{13} \times U(1), A_6 \times D_{15} \times U(1), A_4 \times D_{17} \times U(1), A_2 \times D_{19} \times U(1), D_{21} \times U(1) $ \end{tabular} \\
\hline
\end{tabular}
\caption[smallcaption]{$D_n (n=4 \ldots 22)$ group breaking by a ${\bf Z}_3$ shift action.}
\label{Tab.DnGroupBreaking}
\end{center}
\end{table}


\begin{table}[h]
\begin{center}
\scriptsize
\begin{tabular}{|c|c|c|c|c|c|}
\hline
 Group &  SM  & Flipped $ SO(10) $ & Flipped $ SU(5) $ & Pati-Salam & Left-right symmetric  \\
\hline
\hline
 \#1 &  & \checkmark & \checkmark &  &  \\
\hline
 \#2 & \checkmark & \checkmark & \checkmark &  & \checkmark \\
\hline
 \#3 & \checkmark & \checkmark & \checkmark &  & \checkmark \\
\hline
 \#4 &  &  &  &  &  \\
\hline
 \#5 & \checkmark &  & \checkmark &  &  \\
\hline
 \#6 & \checkmark & \checkmark & \checkmark & \checkmark & \checkmark \\
\hline
 \#7 & \checkmark & \checkmark & \checkmark &  & \checkmark \\
\hline
 \#8 & \checkmark &  & \checkmark & \checkmark & \checkmark \\
\hline
 \#9 & \checkmark & \checkmark & \checkmark & \checkmark & \checkmark \\
\hline
 \#10 & \checkmark & \checkmark & \checkmark & \checkmark & \checkmark \\
\hline
 \#11 & \checkmark & \checkmark & \checkmark & \checkmark & \checkmark \\
\hline
 \#12 & \checkmark & \checkmark & \checkmark & \checkmark & \checkmark \\
\hline
 \#13 & \checkmark & \checkmark & \checkmark & \checkmark & \checkmark \\
\hline
 \#14 & \checkmark &  & \checkmark & \checkmark & \checkmark \\
\hline
 \#15 & \checkmark & \checkmark & \checkmark & \checkmark & \checkmark \\
\hline
 \#16 & \checkmark & \checkmark & \checkmark & \checkmark & \checkmark \\
\hline
 \#17 & \checkmark & \checkmark & \checkmark & \checkmark & \checkmark \\
\hline
 \#18 & \checkmark & \checkmark & \checkmark &  & \checkmark \\
\hline
 \#19 & \checkmark & \checkmark & \checkmark & \checkmark & \checkmark \\
\hline
 \#20 & \checkmark &  & \checkmark & \checkmark & \checkmark \\
\hline
 \#21 & \checkmark & \checkmark & \checkmark & \checkmark & \checkmark \\
\hline
 \#22 & \checkmark &  & \checkmark & \checkmark & \checkmark \\
\hline
 \#23 & \checkmark & \checkmark & \checkmark & \checkmark & \checkmark \\
\hline
 \#24 & \checkmark & \checkmark & \checkmark & \checkmark & \checkmark \\
\hline
 \#25 & \checkmark &  & \checkmark & \checkmark & \checkmark \\
\hline
 \#26 & \checkmark &  & \checkmark & \checkmark & \checkmark \\
\hline
 \#27 & \checkmark &  & \checkmark & \checkmark & \checkmark \\
\hline
 \#28 & \checkmark &  &  & \checkmark & \checkmark \\
\hline
 \#29 & \checkmark &  & \checkmark & \checkmark & \checkmark \\
\hline
 \#30 & \checkmark &  &  & \checkmark & \checkmark \\
\hline
 \#31 & \checkmark &  &  &  & \checkmark \\
\hline
 \#32 & \checkmark & \checkmark & \checkmark &  & \checkmark \\
\hline
 \#33 &  &  &  &  &  \\
\hline
 \#34 & \checkmark &  & \checkmark &  & \checkmark \\
\hline
 \#35 & \checkmark & \checkmark & \checkmark &  & \checkmark \\
\hline
 \#36 & \checkmark & \checkmark & \checkmark & \checkmark & \checkmark \\
\hline
 \#37 & \checkmark & \checkmark & \checkmark &  & \checkmark \\
\hline
 \#38 & \checkmark & \checkmark & \checkmark &  & \checkmark \\
\hline
 \#39 & \checkmark & \checkmark & \checkmark &  & \checkmark \\
\hline
 \#40 & \checkmark & \checkmark & \checkmark &  & \checkmark \\
\hline
 \#41 &  &  &  &  &  \\
\hline
 \#42 & \checkmark &  & \checkmark &  & \checkmark \\
\hline
 \#43 & \checkmark & \checkmark & \checkmark & \checkmark & \checkmark \\
\hline
 \#44 & \checkmark & \checkmark & \checkmark & \checkmark & \checkmark \\
\hline
 \#45 & \checkmark & \checkmark & \checkmark &  & \checkmark \\
\hline
 \#46 & \checkmark &  & \checkmark & \checkmark & \checkmark \\
\hline
 \#47 & \checkmark & \checkmark &  & \checkmark & \checkmark \\
\hline
 \#48 & \checkmark & \checkmark & \checkmark & \checkmark & \checkmark \\
\hline
 \#49 & \checkmark & \checkmark & \checkmark & \checkmark & \checkmark \\
\hline
 \#50 & \checkmark & \checkmark & \checkmark & \checkmark & \checkmark \\
\hline
 \#51 & \checkmark &  & \checkmark & \checkmark & \checkmark \\
\hline
 \#52 & \checkmark & \checkmark & \checkmark & \checkmark & \checkmark \\
\hline
 \#53 & \checkmark & \checkmark & \checkmark & \checkmark & \checkmark \\
\hline
 \#54 & \checkmark & \checkmark & \checkmark & \checkmark & \checkmark \\
\hline
 \#55 & \checkmark &  & \checkmark & \checkmark & \checkmark \\
\hline
\end{tabular}
\caption[smallcaption]{The SM group and typical grand unified groups from (22,6)-dimensional Narain lattices in ${\bf Z}_3$ asymmetric orbifolds.}
\label{Tab:SMGUT1}
\end{center}
\end{table}

\begin{table}[h]
\begin{center}
\scriptsize
\begin{tabular}{|c|c|c|c|c|c|}
\hline
 Group &  SM  & Flipped $ SO(10) $ & Flipped $ SU(5) $ & Pati-Salam & Left-right symmetric  \\
\hline
\hline
 \#56 & \checkmark &  & \checkmark & \checkmark & \checkmark \\
\hline
 \#57 & \checkmark & \checkmark & \checkmark & \checkmark & \checkmark \\
\hline
 \#58 & \checkmark & \checkmark & \checkmark & \checkmark & \checkmark \\
\hline
 \#59 & \checkmark & \checkmark & \checkmark & \checkmark & \checkmark \\
\hline
 \#60 & \checkmark &  & \checkmark & \checkmark & \checkmark \\
\hline
 \#61 & \checkmark & \checkmark & \checkmark & \checkmark & \checkmark \\
\hline
 \#62 & \checkmark & \checkmark & \checkmark &  & \checkmark \\
\hline
 \#63 & \checkmark & \checkmark & \checkmark & \checkmark & \checkmark \\
\hline
 \#64 & \checkmark &  & \checkmark & \checkmark & \checkmark \\
\hline
 \#65 & \checkmark & \checkmark & \checkmark & \checkmark & \checkmark \\
\hline
 \#66 & \checkmark &  & \checkmark & \checkmark & \checkmark \\
\hline
 \#67 & \checkmark & \checkmark & \checkmark & \checkmark & \checkmark \\
\hline
 \#68 & \checkmark & \checkmark & \checkmark & \checkmark & \checkmark \\
\hline
 \#69 & \checkmark &  & \checkmark & \checkmark & \checkmark \\
\hline
 \#70 & \checkmark &  & \checkmark & \checkmark & \checkmark \\
\hline
 \#71 & \checkmark & \checkmark & \checkmark & \checkmark & \checkmark \\
\hline
 \#72 & \checkmark & \checkmark & \checkmark & \checkmark & \checkmark \\
\hline
 \#73 & \checkmark &  & \checkmark & \checkmark & \checkmark \\
\hline
 \#74 & \checkmark & \checkmark & \checkmark & \checkmark & \checkmark \\
\hline
 \#75 & \checkmark &  & \checkmark & \checkmark & \checkmark \\
\hline
 \#76 & \checkmark &  & \checkmark & \checkmark & \checkmark \\
\hline
 \#77 & \checkmark &  & \checkmark & \checkmark & \checkmark \\
\hline
 \#78 & \checkmark &  & \checkmark & \checkmark & \checkmark \\
\hline
 \#79 & \checkmark &  & \checkmark & \checkmark & \checkmark \\
\hline
 \#80 & \checkmark &  &  & \checkmark & \checkmark \\
\hline
 \#81 & \checkmark &  & \checkmark & \checkmark & \checkmark \\
\hline
 \#82 & \checkmark &  &  & \checkmark & \checkmark \\
\hline
 \#83 & \checkmark &  &  &  & \checkmark \\
\hline
 \#84 & \checkmark & \checkmark & \checkmark &  & \checkmark \\
\hline
 \#85 & \checkmark & \checkmark & \checkmark &  & \checkmark \\
\hline
 \#86 & \checkmark & \checkmark & \checkmark & \checkmark & \checkmark \\
\hline
 \#87 & \checkmark & \checkmark & \checkmark & \checkmark & \checkmark \\
\hline
 \#88 & \checkmark & \checkmark & \checkmark & \checkmark & \checkmark \\
\hline
 \#89 & \checkmark & \checkmark & \checkmark & \checkmark & \checkmark \\
\hline
 \#90 & \checkmark & \checkmark & \checkmark & \checkmark & \checkmark \\
\hline
\end{tabular}
\caption[smallcaption]{The SM group and typical grand unified groups from (22,6)-dimensional Narain lattices in ${\bf Z}_3$ asymmetric orbifolds(continued).}
\label{Tab:SMGUT2}
\end{center}
\end{table}

\end{document}